\documentstyle[epsfig]{elsart}

\newcommand{\at}{\overline{10}}

\begin{document}
\begin{frontmatter}
%\vspace*{-2cm}
\begin{flushright}
\bf 
RUB-TPII-20/2005
\end{flushright}

\vspace*{2cm}

\title{SU(3) systematization  of baryons}

\author{V. Guzey} 
\address{Institut f{\"u}r Theoretische Physik II, Ruhr-Universit{\"a}t
Bochum, D-44780 Bochum, Germany}
\ead{vadim.guzey@tp2.ruhr-uni-bochum.de}

\author{M.V. Polyakov}
\address{Institut f{\"u}r Theoretische Physik II, Ruhr-Universit{\"a}t
Bochum, D-44780 Bochum, Germany, and \\
Petersburg Nuclear Physics Institute, Gatchina, St. Petersburg
188300, Russia}
\ead{maxim.polyakov@tp2.ruhr-uni-bochum.de}

\bigskip
\bigskip
\bigskip
\bigskip

\begin{abstract} 
\noindent

We review the spectrum of all baryons with the mass less than 
approximately 2000-2200 MeV using methods based on  
the approximate flavor SU(3) symmetry of the strong interaction.
The application of the Gell-Mann--Okubo mass formulas and SU(3)-symmetric 
predictions for two-body hadronic decays allows us to
successfully catalogue almost all known baryons in twenty-one 
SU(3) multiplets.
In order to have complete multiplets,
we predict the existence of several strange particles,
 most notably the $\Lambda$ hyperon with $J^P=3/2^-$, the mass around
1850 MeV, the total width approximately 130 MeV,
significant branching into the
$\Sigma \pi$ and $\Sigma(1385) \pi$ states
 and a very small coupling to
the $N \overline{K}$  state.
Assuming that the antidecuplet exists, we show how a simple scenario,
 in which the antidecuplet mixes with an octet, allows to understand the
 pattern of the 
antidecuplet decays and make predictions for the unmeasured decays.

\end{abstract}

\end{frontmatter}
%
%\vspace*{.cm}
%\vspace*{\fill}
%\noindent $^{*}$) Work supported by the Sofia Kovalevskaya Program 
%of the Alexander von Humboldt Foundation.

     \newpage
\tableofcontents \newpage

\section{Introduction}
\label{sec:intro}

After some thirty years of recess, hadron spectroscopy becomes again 
a lively and exciting field of research in hadronic physics. 
The recent wave of interest in light baryon spectroscopy was initiated
by the report of the discovery of the explicitly exotic baryon with
strangeness +1, now called the $\Theta^+$~\cite{Nakano:2003qx}.
The properties of the $\Theta^+$ and the other members of the 
antidecuplet were predicted in 
the chiral quark soliton model~\cite{Diakonov:1997mm}.
While at the time of writing of this report, the fate of the $\Theta^+$ and
$\Xi^{--}$~\cite{Alt:2003vb}, yet another explicitly exotic member 
of the antidecuplet,
is uncertain and is a subject of controversy, the $\Theta^+$ has 
nevertheless already contributed to hadron physics by popularizing it and 
forcing critical assessments and advances of the used experimental
methods and theoretical models.

The situation with the $\Theta^+$ has revealed that many more surprises
in the spectrum of light baryons could be hiding. In particular, the 
existence of new light and very narrow nucleon resonances was recently
suggested~\cite{Azimov:2003bb}. In addition,
the problem of missing resonances predicted in the constituent quark model 
as well as the explanation of the light mass of the Roper resonance $N^{\ast}(1440)$
and the $\Lambda(1405)$ still await their solutions,
 see~\cite{Capstick:2000qj} for a recent review.

In light of the expected advances in the baryon spectroscopy,
it is topical to systematize the spectrum of presently known baryons
according to the flavor SU(3) group.
Last time this was done in 1974~\cite{Samios:1974tw} when
many baryons were not yet known.
In this work, we review the spectrum of all baryons with the mass less than 
approximately 2000-2200 MeV using general and almost  model-independent 
methods based on the approximate flavor SU(3) symmetry of strong interactions.
We successfully place almost all known baryons in twenty-one SU(3) multiplets and,
thus, confirm the prediction~\cite{Samios:1974tw} that the approximate
SU(3) symmetry works remarkably well. In order to complete the multiplets,
we predict the existence of several strange particles.
They appear underlined in this review, see e.g.~the Table of Contents.
 Among them, the most
remarkable is the $\Lambda$ hyperon with
$J^P=3/2^-$, the mass around 1850 MeV, the total width $\approx 130$ MeV,
significant branching into the
$\Sigma \pi$ and $\Sigma(1385) \pi$ states
and a very small coupling to the $N \overline{K}$ state.
Our analysis gives a model-independent confirmation of the constituent quark model prediction
that  there should exist a new $\Lambda$ baryon with the mass between 1775 MeV and 1880 MeV,
which almost decouples from the $N {\overline K}$ 
state~\cite{Isgur:1978xj,Loring:2001ky,Glozman:1997ag}.

The hypothesis of the approximate SU(3) symmetry of strong interactions put
 forward by Gell-Mann and Ne'eman in the early 60's was probably the 
most successful and fruitful idea for the systematization of elementary 
particles, see a classic compilation of original papers~\cite{Eightfoldway}.
A natural assumption that the part of the strong interaction that violates
SU(3) symmetry is proportional to the mass of the strange 
quark\footnote{The original formulation used fictitious leptons to build
the fundamental SU(3) representation. The quarks would be invented shortly 
after.} 
resulted in very successful Gell-Mann--Okubo (GMO) mass
formulas describing the mass splitting inside a given SU(3) multiplet. 
The approximate SU(3) symmetry has not only enabled to bring order to
 the spectroscopy of hadrons
but has also allowed to predict new particles, which were later confirmed
 experimentally.  
The most  famous prediction in the baryon sector is
the last member of the ground-state decuplet, $\Omega (1672)$,
 see~\cite{Eightfoldway}, whose spin and parity are still (!) not measured
 but rather predicted using SU(3).

Because of their simplicity and almost model-independence, 
the approximate SU(3) symmetry of strong interactions and the 
resulting systematics of strongly interacting particles have become 
a classic textbook subject, see~\cite{Kokkedee,Lichtenberg,Close}.
Theoretical methods based on the approximate flavor SU(3) symmetry
work very efficiently as a bookkeeping tool. For instance,
 a few still missing baryon resonances can be identified
and their masses and partial decay widths can be predicted, as we
 shall demonstrate.
Since the approach  
does not involve internal degrees of freedom of QCD and makes only very   
general assumptions, its applicability is 
mostly limited to  the description of the
 baryon spectrum, certain gross features of
 strong decays and 
static properties of baryons.
As soon as the microscopic structure of hadrons is of interest,
one needs to use other (dynamical) approaches such as quark models,
 effective theories or lattice gauge calculations.

\underline{{\bf Basics of flavor  SU(3)}}

The exact flavor SU(3) symmetry of strong interactions  predicts the existence
of definite representations or multiplets: 
singlets (\textbf{1}), octets  (\textbf{8}), decuplets (\textbf{10}),
antidecuplets  ($\bf{\overline{10}}$)
twenty-seven-plets (\textbf{27}), thirty-five-plets (\textbf{35}), etc.,
where the numbers in the bold face denote the dimension 
of the representation (the number of particles in the multiplet).
A common feature of these multiplets is that they all have zero
triality~\cite{Lichtenberg}.
Thus, from the pure SU(3) point of view, there is nothing mysterious in the 
existence of the antidecuplet. Actually, exotic baryons with positive
strangeness (then called $Z$-resonances) have been mentioned in the literature
dealing with SU(3) multiplets since the late 60's~\cite{Kokkedee}.

Except for the  antidecuplet, all known hadrons belong to singlet, octet and
decuplet representations, which naturally follows from the Clebsh-Gordan
series for mesons and baryons, respectively,
\begin{eqnarray}
&&{\bf 3} \otimes {\bar {\bf 3}}={\bf 1} \oplus {\bf 8} \,,  \nonumber\\
&&{\bf 3} \otimes {\bf 3} \otimes {\bf 3}=
{\bf 1} \oplus {\bf 8} \oplus {\bf 8} \oplus {\bf 10} \,.
\label{eq:clebsh_gordan}
\end{eqnarray}
All states, which do not belong to singlets, octets or decuplets and have the baryon
number $|B| \leq 1$, are called exotic~\cite{Kokkedee,Lichtenberg}.
For instance, the exotic antidecuplet representation
can only be constructed out of at least
 four quarks and one antiquark
\begin{equation}
\bf{3} \otimes  \bf{3} \otimes \bf{3} \otimes \bf{3} \otimes \bf{{\bar 3}}=(3)\bf{1}\oplus
(8) \bf{8} \oplus (4) \bf{10} \oplus (2) \bf{\overline{10}} \oplus (3) \bf{27}\oplus \bf{35}
 \,.
\label{eq:multiplication}
\end{equation}

\begin{figure}[h]
\begin{center}
\epsfig{file=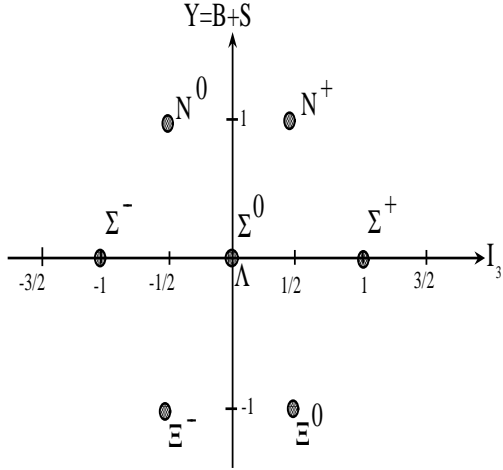,width=15cm,height=15cm}
\vskip -6cm
\caption{SU(3) octet.}
\label{fig:octet}
\end{center}
\end{figure}

\begin{figure}[h]
\begin{center}
\epsfig{file=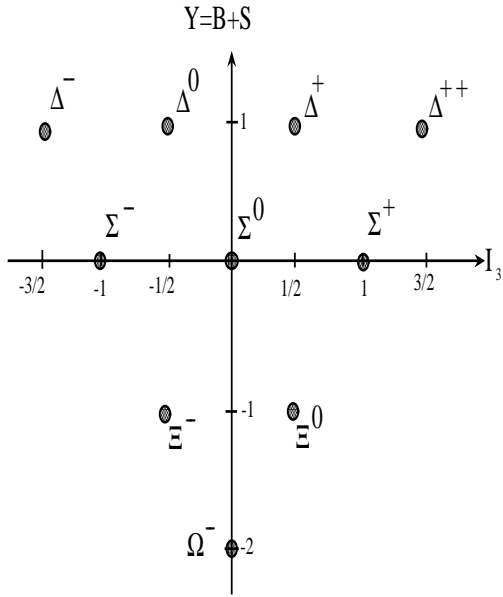,width=15cm,height=15cm}
\vskip -6cm
\caption{SU(3) decuplet.}
\label{fig:decuplet}
\end{center}
\end{figure}

\begin{figure}[h]
\begin{center}
\epsfig{file=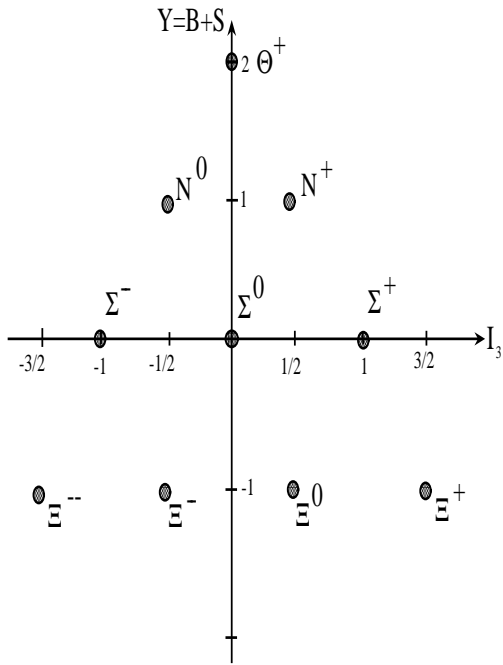,width=15cm,height=15cm}
\vskip -6cm
\caption{SU(3) antidecuplet.}
\label{fig:antidecuplet}
\end{center}
\end{figure}

Every particle in a given SU(3) multiplet is uniquely characterized by its
isospin $I$, the $z$-component of the isospin $I_3$ and hypercharge $Y$. Therefore, 
it is customary to represent the particle content of SU(3) multiplets
in the $I_3-Y$ axes. The octet, decuplet and antidecuplet are presented in 
Figs.~\ref {fig:octet}, \ref{fig:decuplet} and  \ref{fig:antidecuplet}.

Today one groups all experimentally known baryons into 
singlets, octets, decuplets and the antidecuplet (if it is 
experimentally confirmed). 
There is no experimental evidence for the need in other (higher)
 SU(3) representations such as  \textbf{27} and \textbf{35}.
However, these representations are discussed in the literature in relation 
to the antidecuplet~\cite{Ellis:2004uz,Praszalowicz:2004dn,Bijker:2003pm}.

In Nature and in QCD, the flavor SU(3) symmetry is broken by non-equal masses of 
the up and down quarks  and the strange quark. This justifies the hypothesis
 that the SU(3)-violating part of the Hamiltonian
 transforms like
 the eighths component of an octet SU(3) 
representation with zero isospin and hypercharge.
This can be understood by noticing that the mass term in the Hamiltonian,
$m_s {\bar s} s$, where $m_s$ is the mass of the strange quark,
which is responsible for the flavor SU(3) breaking, has no net strangeness and
isospin and transforms under SU(3) like the direct sum of the octet 
and singlet representations, $\bf{3} \otimes \bf{{\bar 3}}=\bf{1} \oplus \bf{8}$.
Then the symmetry breaking is induced by 
the octet component.

For practical applications, it is useful to work in the tensor
notation of deSwart~\cite{deSwart:1963gc}. In this notation,  
each operator has 
the form $T^{\mu}_{Y I}$, where 
$\mu$ labels the SU(3) representation, $Y$ denotes the hypercharge and
$I$ denotes the isospin. Therefore, 
the SU(3) breaking part of the Hamiltonian  is
proportional to $T^8_{00}$.

This observation allows one to write down
a general formula for the mass of the baryon, which belongs 
to the SU(3) multiplet of the
dimension ${\bf \mu}$ and has the hypercharge $Y$  and the isospin $I$, 
as a perturbative expansion in terms of powers of the SU(3)-violating 
Hamiltonian
\begin{equation}
 M_B^\mu=M_0^\mu + \sum_\gamma A_\gamma^\mu \left(
\begin{array}{cc}
\mu & 8 \\
Y I & 0 0
\end{array}\right|\left.\begin{array}{c}
          \mu_\gamma \\ Y I
          \end{array}\right)\, .
          \label{GOgeneral}
\end{equation}
In this equation, we retained only the SU(3)-symmetric term ($M_0^\mu$)
and the term linear in the SU(3) symmetry breaking operator --
the phenomenological constants $A_\gamma^\mu$ are proportional to the 
difference between the
strange quark mass $m_s$ and the mass of the $u$ and $d$ quarks.
Since the resulting representation ${\bf \mu}$ can appear several times in the 
tensor product of the initial ${\bf \mu}$ and ${\bf 8}$, the final 
representation contains the degeneracy label $\gamma$.
The factors in the brackets are the so-called SU(3) isoscalar factors, which
are known for all SU(3) multiplets and transitions~\cite{deSwart:1963gc}.
Equation~(\ref{GOgeneral}) is the most general form  of the Gell-Mann--Okubo 
mass formula for an arbitrary baryon multiplet ${\bf \mu}$.

The tensor product 
${\bf 8} \otimes {\bf 8}$ contains two octet representations,
\begin{equation}
{\bf 8} \otimes {\bf 8}={\bf 1} \oplus {\bf 8} \oplus {\bf 8} \oplus {\bf 10}\oplus \bf{\overline{10}} \oplus {\bf 27} \,,
\label{eq:8plus8}
 \end{equation}
 which
means that,
according to Eq.~(\ref{GOgeneral}), the masses of the $N$, $\Lambda$, $\Sigma$
and $\Xi$ members of any octet can be expressed in terms of three constants:
the overall mass of the octet  $M_0^8$ and two constants $A_1^8$ and $A_2^8$. Therefore, 
evaluating the corresponding isoscalar factors and eliminating the unknown constants,
 one obtains the Gell-Mann--Okubo mass formula 
for the octet
 \begin{equation}
\frac{1}{2}\left( m_N+m_{\Xi} \right)=\frac{1}{4}\left(3\, m_{\Lambda}+m_{\Sigma} \right) \,.
\label{eq:gmo8}
\end{equation}

In the case of decuplets, the tensor product ${\bf 8} \otimes {\bf 10}$ 
contains only one decuplet representation, 
\begin{equation}
{\bf 8} \otimes {\bf 10}={\bf 8} \oplus {\bf 10} \oplus {\bf 27} \oplus {\bf 35} \,,
\label{eq:8plus10}
\end{equation}
which means that the masses 
of the $\Delta$, $\Sigma$, $\Xi$ and $\Omega$ states of any decuplet 
can be expressed in terms of two free constants: $M_0^{10}$ and $A^{10}$.
Consequently, there are two independent relations among the masses --
the so-called equal spacing rule
\begin{equation}
m_{\Sigma}-m_{\Delta}=m_{\Xi}-m_{\Sigma}=m_{\Omega}-m_{\Xi}  \,.
\label{eq:gmo10}
\end{equation}
Along the similar lines,  the equal spacing rule can be derived for the antidecuplet
\begin{equation}
m_{N_{\overline{10}}}-m_{\Theta^+}=m_{\Sigma_{\overline{10}}}-m_{N_{\overline{10}}}= m_{\Xi_{\overline{10}}}-m_{\Sigma_{\overline{10}}} \,.
\label{eq:gmoanti10}
\end{equation}

\underline{{\bf $U$-spin}}

\begin{figure}[h]
\begin{center}
\epsfig{file=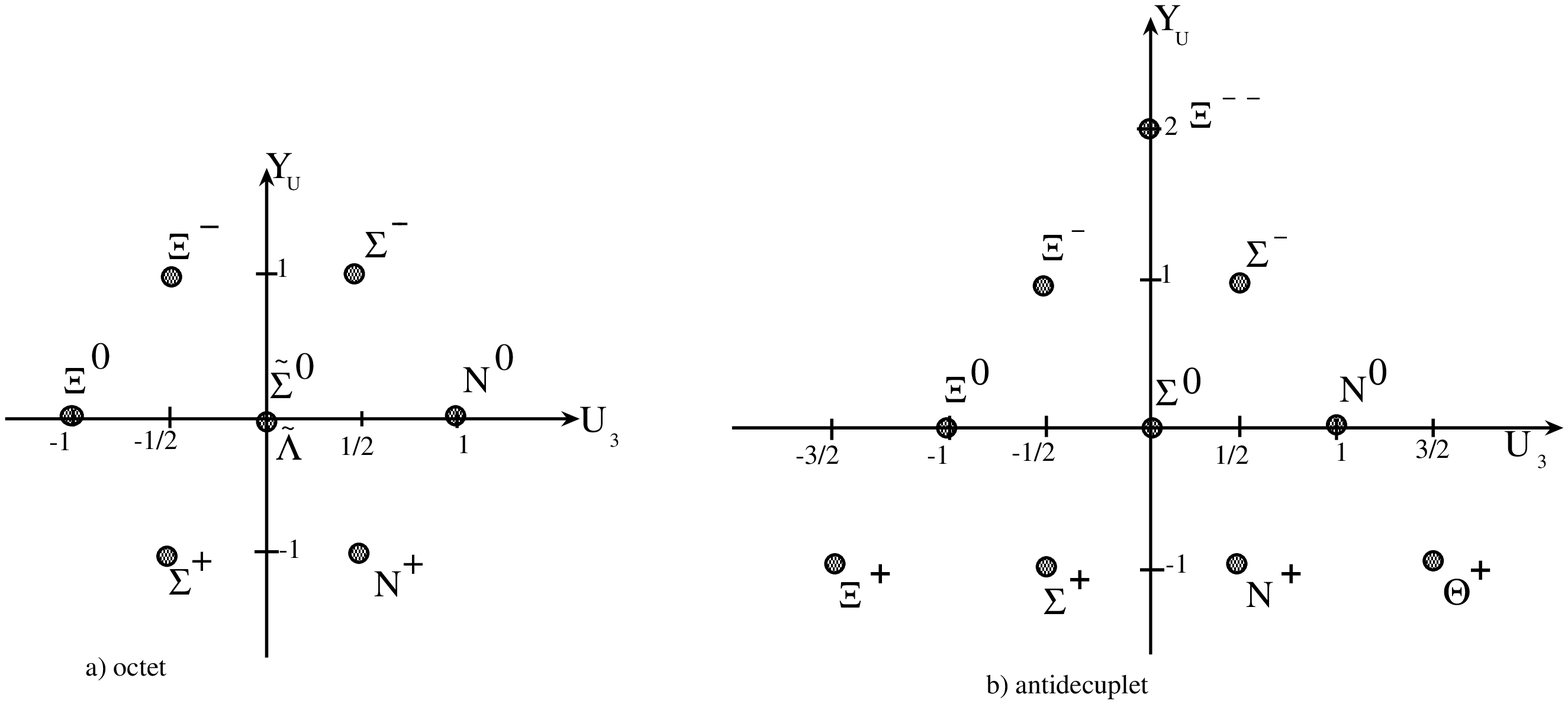,width=15cm,height=15cm}
\vskip -6cm
\caption{An octet and the antidecuplet in the $(U_3,Y_U)$-basis.}
\label{fig:uspin}
\end{center}
\end{figure}

We now briefly discuss a useful concept of the $U$-spin, which makes very
transparent the derivation of SU(3) predictions for the relations among
 the magnetic moments of a given multiplet~\cite{Coleman:1961jn} as well 
as  other SU(3) electromagnetic predictions.
The operator of the electromagnetic current,
\begin{equation}
J_{\mu}=\frac{2}{3} {\bar u} \gamma_{\mu} u-\frac{1}{3} {\bar d} \gamma_{\mu} d-
\frac{1}{3} {\bar s} \gamma_{\mu} s \,,
\label{eq:em}
\end{equation}
contains the $I=0$ and $I=1$ components and, thus, transforms in a complicated 
way under isospin rotations.
 However, if instead of the 
$(I_3,Y)$-basis, one characterizes a given SU(3) multiplet by the 
third component of the so-called $U$-spin and the corresponding 
hypercharge $Y_U$, see e.g.~\cite{Lichtenberg,Novozhilov},
\begin{eqnarray}
U_3&=&-\frac{1}{2} I_3 +\frac{3}{4} Y \,, \nonumber\\
Y_U&=&-Q \,,
\end{eqnarray}
the operator of the electromagnetic current $J_{\mu}$
becomes proportional to $T^{8}_{00}$ since ${\bar u} \gamma_{\mu} u$ is the $U$-spin
singlet and ${\bar d} \gamma_{\mu} d+{\bar s} \gamma_{\mu} s$ is invariant under
$U$-spin rotations.
 Therefore,
in the $(U_3,Y_U)$-basis the derivation of SU(3) relations among 
electromagnetic transition amplitudes proceeds as in the case of the
Gell-Mann--Okubo mass formula. The decomposition of an SU(3) octet and
the antidecuplet in the $(U_3,Y_U)$-basis is presented in Fig.~\ref{fig:uspin}.
In the figure, $\tilde{\Sigma}^0=-(1/2) \Sigma^0+(\sqrt{3}/2) \Lambda$ and
$\tilde{\Lambda}=-(\sqrt{3}/2) \Sigma^0-(1/2) \Lambda$~\cite{Novozhilov}. 

Below we consider two applications of the $U$-spin: SU(3) predictions
for the magnetic moments
of the octet and the transition magnetic moments of the antidecuplet.

It follows from the assumption of the $U$-spin conservation that the
magnetic moments (and electric charges) of all members of the same
$U$-spin multiplet are equal. From the left panel of Fig.~\ref{fig:uspin}, one
then immediately obtains that
\begin{eqnarray}
&&\mu_{\Sigma^-}  =  \mu_{\Xi^-} \,, \nonumber\\
&& \mu_{\Xi^0}  =  \mu_n \,, \nonumber\\
&& \mu_p  =  \mu_{\Sigma^+} \,.
\label{eq:uspin_multiplets}
\end{eqnarray}
Additional relations are obtained from the observation that the operator of
 the electromagnetic current
is proportional to $T^{8}_{00}$ in the $(U_3,Y_U)$-basis. This allows one to 
write the magnetic moments of an SU(3) octet in the form, which closely 
resembles Eq.~(\ref{GOgeneral}) for the octet mass splitting,
\begin{equation}
 \mu_B=\sum_\gamma C_\gamma \left(
\begin{array}{cc}
8 & 8 \\
Y_U U & 0 0
\end{array}\right|\left.\begin{array}{c}
          8_\gamma \\ Y_U U
          \end{array}\right)\, ,
        \label{eq:magnetic_moments}
\end{equation}
where $C_\gamma=1,2$ are two free coefficients.
 A direct evaluation of
 the isoscalar factors in Eq.~(\ref{eq:magnetic_moments}) and the use
of the auxiliary isospin relation $\Sigma^{0}=(\Sigma^+ + \Sigma^-)/2$
allows one to derive the following relations among
the  eight magnetic moments of the octet and the 
$\mu_{\Sigma^0 \Lambda}$ matrix element for the 
$\Sigma^0 \to \Lambda \, \gamma$ transition (we also list the relations
of Eq.~(\ref{eq:uspin_multiplets}) for completeness)
\begin{eqnarray}
&&\mu_{\Sigma^-}  =  \mu_{\Xi^-}=-\mu_p-\mu_n \,, \nonumber\\
&&\mu_{\Xi^0}  =  \mu_n  \,, \nonumber\\
&&\mu_{\Lambda}  =  -\mu_{\Sigma^0}=\frac{1}{2}\, \mu_n \,, \nonumber\\
&&\mu_p  =  \mu_{\Sigma^+} \,, \nonumber\\ 
&&\mu_{\Sigma^0 \Lambda}  =\frac{\sqrt{3}}{2} \, \mu_n \,.
\end{eqnarray}

Turning to the antidecuplet-octet transition magnetic moments, we
immediately see from Fig.~\ref{fig:uspin} that, in the SU(3) limit,
the transition magnetic moments involving $\Xi^{--}$, $\Theta^{+}$,
$N_{\at}^+$, $\Sigma_{\at}^+$ and $\Xi_{\at}^+$ are exactly zero since
the transition is prohibited by the $U$-spin conservation. The remaining
non-zero transition magnetic moments can be parameterized in the following
form 
\begin{equation}
 \mu_{B_1 B_2}=C_{\at \to  8} \left(
\begin{array}{cc}
8 & 8 \\
Y_U U & 0 0
\end{array}\right|\left.\begin{array}{c}
          \at \\ Y_U U
          \end{array}\right)\, ,
        \label{eq:magnetic_moments_transtions}
\end{equation}
where $C_{\at \to  8}$ is a free constant. Since the two relevant (for
the transitions between the states with $Y_U=0,1$ of the antidecuplet
and the octet)
isoscalar factors in  Eq.~(\ref{eq:magnetic_moments_transtions})
are equal, we have the following
simple relations between the transition magnetic moments
\begin{eqnarray}
&&\mu_{\Sigma_{\at}^- \Sigma^-}=\mu_{\Xi_{\at}^- \Xi^-}=\mu_{N_{\at}^0 N^0}=
\mu_{\Xi_{\at}^0 \Xi^0} \,, \nonumber\\
&&\mu_{\Sigma_{\at}^0 \Sigma^0}=-\frac{1}{2} \,\mu_{N_{\at}^0 N^0} \,, \nonumber\\
&&\mu_{\Sigma_{\at}^0 \Lambda}=\frac{\sqrt{3}}{2} \,\mu_{N_{\at}^0 N^0} \,.
\end{eqnarray}
Effects of the SU(3) breaking in the transition magnetic moments were
considered in~\cite{Kim:2005gz}.

It is important to note that the simple consequence of the $U$-spin 
conservation that $\mu_{p_{\at} p}=0$ and  $\mu_{n_{\at} n} \neq 0$ leads to
the dramatic prediction that the photoexcitation of the 
nucleon-like  member of the antidecuplet,
$N_{\at}$, is suppressed on the proton target and, hence, predominantly takes
place on the neutron target~\cite{Polyakov:2003dx}, see also the discussion
in Sect.~\ref{sec:anti10}.

\underline{{\bf SU(3) systematization: main objectives, methods and results}}

The goal of the SU(3) systematization of baryons is to group known baryons
 into SU(3)
singlets, octets and decuplets. Besides the requirement that all particles in the
same multiplet must have the same spin and parity, one subjects the baryons to two
tests. The first and the basic test is the GMO mass formulas, 
see Eqs.~(\ref{eq:gmo8}) and (\ref{eq:gmo10}), which are indeed
 very useful for the systematization of baryons.

For the fine tuning of the multiplets, however, the GMO mass relations
may be useless. 
First, it might happen  that there are several candidates
 (there are several
baryons with the correct or unmeasured  spin-parity within a given mass range) for
a member of a given multiplet. The mass formulas will work equally well
for all candidates and, thus, will not be able to pick out the right one.
Second, states from different multiplets  with the same spin-parity can mix.
The mass formulas are not particularly
sensitive to the typically small mixing since 
mixing parameters will enter the GMO mass relation as a second order in $m_s$
 correction. Therefore, the multiplet decomposition of a given physical
 state cannot be determined from the  GMO mass formulas. 
Third, if a multiplet is incomplete, i.e.~some of its members are yet
unknown, the mass formulas may have no predictive power.  
Fourth, there is no objective criterion of the accuracy of the GMO formulas for octets.

The next test, which the group of baryons has to pass in order 
to be placed in the same SU(3) multiplet, is based on the SU(3) analysis of
 two-hadron partial decay widths~\cite{Samios:1974tw}. 
The requirement is that the SU(3) predictions for 
the partial decay widths of the multiplet in question should describe (e.g.~in 
the sense of the $\chi^2$ fit) the corresponding experimental values.
The SU(3) prediction for the partial decay widths has the form
\begin{equation}
\Gamma\left(B_1\to B_2+P \right)=\left|g_{B_1 B_2 P} \right|^2 \cdot {\rm barrier \ factor}
 \cdot {\rm phase \ space \ factor} \,,
\label{eq:width}
\end{equation}
where $B_1$ is the initial baryon; $B_2$ and $P$ are the final baryon and 
pseudoscalar meson, which are normally stable and belong to the ground-state 
multiplets; $g_{B_1 B_2 P}$ are the SU(3) symmetric coupling constants;
``${\rm barrier \ factor}$'' takes into account spins and parity of the involved hadrons
and the relative orbital moment of the final $B_2+P$ system;
''${\rm phase \ space \ factor}$'' is the usual kinematic phase space factor.
In Eq.~(\ref{eq:width}), one explicitly assumes that
 the only source of the SU(3) symmetry breaking is the different physical masses 
entering the barrier and phase space factors.

Another possibility would be to 
assume that the part of the strong interaction Hamiltonian
 responsible for the decays contains an explicit SU(3) symmetry breaking 
term~\cite{Dobson:1978gh}. However, in this case, 
the beauty, simplicity and predictive power of the whole approach will
be largely lost since one will have to introduce a number of
unknown parameters.

The main attraction of the assumption of the SU(3)-symmetric coupling 
constants $g_{B_1 B_2 P}$ is the possibility to relate all decays
of a given multiplet using only a few phenomenological constants.
This assumption can be tested by performing the $\chi^2$ analysis of
the measured decays.
Below we give two important examples, which we shall use in our 
numerical analysis. In this work, we
adopt the notation of~\cite{Samios:1974tw}.

For the $\bf{8} \to \bf{8} + \bf{8}$ decays ($B_{1}$, $B_{2}$ and $P$ belong to the 
octet representations),
 the $g_{B_1 B_2 P}$  coupling
constants in the SU(3) symmetric limit are parameterized in
terms of two constants $A_s$ and $A_a$ (the tensor product ${\bf 8} \otimes {\bf 8}$
contains two octets labelled ${\bf 8}_S$ and ${\bf 8}_A$)
\begin{equation}
g_{B_1 B_2 P }=A_s \left(
\begin{array}{cc}
8 & 8 \\
Y_2 I_2 & Y_P I_P
\end{array}\right|\left.\begin{array}{c}
          8_S\\Y_1 I_1
          \end{array}\right)
+A_a \left(
\begin{array}{cc}
8 & 8 \\
Y_2 I_2 & Y_P I_P
\end{array}\right|\left.\begin{array}{c}
          8_A\\Y_1 I_1
          \end{array}\right)  \,.
\label{eq:888}
\end{equation}
 In this equation, $Y_{1,2}$ and $I_{1,2}$ are hypercharges
and isospins of the baryons $B_{1,2}$; the corresponding symbols
with the subscript $P$ refer to the pseudoscalar meson.
In practice, it is more convenient to perform the $\chi^2$ fit using  an 
alternative pair of parameters $A_8$ and $\alpha$~\cite{Samios:1974tw}
\begin{equation}
A_8=\frac{\sqrt{15}}{10} A_s+\frac{\sqrt{3}}{6} A_a \,, \quad \quad \alpha=\frac{\sqrt{3}}{6} \frac{A_a}{A_8} \,.
\end{equation}
With this parameterization, $A_8$ is directly determined by the generally well-measured 
$N \to N \, \pi$ partial decay width and $\alpha$ is determined by the often 
measured $\Sigma \to \Sigma \, \pi$ partial decay width.

For the ${\bf 8} \to {\bf 10}+{\bf 8}$ decays 
($B_2$ belongs to the ground-state decuplet),
the corresponding
coupling constants can be expressed in terms of a single universal 
 coupling constant $A_{8}^{\prime}$
\begin{equation}
 g_{B_1 B_2 P }=A_{8}^{\prime}\ \left(
\begin{array}{cc}
10 & 8 \\
Y_2 I_2 & Y_P I_P
\end{array}\right|\left.\begin{array}{c}
          8 \\Y_1 I_1
          \end{array}\right) \,.
\label{eq:8108}
\end{equation}
Similarly to Eq.~(\ref{eq:8108}), the coupling constants for the
 ${\bf 10} \to {\bf 8}+{\bf 8}$
and ${\bf 10} \to {\bf 10}+{\bf 8}$ decays are expressed in terms of just one
coupling constant, see Sect.~\ref{sec:decuplets}.

It is well-known that states with the same spin and parity from different
unitary multiplets can mix.
The GMO mass relations and the SU(3) relations among the coupling constants
are modified in the presence of the mixing. Let us illustrate this
with the physically important  example of mixing between an octet and a singlet
 $\Lambda$ baryon.
Introducing the mixing angle $\theta$, the physical mostly octet state
$|\Lambda_8 \rangle$ and the mostly singlet state $|\Lambda_1 \rangle$ can be 
written as linear superpositions of the bare $|\Lambda_8^0 \rangle$ and
$|\Lambda_1^0 \rangle$ states
\begin{eqnarray}
|\Lambda_8 \rangle &=& \cos \theta \,|\Lambda_8^0 \rangle+\sin \theta \,|\Lambda_1^0 \rangle \,, \nonumber\\
|\Lambda_1 \rangle &=& -\sin \theta \, |\Lambda_8^0 \rangle+\cos \theta \,|\Lambda_1^0 \rangle \,.
\label{eq:mixing}
\end{eqnarray}
While the bare states are eigenstates of the (idealized) SU(3)-symmetric Hamiltonian,
the physical states are eigenstates of the Hamiltonian with SU(3) symmetry breaking
terms (mass eigenstates).
The GMO mass relation for the octet becomes
\begin{equation}
\frac{1}{2}\left( m_N+m_{\Xi} \right)=\frac{1}{4}\left(3\, m_8^0+m_{\Sigma} \right)
=\frac{1}{4}\left(3\, \left(m_8 \cos^2 \theta+m_1 \sin^2 \theta \right)
+m_{\Sigma} \right) \,,
\label{eq:mixing:masses}
\end{equation}
where the mass of $|\Lambda_8^0 \rangle$ denoted as 
$m_8^0$ is expressed in terms  of the physical masses of the octet ($m_8$)
and singlet ($m_1$) $\Lambda$ baryons.

The predictive power of the mass relation has reduced since a new free 
parameter, the mixing angle $\theta$, has been introduced.
Equation~(\ref{eq:mixing:masses}) also illustrates that, if
the mixing angle is small, it is inconsistent and also
impractical to determine it from the 
modified GMO mass formula because the mixing angle enters 
Eq.~(\ref{eq:mixing:masses}) as a second order
correction in the mass of the strange quark, $\theta \propto {\cal O}(m_s)$,
which was neglected in the derivation of the GMO mass
relations.  

The self-consistent and practical way to establish whether the mixing takes place
and to determine the value of the mixing angle(s) is to consider decays.
In the context of the considered example,
the physical octet and singlet coupling constants are
\begin{eqnarray}
g_{\Lambda_8 B_2 P}&=& \cos \theta \, g_{\Lambda_8^0 B_2 P}+\sin \theta \, g_{\Lambda_1^0 B_2 P} \,, \nonumber\\
g_{\Lambda_1 B_2 P}&=& -\sin \theta \, g_{\Lambda_8^0 B_2 P}+\cos \theta \, g_{\Lambda_1^0 B_2 P}  \,,
\label{eq:mixing:cc}
\end{eqnarray}
where the SU(3) universal coupling constants $g_{\Lambda_8^0 B_2 P}$ are given 
by Eq.~(\ref{eq:888}); the ${\bf 1} \to {\bf 8}+{\bf 8}$ coupling constants
$g_{\Lambda_1^0 B_2 P}$ will be discussed later.
If the introduction of the mixing improves
the description of the data on decays of the $\Lambda_8$  and $\Lambda_1$ hyperons, then this
serves as an unambiguous confirmation of the mixing hypothesis. 
This happens in the case of the mixing of the octet $\Lambda(1670)$ with 
the singlet $\Lambda(1405)$. Another example, when the mixing with a singlet $\Lambda$
baryon is established, is the mostly singlet $\Lambda(1520)$, which decays into 
the $\Sigma(1385) \pi$ final state. Since the 
${\bf 1} \to {\bf 10}+\bf{8} $ decay is forbidden by SU(3),
the decay can take place only due to the mixing, presumably with the
octet $\Lambda(1690)$.

It is important to note that
due to the interference between the terms proportional to
$g_{\Lambda_8^0 B_2 P}$ and  $g_{\Lambda_1^0 B_2 P}$,  the mixing angle 
enters the partial decay width of Eq.~(\ref{eq:width}) in the first power
 and, hence, it can be determined consistently and more reliably 
than via the modified GMO mass 
relation.

To the best of our knowledge, the most recent and thorough SU(3)
systematization of hadrons using the Gell-Mann--Okubo mass relations and
the $\chi^2$ fit to the experimentally
measured decays was performed  by Samios, Goldberg and 
Meadows in 1974~\cite{Samios:1974tw}.
We quote the main conclusion of that work: {\it The detailed study of mass
relationships, decay rates, and interference phenomena shows remarkable
agreement  with that expected from the most simple unbroken SU(3) symmetry
scheme}.

In addition to the well-known GMO mass formulas, there exist almost unknown
similar relations for the total widths of particles in  a given SU(3) 
multiplet derived by Weldon~\cite{Weldon:1977wf}. Using the perturbation theory for
 non-stationary
 states, it was shown that, except for the ground-state
decuplet, the pattern of the total width splitting inside a given multiplet
 is the same as for the mass splitting. This means that the total widths of octet
members obey the relationship
\begin{equation}
\frac{1}{2}\left( \Gamma_N+\Gamma_{\Xi} \right)=\frac{1}{4}\left(3\, \Gamma_{\Lambda}+\Gamma_{\Sigma} \right) \,.
\label{eq:weldon1}
\end{equation}
For decuplets, the total widths should obey the equal spacing rule
\begin{equation}
\Gamma_{\Sigma}-\Gamma_{\Delta}=\Gamma_{\Xi}-\Gamma_{\Sigma}
=\Gamma_{\Omega}-\Gamma_{\Xi} \,.
\label{eq:weldon2}
\end{equation}
A similar equal-space relation holds for the antidecuplet.
For the ground-state decuplet, the relation among the total widths is 
different~\cite{Weldon:1977wf}.
For unstable particles, the total width is as 
intrinsic and fundamental as the mass and, therefore, it is not 
surprising that there is a
relation among the total widths of baryons from the same multiplet.

Below we sketch the derivation of  Weldon's relations.
Using Eq.~(\ref{eq:width}) for the partial decay widths, one can sum over all possible 
decay modes of the initial baryon~\cite{Guzey:2005rx}
\begin{eqnarray}
 &&\sum_{Y_2,I_2,Y_P,I_P} \Gamma\left(B_1\to
B_2+P\right)=
\Gamma_0+\\
\nonumber &&\sum_{i=1,2,P,\gamma}C_i\ \left[
\sum_{Y_2,I_2,Y_P,I_P,\delta} \left(\sum_{\delta} A_\delta
\left(
\begin{array}{cc}
\mu_2 & \mu_P \\
Y_2 I_2 & Y_\phi I_P
\end{array}\right|\left.\begin{array}{c}
          \mu_{1 \delta} \\Y_1 I_1
          \end{array}\right)\right)^2
          \left(
\begin{array}{cc}
\mu_i & 8 \\
Y_i I_i & 0 0
\end{array}\right|\left.\begin{array}{c}
          \mu_{i \gamma }\\ Y_i I_i
          \end{array}\right)\right]\, .
\label{eq:wsum}
\end{eqnarray}
In this equation, $\Gamma_0$ is the baryon width in the limit of unbroken
SU(3) symmetry. The second term is proportional to
the SU(3)-symmetric $|g_{B_1 B_2 P }|^2$ multiplied by the SU(3)-violating
term coming from the Taylor expansion of the masses entering
 the barrier  and phase  volume factors about the central mass $M_0^{\mu}$ of
the corresponding multiplets.
A direct evaluation shows that,
provided that all decay channels  are open, Eq.~(\ref{eq:wsum}) can be written in the 
following form
\begin{equation}
 \sum_{Y_2,I_2,Y_P,I_P} \Gamma\left(B_1\to
B_2+P\right)=
\Gamma_0+\sum_\gamma D_\gamma \left(
\begin{array}{cc}
\mu_1 & 8 \\
Y_1 I_1 & 0 0
\end{array}\right|\left.\begin{array}{c}
          \mu_{1 \gamma } \\ Y_1 I_1
          \end{array}\right)\, .
\end{equation}
A comparison to Eq.~(\ref{GOgeneral}) demonstrates that the sums of all 
two-body hadronic partial decay widths satisfy the same relations as the baryon masses.
 Note that we have derived a more specific form of Weldon's relations, which are valid not
only for the total widths, but also for the sum of the partial decay widths
into the fixed final representations.
Summing over all possible final state SU(3) representations, one obtains Weldon's relations
for the total widths, Eqs.~(\ref{eq:weldon1}) and (\ref{eq:weldon2}).

Our derivation is based
on the assumption that all possible decay channels for a given set
of multiplets $ \mu_1 \to   \mu_2+ \mu_P$ are
kinematically open. Therefore, the Weldon relations are especially useful
 for multiplets with heavy
 baryons for which enough decay modes are open.

In practice, however, Weldon's formulas are of little use in 
the SU(3) systematization of baryons. The derivation of Eqs.~(\ref{eq:weldon1})
and (\ref{eq:weldon2}) implies that all two-body decay channels of a given baryon
are open. For light baryons, some decays are kinematically
prohibited and  Weldon's formulas are not expected to hold. In the opposite case when
all decay channels are open, Weldon's relations is a mere consequence of the 
approximate SU(3) symmetry and they do not supply any extra information, which
would  not be already present in the used formalism.

The list of SU(3) multiplets of baryons with 
the mass less than approximately 2000-2200 MeV, which one could find in the literature
\cite{Kokkedee,Samios:1974tw,Martin:1964,Tripp:1967kj,Pakvasa:1969xe,Dobson:1969xk,Plane:1970yf},
is summarized in Table~\ref{table:before}.
The first column indicates SU(6)$\times$O(3) supermultiplets; the second and third columns
enumerate and indicate the type of the SU(3) representation, its spin and parity.
 The masses in the parenthesis
are for $(N, \Lambda, \Sigma, \Xi)$ members of the octets and for
$(\Delta, \Sigma, \Xi, \Omega)$ members of  the decuplets. 
We give the modern values of the masses~\cite{Eidelman:2004wy}.
\begin{table}[h]
\begin{center}
\begin{tabular}{|c|c|c|c|}
\hline
  & 1 & $(8,\frac{1}{2}^+)$ & (939, 1115, 1189, 1314) \\
 & 2 & $(10,\frac{3}{2}^+)$ & (1232, 1385, 1530, 1672) \\ 
$(56, L=0)$  & 3 & $(8,\frac{1}{2}^+)$ & (1440, $\dots$, $\dots$, $\dots$) \\
& 4 &$(8,\frac{1}{2}^+)$ & (1710, $\dots$ , $\dots$ ,  $\dots$) \\ 
& 5 & $(10,\frac{3}{2}^+)$ &  (1600, $\dots$, $\dots$, $\dots$) \\
\hline \hline
& 6 & $(1,\frac{1}{2}^-)$ & $\Lambda(1405)$ \\
& 7 & $(1,\frac{3}{2}^-)$ & $\Lambda(1520)$ \\
& 8 & $(8,\frac{3}{2}^-)$ & (1520, 1690, 1670, 1820) \\
$(70, L=1)$& 9 & $(8,\frac{1}{2}^-)$ & (1535, 1670, 1750, \underline{1835}) \\
& 10 & $(10,\frac{1}{2}^-)$ & (1620, $\dots$, $\dots$, $\dots$) \\
 & 11 & $(8,\frac{3}{2}^-)$ & (1700, $\dots$, $\dots$, $\dots$) \\
& 12 & $(8,\frac{5}{2}^-)$ & (1675, 1830, 1775, 1950) \\
& 13 & $(10,\frac{3}{2}^-)$ & (1700, $\dots$, $\dots$, $\dots$) \\
& 14 & $(8,\frac{1}{2}^-)$ & (1650, $\dots$, $\dots$, $\dots$) \\
\hline \hline
& 15 & $(8,\frac{5}{2}^+)$ & (1680, 1820, 1915, 2030) \\
& 16 & $(10,\frac{3}{2}^+)$ & (\underline{1920}, $\dots$, $\dots$, $\dots$) \\
$(56, L=2)$ & 17 & $(8,\frac{3}{2}^+)$ & (1720, $\dots$, $\dots$, $\dots$) \\
& 18 & $(10,\frac{5}{2}^+)$ & (1905, $\dots$, $\dots$, $\dots$) \\
& 19 & $(10,\frac{1}{2}^+)$ & (1910, $\dots$, $\dots$, $\dots$) \\
& 20 & $(10,\frac{7}{2}^+)$ & (1950, 2030, \underline{2120}, \underline{2250}) \\
\hline
\end{tabular}
\caption{SU(3) multiplets known by 1974.}
\label{table:before}
\end{center}
\end{table}

In order to determine the minimal number of possible SU(3) multiplets, it is useful
to use as guide the SU(6) classification scheme, which combines the flavor SU(3) group 
with the spin SU(2) group~\cite{Kokkedee,Lichtenberg}.
SU(6) implies that quarks are non-relativistic and, hence, there is no reason that
SU(6) is a symmetry of QCD. However, phenomenologically SU(6) works surprisingly well.
 Starting with three constituent
quarks in the fundamental SU(6) representation, one obtains the following 
allowed multiplets
\begin{equation}
{\bf 6} \otimes {\bf 6} \otimes {\bf 6}={\bf 20} \oplus {\bf 56} \oplus {\bf 70} \oplus {\bf 70} \,,
\end{equation}
where the ${\bf 20}$ representation is totally antisymmetric, 
${\bf 56}$ is totally symmetric and ${\bf 70}$ has mixed symmetry. 
It is a phenomenological observation that most likely only the ${\bf 56}$ and ${\bf 70}$
SU(6) representations are realized in Nature~\cite{Samios:1974tw}.
 They have the following decomposition 
in terms of SU(3) representations~\cite{Kokkedee}
\begin{eqnarray}
{\bf 56}&=&\left({\bf 8},\ \frac{1}{2} \right) \oplus \left({\bf 10},\ \frac{3}{2} \right) \,, \nonumber\\
{\bf 70}&=&\left({\bf 1},\ \frac{1}{2} \right) \oplus \left({\bf 8},\ \frac{1}{2} \right)
\oplus \left({\bf 8},\ \frac{3}{2} \right) \oplus \left({\bf 10},\ \frac{1}{2} \right)
 \,,
\label{eq:su6tosu3}
\end{eqnarray} 
where the second number in the parenthesis denotes the total spin $S$ of the multiplet.

Since parity is the quantum number beyond SU(6), in order to obtain multiplets of 
different parities, one couples the total spin $S$ to the total orbital moment
of the three quarks $L$ (it
corresponds to the group of spacial rotations O(3)]. The resulting baryon wave function
must be symmetric [we always keep in mind the final antisymmetrization under color SU(3)].
Therefore, assuming that the radial part of the wave function is symmetric, one
sees that the 
${\bf 56}$ representation admits only even $L$, while
the ${\bf 70}$ representation accepts all $L$. Then, coupling $L=0,2$ to the  total spin
$S$ of the SU(3) multiplets that belong ${\bf 56}$ and coupling $L=1$ to $S$
of the  SU(3) multiplets that belong ${\bf 70}$, see Eq.~(\ref{eq:su6tosu3}),
one obtains the following decomposition in terms of SU(3) multiplets with different $J^P$ 
\begin{eqnarray}
({\bf 56},L=0)&=&\left({\bf 8},\ \frac{1}{2}^+ \right) \oplus \left({\bf 10},\ \frac{3}{2}^+ \right) \,, \nonumber\\
({\bf 70},L=1)&=&\left({\bf 1},\ \frac{1}{2}^- \right) \oplus \left({\bf 1},\ \frac{3}{2}^-\right) \oplus2  \left({\bf 8},\ \frac{1}{2}^- \right) 
\oplus 2  \left({\bf 8},\ \frac{3}{2}^- \right) \oplus \left({\bf 8},\ \frac{5}{2}^- \right)
\nonumber\\
& \oplus & \left({\bf 10},\ \frac{1}{2}^- \right)\oplus  \left({\bf 10},\ \frac{3}{2}^- \right)\,,  \nonumber\\
({\bf 56},L=2)&=&\left({\bf 8},\ \frac{3}{2}^+ \right) \oplus \left({\bf 8},\ \frac{5}{2}^+ \right) \oplus \left({\bf 10},\ \frac{1}{2}^+ \right)
 \oplus \left({\bf 10},\ \frac{3}{2}^+ \right)
\oplus \left({\bf 10},\ \frac{5}{2}^+ \right)\nonumber\\
&\oplus& \left({\bf 10},\ \frac{7}{2}^+ \right)
 \,.
\label{eq:counting}
\end{eqnarray}
Therefore, the assumption that only the $({\bf 56},L=0)$, $({\bf 70},L=1)$ and 
$({\bf 56},L=2)$ supermultiplets of SU(6)$\times$O(3) are possible indicates the
existence of the seventeen SU(3) multiplets listed in Eq.~(\ref{eq:counting}).
The actual number of multiplets is larger because new multiplets can be formed by 
radial  excitations.
For instance, Table~\ref{table:before} contains twenty SU(3) multiplets: seventeen
listed in Eq.~(\ref{eq:counting}) and additional multiplets 3, 4 and 5, 
which can be thought of as radial excitations of the corresponding ground-state
octet and decuplet multiplets.

We stress that without attempting to model
the dynamics of the quark interaction, one can only accept as an empirical 
fact that Nature seems to use only the $({\bf 56},L=0,2)$ and $({\bf 70},L=1)$ 
supermultiplets for the baryons with the mass less than approximately 2000-2200 MeV, 
as indicated in Table~\ref{table:before}. Note that for heavier baryons,
other, higher supermultiplets are needed. For instance, an octet and singlet
 with $J^P=7/2^{-}$, which do not fit to Eq.~(\ref{eq:counting}) and which
supposedly belong to the  $({\bf 70},L=3)$ supermultiplet,
were considered in~\cite{Samios:1974tw}.

The problem of missing resonances in constituent quark models partially is a consequence of
the fact that in an attempt to derive the spectrum of baryons,
 the quark models have no reasons to prefer some SU(6)$\times$O(3) supermultiplets 
the others. For example, the prediction for positive-parity 
baryons~\cite{Isgur:1978wd} include
five low-lying SU(6)$\times$O(3) supermultiplets: $({\bf 56}, L=0,2)$,
$({\bf 70}, L=0,2)$ and $({\bf 20}, L=1)$, which naturally makes the number of the 
predicted baryons larger than found (so far) in Nature.
In contrast to the positive-parity baryons, the quark models predict
the correct number of low-lying negative-parity baryons (the same as in
Table~\ref{table:before}) because they
essentially employ only the $({\bf 70}, L=1)$ 
supermultiplet~\cite{Isgur:1978xj,Loring:2001ky,Glozman:1997ag}.

Now we discuss Table~\ref{table:before} in some detail.
Multiplets 1, 2, 6-9, 12, 15 and 20 were considered in~\cite{Samios:1974tw}.
While all the multiplets are mentioned in~\cite{Kokkedee}, no SU(3) analysis of 
the decays was performed.
In  Table~\ref{table:before},  we underlined genuine predictions
of new particles: Kokkedee~\cite{Kokkedee} predicted $\Delta(1920)$
 with $J^P=3/2^+$; Samios, Goldberg and Meadows~\cite{Samios:1974tw} 
predicted $\Xi(2120)$ and
$\Omega(2250)$ with $J^P=7/2^+$ and $\Xi(1835)$ with $J^P=1/2^-$. All these
particles, except for $\Xi(1835)$, were later discovered.
In addition, it was supposed in~\cite{Kokkedee} that $N(1440)$ and  
$N(1710)$ with  $J^P=1/2^+$ and $\Delta(1600)$ with $J^P=3/2^+$  represent radial 
excitations of the corresponding ground-state multiplets and thus, 
belong to the $({\bf 56},L=0)$ supermultiplet. Our analysis confirms this assumption.

The Review of Particle Physics 2004 (RPP)~\cite{Eidelman:2004wy}, 
in Sect.~{\it Quark model}
also gives the list of SU(3) multiplets, which we present in Table~\ref{table:pdg}.
Tables~\ref{table:before} and \ref{table:pdg} are very similar except for
the following differences.
Multiplets 5, 16 and 19 are not mentioned in the RPP; octet 9 contains $\Sigma(1620)$ 
instead of $\Sigma(1750)$, which is assumed to belong to octet 14;
octet 4 is assumed to belong to  the $({\bf 70},L=0)$ supermultiplet, which 
is not present in  Table~\ref{table:before}.
Throughout this work, we shall refer to the multiplets as they are numbered in
Table~\ref{table:before}.

\begin{table}[h]
\begin{center}
\begin{tabular}{|c|c|c|c|}
\hline
  & 1 & $(8,\frac{1}{2}^+)$ & (939, 1116, 1193, 1318) \\
 & 2 & $(10,\frac{3}{2}^+)$ & (1232, 1385, 1530, 1672) \\ 
$(56, L=0)$  & 3 & $(8,\frac{1}{2}^+)$ & (1440, 1600, 1660, $\dots$) \\
\hline \hline
$(70, L=0)$ & 4 &$(8,\frac{1}{2}^+)$ & (1710, 1810, 1880, $\dots$) \\ 
\hline \hline
& 6 & $(1,\frac{1}{2}^-)$ & $\Lambda(1405)$ \\
& 7 & $(1,\frac{3}{2}^-)$ & $\Lambda(1520)$ \\
& 8 & $(8,\frac{3}{2}^-)$ & (1520, 1690, 1670, 1820) \\
$(70, L=1)$& 9 & $(8,\frac{1}{2}^-)$ & (1535, 1670, 1620, $\dots$ ) \\
& 10 & $(10,\frac{1}{2}^-)$ & (1620, $\dots$, $\dots$, $\dots$) \\
 & 11 & $(8,\frac{3}{2}^-)$ & (1700, $\dots$, $\dots$, $\dots$) \\
& 12 & $(8,\frac{5}{2}^-)$ & (1675, 1830, 1775, $\dots$) \\
& 13 & $(10,\frac{3}{2}^-)$ & (1700, $\dots$, $\dots$, $\dots$) \\
& 14 & $(8,\frac{1}{2}^-)$ & (1650, 1800, 1750, $\dots$) \\
\hline \hline
& 15 & $(8,\frac{5}{2}^+)$ & (1680, 1820, 1915, 2030) \\
$(56, L=2)$ & 17 & $(8,\frac{3}{2}^+)$ & (1720, 1890, $\dots$, $\dots$) \\
& 18 & $(10,\frac{5}{2}^+)$ & (1905, $\dots$, $\dots$, $\dots$) \\
& 20 & $(10,\frac{7}{2}^+)$ & (1950, 2030, $\dots$, $\dots$) \\
\hline
\end{tabular}
\caption{SU(3) multiplets from the Review of Particle Physics 2004.}
\label{table:pdg}
\end{center}
\end{table}

The aim of this review is to update the picture of SU(3) multiplets
presented in Tables~\ref{table:before} and \ref{table:pdg} following
the approach of~\cite{Samios:1974tw}. First, Samios {\it et al.} chose not to
use a number of baryons already present in the 1972 edition of the Review of
Particle Physics~\cite{RPP:1972} partially because of 
relatively weak evidence of their existence and partially because they
corresponded to rather incomplete multiplets.
Those of them, which are not listed in  Tables~\ref{table:before}, include 
(we give the values of the masses as they were in 1972)
\begin{eqnarray}
&& N(1860) \frac{3}{2}^+ \,,\  \Lambda(1750) \frac{1}{2}^+ \,,\ \Lambda(1860)\frac{3}{2}^+
 \,,\ \Lambda(1870)\frac{1}{2}^-  \,, \ \Sigma(1480)? \,, \ \Sigma(1620)\frac{1}{2}^- \,,
\nonumber\\
&& \Sigma(1620)\frac{1}{2}^+ \,, \  \Sigma(1690)? \,, \ \Sigma(1880)\frac{1}{2}^+ \,, 
\ \Sigma(1940)\frac{3}{2}^- \,, \ \Sigma(2070)\frac{5}{2}^+ \,, \ \Sigma(2080)\frac{3}{2}^+
\,, \nonumber\\
&& \Xi(1630)? \,,
\end{eqnarray} 
where the number next to the particle's mass denotes its $J^P$.

Second, a number of new particles since 1972 were reported. These include
\begin{eqnarray}
&&\Delta(1920)\frac{3}{2}^+ \,, \ \Lambda(1600)\frac{1}{2}^+  \,, \
\Sigma(1560)? \,, \ \Sigma(1840)\frac{3}{2}^+ \,, \nonumber\\
&&\Xi(1690)? \,, \ \Xi(2120)? \,, \
\Omega(2250)? \,, \ \Omega(2280)? \,, \ \Omega(2470)? \,. 
\end{eqnarray}

In addition, the values of many measured partial decay widths
have changed their values and, in general, have became more precise. In our analysis, 
whenever possible, we try to use uniform sources of information on
the partial decays widths.
The partial decay widths of $N$ and $\Delta$ baryons are predominantly taken 
from~\cite{Manley:1992yb}, 
which appears to be the analysis preferred by the authors of the Review of Particle 
Physics~\cite{Eidelman:2004wy};
 the partial decay widths of $\Lambda$ and $\Sigma$ 
hyperons are taken from~\cite{Gopal:1976gs,Cameron:1977jr,Gopal:1980ur}, which is by far 
the most recent and comprehensive analysis of strange particles.

For reader's convenience, the final results of our SU(3) systematization 
of all baryons with the mass less than
approximately 2000-2200 MeV are presented in Table~\ref{table:final_intro}.
The underlined entries in the table
are predictions of new particles, which are absent in the Review of Particle 
Physics. 
\begin{table}[t]
\begin{center}
\begin{tabular}{|c|c|c|c|}
\hline
  & 1 & $(8,\frac{1}{2}^+)$ & (939, 1115, 1189, 1314) \\
 & 2 & $(10,\frac{3}{2}^+)$ & (1232, 1385, 1530, 1672) \\ 
$(56, L=0)$  & 3 & $(8,\frac{1}{2}^+)$ & (1440, 1600, 1660, 1690) \\
& 4 &$(8,\frac{1}{2}^+)$ & (1710, 1810, 1880, \underline{1950}) \\ 
& 5 & $(10,\frac{3}{2}^+)$ &  (1600, 1690, \underline{1900}, \underline{2050}) \\
\hline \hline
& 6 & $(1,\frac{1}{2}^-)$ & $\Lambda(1405)$ \\
& 7 & $(1,\frac{3}{2}^-)$ & $\Lambda(1520)$ \\
& 8 & $(8,\frac{3}{2}^-)$ & (1520, 1690, 1670, 1820) \\
$(70, L=1)$& 9 & $(8,\frac{1}{2}^-)$ & (1535, 1670, 1560, \underline{1620-1725}) \\
& 10 & $(10,\frac{1}{2}^-)$ & (1620, 1750, \underline{1900}, \underline{2050}) \\
 & 11 & $(8,\frac{3}{2}^-)$ & (1700, \underline{1850}, 1940, \underline{2045}) \\
& 12 & $(8,\frac{5}{2}^-)$ & (1675, 1830, 1775, 1950) \\
& 13 & $(10,\frac{3}{2}^-)$ & (1700, \underline{1850}, \underline{2000}, \underline{2150}) \\
& 14 & $(8,\frac{1}{2}^-)$ & (1650, 1800, 1620, \underline{1860-1915}) \\
\hline \hline
& 15 & $(8,\frac{5}{2}^+)$ & (1680, 1820, 1915, 2030) \\
& 16 & $(10,\frac{3}{2}^+)$ & (1920, 2080, \underline{2240}, 2470) \\
$(56, L=2)$ & 17 & $(8,\frac{3}{2}^+)$ & (1720, 1890, 1840, \underline{2035}) \\
& 18 & $(10,\frac{5}{2}^+)$ & (1905, 2070, 2250, 2380) \\
& 19 & $(10,\frac{1}{2}^+)$ & (1910, \underline{2060}, \underline{2210},
 \underline{2360} ) \\
& 20 & $(10,\frac{7}{2}^+)$ & (1950, 2030, 2120, 2250) \\
\hline \hline
& 21 & $(\at,\frac{1}{2}^+)$ & (1540, 1670, \underline{1760}, 1862) \\
\hline
\end{tabular}
\caption{The final list  of SU(3) multiplets.}
\label{table:final_intro}
\end{center}
\end{table}

This review is organized as follows. The SU(3) systematization of octets is 
presented in Sect.~\ref{sec:octets}. 
The analysis is then repeated for decuplets
in Sect.~\ref{sec:decuplets}. 
In Sect.~\ref{sec:anti10}, we discuss the SU(3) analysis of the antidecuplet
and consider its mixing with octet 3, which allows us to obtain a
 self-consistent picture of the antidecuplet decays.
We discuss our results in Sect.~\ref{sec:conclusions}.

In summary, we have succeeded in placing nearly all known baryons with the 
mass less than approximately
2000-2200 MeV into  twenty-one SU(3) multiplets. In order to have complete multiplets,
we predict the existence of a number of strange particles.
The most remarkable among them is the $\Lambda$ hyperon with $J^P=3/2^-$, 
the mass around 1850 MeV, the total width $\approx 130$ MeV,
significant branching into the
$\Sigma \pi$ and $\Sigma(1385) \pi$ states
and a very small coupling to the $N \overline{K}$ state.
This is remarkable because all other eleven $\Lambda$ hyperons,
which are required for the consistency of our SU(3) picture, 
are known and have 
very high (three and four stars) status in the RPP~\cite{Eidelman:2004wy}.
Our prediction model-independently confirms the constituent quark model prediction that 
 there should exist a new $\Lambda$ baryon with $J^P=3/2^-$ in the
$1775-1880$ MeV mass range~~\cite{Isgur:1978xj,Loring:2001ky,Glozman:1997ag}.
In addition, we show how SU(3) can be effectively applied for the
 systematization and predictions of 
 the decays of the antidecuplet.

\section{SU(3) classification of octets}
\label{sec:octets}

In this section, we review the picture of  SU(3) octets 
of baryons with the mass less than approximately 2000-2200 MeV using the 
the Gell-Mann--Okubo mass formulas and the $\chi^2$ fit to the measured
partial decay widths. We scrutinize one octet at a time, starting
with the established octets considered in~\cite{Samios:1974tw}.

\subsection{Accuracy of the Gell-Mann--Okubo formula}
\label{ss:accuracy}

We mentioned in the Introduction that there is no objective criterion of 
the accuracy of the Gell-Mann--Okubo mass formula for octets.
Below we define an estimator for the GMO relations for decuplets and
discuss how the accuracy of the GMO relation for octets can be estimated.
The same reasoning applies to Weldon's formulas for total widths.

The accuracy of the GMO formula for decuplets can be estimated as follows.
Introducing the average mass splitting, $\langle z \rangle$, and the standard 
deviation, $\Delta z$,
\begin{eqnarray}
\langle z \rangle & = & \frac{1}{3} \left(m_{\Sigma}-m_{\Delta}+m_{\Xi}-m_{\Sigma}+m_{\Omega}-m_{\Xi}\right) =\frac{1}{3}\left(m_{\Omega}- m_{\Delta} \right) \,,
 \nonumber\\
\Delta z & = & \frac{1}{\sqrt{2}}\left((m_{\Sigma}-m_{\Delta}-z)^2+(m_{\Xi}-m_{\Sigma}-z)^2+
(m_{\Omega}-m_{\Xi}-z)^2 \right)^{1/2} \,,
\label{spld2}
\end{eqnarray}
one can estimate the accuracy of the equal spacing rule by comparing $\Delta z$ to 
$\langle z \rangle$, e.g.
\begin{equation}
{\rm accuracy} = \frac{\Delta z}{\langle z \rangle} \,.
\label{spld3}
\end{equation}
Note that the numerator has the order ${\cal O}(m_s^2)$ and the denominator
 has the order ${\cal O}(m_s)$. Thus defined ${\rm accuracy}$
does not depend on the overall position of the decuplet (the 
mass $M_0^{10}$).

Turning to the accuracy of the GMO formula for octets, one observes that
 it is impossible to suggest an 
estimator of the accuracy which would have the form similar to 
Eqs.~(\ref{spld2}) and
(\ref{spld3}), since there is just one equation relating four masses.
Therefore, one can estimate the accuracy of the octet GMO mass formula
 only qualitatively, for example, by comparing the mismatch between the
 right and left hand sides of Eq.~(\ref{eq:gmo8}),
\begin{equation}
\Delta M \equiv \frac{1}{2}\left(m_N+m_{\Xi}\right)-\frac{1}{4}\left(3\, m_{\Lambda}
+m_{\Sigma}\right) \,,
\label{eq:mismatch}
\end{equation}
 to the typical hadronic
 scale of approximately 1 GeV
or to some average mass of the octet in question.

Of course, one might argue that using three free parameters of the GMO mass 
formula for the octet~(\ref{GOgeneral}), one can perform a $\chi^2$ fit to the
four experimental masses and, thus, one can objectively judge how well the GMO mass
formula works. However, the SU(3)-symmetric mass $M_0^{\mu}$ has nothing to do
with the mass splitting and the original idea of Gell-Mann and Okubo and, hence,
should not be used in assessing the accuracy of the GMO formula.

Replacing the masses by the total widths in 
Eqs.~(\ref{spld2}) and (\ref{spld3}), 
one obtains an estimate for the accuracy of
 Weldon's formula for decuplets, see Eq.~(\ref{eq:weldon2}).
Similarly to the GMO formula, there is no prescription for the estimation of 
the accuracy of Weldon's formula for  octets, see  Eq.~(\ref{eq:weldon1}).
It is not even clear what hadronic scale the mismatch between the left
and right hand sides of Eq.~(\ref{eq:weldon1}) should be compared to since
1 GeV cannot be used. 
Hence, in this work we will not actively use Weldon's relations.

\subsection{Universal SU(3) coupling constants and the barrier and phase space factors}
\label{ss:cc}

The assumption that the approximate flavor SU(3) symmetry is violated
 only by the different
 masses of the baryons and is exact for their decays, 
gives the possibility to relate all decays
of a given multiplet using only a few phenomenological constants.
For the decays of octets, these are $A_8$ and $\alpha$ of Eq.~(\ref{eq:888})
and $A_8^{\prime}$ of Eq.~(\ref{eq:8108}).

In addition to octets, we will consider decays of 
two SU(3) singlets, which are
 represented by the $\Lambda(1405)$ and $\Lambda(1520)$ hyperons.
The ${\bf 1} \to {\bf 8}+{\bf 8}$ coupling constants have the form
 \begin{equation}
g_{B_1 B_2 P }=A_1 \left(
\begin{array}{cc}
8 & 8 \\
Y_2 T_2 & Y_P T_P
\end{array}\right|\left.\begin{array}{c}
          1\\00 
          \end{array}\right)   \,,
\label{eq:188}
\end{equation}
where $A_1$ is a free parameter to be determined from the $\chi^2$ fit to the decay rates.
The coupling constants for all decay modes of octets and singlets in terms
of $A_8$, $\alpha$, $A_8^{\prime}$ and $A_1$ are summarized in 
Tables~\ref{table:octets} and \ref{table:octetsb}.

\begin{table}[h]
\begin{center}
\begin{tabular}{|c c c c|}
\hline
 & ${\bf 8} \to {\bf 8}+{\bf 8}$ & ${\bf 1} \to {\bf 8}+{\bf 8}$
& ${\bf 8} \to {\bf 10}+{\bf 8}$ \\
Decay mode & $g_{B1 B_2 P}$ &  $g_{B1 B_2 P}$ & $g_{B1 B_2 P}$\\
\hline
$N \to N \pi$ & $\sqrt{3}\, A_8$ & & \\
$ \to N \eta$ & $[(4\, \alpha-1)/\sqrt{3}]\, A_8$ & &  \\
$ \to \Sigma K$ & $\sqrt{3}\, (2 \,\alpha-1)\, A_8$ & & \\
$ \to \Lambda K$ & $-[(2 \alpha+1)/\sqrt{3}]\, A_8$ & & \\
$ \to \Delta \pi$ & & & $-2/\sqrt{5} \, A_8^{\prime}$ \\
$ \to \Sigma^{\ast} K$ & & & $1/\sqrt{5} \, A_8^{\prime}$ \\
\hline
$\Lambda \to N {\overline K}$ & $\sqrt{2/3}\,(2\, \alpha+1) \, A_8$ &
 $1/2 \, A_1$ & \\
$\to \Sigma \pi$ & $2\,(\alpha-1) \, A_8$ & $\sqrt{6/4} \, A_1$ &  \\
$\to \Lambda \eta$ & $2/\sqrt{3}\,(\alpha-1) \, A_8$ & $-(\sqrt{2}/4) \, A_1$
 & \\
$\to \Xi K$ & $\sqrt{2/3}\,(4\,\alpha-1) \, A_8$ & $-1/2 \, A_1$ &  \\
$ \to \Sigma^{\ast} \pi$ & & & $-\sqrt{15}/5 \, A_8^{\prime}$  \\
$ \to \Xi^{\ast} K$ & & & $\sqrt{10}/5 \, A_8^{\prime}$ \\
\hline
$\Sigma \to \Sigma \pi$ & $2 \sqrt{2} \, \alpha \,A_8$ & & \\
$\to \Lambda \pi$ &  $-2/\sqrt{3}\,(\alpha-1) \, A_8$ & &  \\
$\to N {\overline K}$ & $\sqrt{2}\,(2 \,\alpha-1) \, A_8$ & &  \\
$\to \Sigma \eta$ & $-2/\sqrt{3}\,(\alpha-1) \, A_8$ & & \\
$\to \Xi K$ & $-\sqrt{2}\, A_8$ & &  \\
$\to \Delta {\overline K}$ & & & $2 \sqrt{30}/15 \, A_8^{\prime}$   \\
$\to \Sigma^{\ast} \pi$ & & & $-\sqrt{30}/15 \, A_8^{\prime}$   \\
$\to \Sigma^{\ast} \eta$ & & & $-\sqrt{5}/5 \, A_8^{\prime}$   \\
$\to \Xi^{\ast} K$ & & & $\sqrt{30}/15 \, A_8^{\prime}$   \\
\hline
\end{tabular}
\caption{The SU(3) universal coupling constants for ${\bf 8} \to {\bf 8}+{\bf 8}$, ${\bf 8} \to {\bf 10}+{\bf 8}$ 
 and  ${\bf 1} \to {\bf 8}+{\bf 8}$ decays.}
\label{table:octets}
\end{center}
\end{table}

\begin{table}[h]
\begin{center}
\begin{tabular}{|c c c c|}
\hline
 & ${\bf 8} \to {\bf 8}+{\bf 8}$ & ${\bf 1} \to {\bf 8}+{\bf 8}$
& ${\bf 8} \to {\bf 10}+{\bf 8}$ \\
Decay mode & $g_{B1 B_2 P}$ &  $g_{B1 B_2 P}$ & $g_{B1 B_2 P}$\\
\hline
$\Xi \to \Xi \pi$ & $\sqrt{3}\,(2\, \alpha-1) \, A_8$ & &  \\
$\to \Lambda {\overline K}$ & $[(4 \,\alpha-1)/\sqrt{3}]\, A_8$ & & \\
$\to \Sigma {\overline K}$ & $\sqrt{3}\, A_8$ & &  \\
$\to \Xi \eta$ & $-[(2 \,\alpha+1)/\sqrt{3}]\, A_8$ & & \\
$ \to \Xi^{\ast} \pi,\  \Xi^{\ast} \eta$ & & & $-\sqrt{5}/5 \, A_8^{\prime}$ \\
$ \to \Sigma^{\ast} {\overline K}$ & & & $\sqrt{5}/5 \, A_8^{\prime}$ \\
$ \to \Omega K$ & & & $\sqrt{10}/5 \, A_8^{\prime}$ \\
\hline
\end{tabular}
\caption{Continuation of Table~\ref{table:octets}.}
\label{table:octetsb}
\end{center}
\end{table}

It is important to note that while the coupling constants 
$A_8$, $A_8^{\prime}$ and $A_1$ are free parameters, 
SU(6) makes unique predictions for $\alpha$, which is related to
the so-called $F/D$ ratio. This can be seen as follows.
Since the tensor product of two ${\bf 8}$ 
representations contains two ${\bf 8}$'s, see Eq.~(\ref{eq:8plus8}), the effective
Lagrangian describing ${\bf 8} \to {\bf 8} + {\bf 8}$ is parameterized in terms of two
free constants $g_0$ and $\alpha$~\cite{Eightfoldway}
\begin{equation}
{\cal L}_{{\rm int}}=2i \,g_0\, \overline{B}_1^j \left[\alpha\, F_i^{jk} +(1-\alpha)D_i^{jk}\right] B_2^k \,P^i \,,
\end{equation}
where $F_i^{jk}=-if_{ijk}$ and $D_i^{jk}=d_{ijk}$ with $f_{ijk}$ and $d_{ijk}$ the antisymmetric
and symmetric SU(3) structure constants; ${B}_{1,2}^j$ and $P^j$ denote the baryon and meson octet 
fields.
 The ratio of the coupling constants in front of 
$F_i^{jk}$ and $D_i^{jk}$ is called the $F/D$ ratio
\begin{equation}
F/D=\frac{\alpha}{1-\alpha} \,.
\end{equation}
From this, one immediately obtains the relation between $\alpha$ and $F/D$
\begin{equation}
\alpha=\frac{F/D}{1+F/D} \,.
\end{equation}
While in SU(3) the ratio $F/D$ is unconstrained, SU(6) makes unique predictions for 
$F/D$~\cite{Gilman:1973kh}.
The SU(6) predictions for $F/D$ and $\alpha$ are summarized in 
Table~\ref{table:alpha}. In the table,
in the second column, $S$ denotes the spin,
which is coupled to the orbital moment $L$ to give the total 
angular moment
of a given SU(3) multiplet. Therefore, SU(6) predictions for $\alpha$ for the
octets with $J=1/2$ and
 $J=3/2$, which belong to the
 $({\bf 70}, L=1)$ representation (octets 8, 9, 11 and 14),
 are ambiguous since $\alpha$ can be either $0.625$ or $-0.5$.
In our analysis, we shall 
use the SU(6) predictions for $\alpha$ as a rough guide.
\begin{table}[h]
\begin{center}
\begin{tabular}{|c c c c |}
\hline
SU(6) & (SU(3),\ S) & $F/D$~\cite{Gilman:1973kh} & $\alpha$ \\
\hline
${\bf 56}$ & $\left({\bf 8}, 1/2 \right)$ & $\frac{2}{3}$ &  $\frac{2}{5}$ \\
${\bf 70}$ & $\left({\bf 8}, 1/2 \right)$ & $\frac{5}{3}$ &  $\frac{5}{8}$ \\
 &           $\left({\bf 8}, 3/2 \right)$ & $-\frac{1}{3}$&  $-\frac{1}{2}$ \\
\hline
\end{tabular}
\caption{SU(6) predictions for $F/D$~\protect\cite{Gilman:1973kh} and $\alpha$. 
In the second column, $S$ denotes the spin of SU(3) multiplets.}
\label{table:alpha}
\end{center}
\end{table}

The second ingredient in the calculation of partial decay widths using
Eq.~(\ref{eq:width}) is the barrier and phase space factors. 
Since we assumed that the only source of SU(3) symmetry breaking is non-equal masses
of baryons in multiplets, the phase space factor, which is well-defined in the 
SU(3)-symmetric limit, can be multiplied by any function of the ratio of the baryon masses.
This introduces ambiguity in the choice  
of the phase space factor. In our analysis, we use the convention of 
Samios {\it et al.}~\cite{Samios:1974tw}, 
which captures the main features of relativistic kinematics and
provides the dimensionless coupling constants
\begin{eqnarray}
&&{\rm barrier \ factor}=\left(\frac{k}{M}\right)^{2l} \,, \nonumber\\
&&{\rm phase \ space \ factor}=\left(\frac{k}{M_1}\right) M \,, 
\label{eq:phase_volume}
\end{eqnarray}
where $k$ is the center-of-mass momentum of the final particles;
$M_1$ is the mass of $B_1$; $M=1000$ MeV is the dimensional parameter;
$l$ is the relative orbital moment of the outgoing $B_2 \, P$ system. 
The orbital moment $l$ is found by requiring the conservation of parity and
the total angular moment in the decay.

\subsection{Multiplet 1: Ground-state octet}

The Gell-Mann--Okubo mass formula for the ground-state octet works with
very high precision,
\begin{equation}
\Delta M =\frac{1}{2}\left(m_N+m_{\Xi}\right)-\frac{1}{4}\left(3\, m_{\Lambda}
+m_{\Sigma}\right) \approx -8 \ {\rm MeV} \,.
\label{eq:m1}
\end{equation}
The mismatch between the left  and the right hand sides of the GMO
mass formula, $\Delta M$,  is less than one per cent of the individual baryon masses.
The ground-state octet is of course stable against two-body hadronic decays
so that the $\chi^2$ analysis of its decays cannot be performed.
However, as discussed in~\cite{Samios:1974tw}, one can attempt to examine
how well SU(3) describes the relation between the phenomenological
$NN\pi$, $\Lambda p K^- $ and $\Sigma p K^-$ coupling constants. Unfortunately,
the two latter  coupling constants are poorly known and  rather ambiguous,
which does not allow one to really test SU(3). 

A better test of SU(3) for the ground-state octet can be performed
by considering SU(3) predictions (the so-called Cabbibo theory) 
for semi-leptonic $B_1 \to B_2 l \nu_l$ decays .
The conclusion of the analysis~\cite{Garcia:1985xz} is that the Cabbibo
 theory works with limited accuracy:
the 3-parameter $\chi^2$ fit to 21 semi-leptonic decays gives
$\chi^2/d.o.f=44.3/18$. The quality of the fit improves when 
SU(3)-breaking and other effects are taken into account~\cite{Garcia:1985xz}.

Therefore, judging by Eq.~(\ref{eq:m1}) and keeping in mind 
the limited success of the Cabbibo
theory of the hadronic semi-leptonic decays, one can conclude that the
approximate SU(3) works well for the 
ground-state octet.

The discussion of the accuracy of the SU(3) predictions for the magnetic
moments of the ground-state octet is a separate subject and
 is beyond the scope of this paper. 
We refer the interested reader to~\cite{Samios:1974tw}.

\subsection{Multiplet 15: (\textbf{8}, $5/2^+$)=(1680, 1820, 1915, 2030)}

This is a well-established octet present both in Tables~\ref{table:before}
and \ref{table:pdg}: 
the $N$, $\Lambda$  and $\Sigma$ members of the octet have a four-star rating
in the RPP; the $\Xi$ state has a three-star rating and $J^P \geq 5/2^?$.

The masses of all the states are known with high accuracy. Taking the RPP
face values for the masses, the mismatch between the left and right hand 
sides of the GMO mass formula is smaller than 0.5\% of the lowest mass
involved, $\Delta M=11.3$ MeV.
Also, the Weldon's relation among the total widths is satisfied with a fair
accuracy (we use the values from Table~\ref{table:m15})
\begin{equation}
\frac{1}{2}\left(\Gamma_N+\Gamma_{\Xi} \right)=80 \ {\rm MeV} \quad  vs. \quad
 \frac{1}{4}\left(3\, \Gamma_{\Lambda}+\Gamma_{\Sigma} \right)=90 
\ {\rm MeV} \,.
\label{eq:weldon_m15}
\end{equation}

The SU(3) predictions for the partial decay widths of the considered octet
are obtained using Eq.~(\ref{eq:width}) with the universal coupling constants
 summarized in Tables~\ref{table:octets} and \ref{table:octetsb} and
with the barrier factor of Eq.~(\ref{eq:phase_volume}) with $l=3$
for the ${\bf 8}  \to {\bf 8}+{\bf 8}$ decays and with $l=1$ for the 
${\bf 8}  \to {\bf 10}+{\bf 8}$ decays.
We perform the $\chi^2$ fit to selected experimentally measured partial decay
widths using the MINUIT program~\cite{MINUIT}. We separately fit the
${\bf 8}  \to {\bf 8}+{\bf 8}$ and ${\bf 8}  \to {\bf 10}+{\bf 8}$ decays
 because, as practice shows, SU(3) works significantly worse for the decays
 involving decuplets. The coupling constants $A_8$, $\alpha$ and $A_8^{\prime}$
are free parameters of the fit. The fit is considered successful, if
the resulting value of the $\chi^2$ function per degree of freedom is
few units. A detailed discussion of the
$\chi^2$ fit and its interpretation can be found in Sect.~{\it Probability} of the RPP~\cite{Eidelman:2004wy}.

As fitted experimental observables we take partial decay widths and
square roots of the product of two partial decay widths. The latter 
quantities can be both positive and negative, depending on the relative
phase between the two involved decay amplitudes. This sign is an important
test for SU(3) which predicts definite relative signs of the coupling
constants, see Tables~\ref{table:octets} and \ref{table:octetsb}.
Note that the use of the interference of two amplitudes as a fitted
observable is an improvement over the analysis of
Samios, Goldberg and Meadows~\cite{Samios:1974tw}, who used only 
decay rates.

Also, if the two partial decay widths that interfere correspond to the
${\bf 8}  \to {\bf 8}+{\bf 8}$ and ${\bf 8}  \to {\bf 10}+{\bf 8}$ decays,
we ignore the sign of the interference and convert it to the 
 ${\bf 8}  \to {\bf 10}+{\bf 8}$ decays width
(provided that the
corresponding  ${\bf 8}  \to {\bf 8}+{\bf 8}$ partial width is known),
 which is then used in the fit.

Throughout this review, we try to use similar sources of experimental
information on decays rates, which are all summarized in the Review of
Particle Physics~\cite{Eidelman:2004wy}. Whenever possible, we do not
use the average values or estimates, but we rather prefer to use the
original references.
For $N$ and $\Delta$ baryons, we predominantly
use the analysis of Manley and Saleski~\cite{Manley:1992yb}, which appears to be preferred
by the RPP. Also,
 some of the decays are taken 
from the analysis of Vrana, Dytman and Lee~\cite{Vrana:1999nt}. 
Most of the decay rates of strange baryons are
taken from the analysis of Gopal {\it et al.}~\cite{Gopal:1976gs,Cameron:1977jr,Gopal:1980ur},
which provides the most recent and comprehensive analysis of strange baryons. 
In several cases, we also use~\cite{Cameron:1978en}.

Table~\ref{table:m15} summarizes the results of our $\chi^2$ fit to
the nine observables of the considered octet. The observables used in the
fit are underlined. The right column presents the SU(3) predictions
for the fitted observables as well as for other observables not used in the
fit. 

As one can see from Table~\ref{table:m15}, the absolute values and
the relative signs of the measured decay rates are reproduced well.
An examination shows that the value of $\chi^2$ is dominated by the 
$\sqrt{\Gamma_{N {\overline K}} \Gamma_{\Sigma \pi}}$  of $\Lambda(1820)$
and $\Sigma(1915)$. In order to lower the $\chi^2$ to the acceptable level,
we increased the experimental error on these two observables by the
 factor $1.5$. 
% Note that in all cases, when we increased experimental errors by hand,
% the value of the multiplication factor was chosen to be the minimal value that
% gives sufficiently low $\chi^2$. 
 
This error manipulation requires an explanation. It is commonly believed that 
the accuracy of SU(3) predictions is approximately 30\%.
In the case of two-body hadronic decays, this means that we expect
that SU(3) predictions can correspond to $\chi^2/{\rm d.o.f.} \approx 1$ only when 
the experimental errors on the fitted observables are about 30\%. For the particular
example of octet 15, this means that some errors can be increased by the factor
of $1.5$ and
still do not exceed 30\%.
Also, an analysis of Bukhvostov~\cite{Bukhvostov:1997nf} shows that the results of physical
 measurements do not
follow the conventional Gaussian distribution -- the tail of the actual probability
distribution is much larger than expected on the basis of the Gaussian distribution.
This effect can be roughly simulated by increasing the dispersion of the Gaussian distribution
(experimental errors) by the factor $2-3$.

It is a phenomenological observation that approximate SU(3) works worse
for decays involving decuplets and, in particular,
for  the ${\bf 8}  \to {\bf 10}+{\bf 8}$ decays~\cite{Samios:1974tw}.
In the considered case, we increase the error on the
fitted $\Gamma_{\Delta \pi}$ by the factor $1.5$. According to our logic (see the 
discussion above), this is 
legitimate procedure since the resulting error is still $\approx 30$\%.

The $\chi^2$ fits to the seven ${\bf 8}  \to {\bf 8}+{\bf 8}$
and two ${\bf 8}  \to {\bf 10}+{\bf 8}$ decays presented in 
Table~\ref{table:m15} give
\begin{eqnarray}
A_8&=&52.0 \pm 1.3 \,, \quad \alpha=0.39 \pm 0.02 \,, \quad \chi^2/{\rm d.o.f.}=7.85/5 \,, \nonumber\\
A_8^{\prime}&=&19.2 \pm 3.4  \,, \quad  \chi^2/{\rm d.o.f.}=3.44/2 \,.
\label{eq:m15}
\end{eqnarray}
The obtained values of the coupling constants are close to those obtained 
in~\cite{Samios:1974tw}. The $\chi^2$ values are larger because the 
experimental errors, which we use in our analysis, are smaller.
The value of $\alpha$ is in excellent agreement with the SU(6) prediction
$\alpha=0.4$, see Table~\ref{table:alpha}.

\begin{table}[h]
\begin{center}
\begin{tabular}{|c c c c |}
\hline
 Mass and width (MeV) & \ Observables \ & Experiment (MeV) & SU(3) pred. (MeV) \\
\hline
$N(1684)$            & $\Gamma_{N \pi}$  &   \underline{$97.3 \pm 7.0$} &  93.2 \\
$\Gamma=139 \pm 8$ &   $\Gamma_{\Delta \pi}$  & \underline{$19.5 \pm 4.3 \times 1.5$} & 9.0 \\
&       $\Gamma_{N \eta}$ & $0.00 \pm 0.01$  & 0.2 \\              
& & & \\
$\Lambda(1823)$ &    $\Gamma_{N {\overline K}}$ & \underline{$44.7 \pm 3.3$} & 44.09 \\
 $\Gamma=77 \pm 5$ & $\sqrt{\Gamma_{N {\overline K}} \Gamma_{\Sigma \pi}}$ & 
\underline{$-21.6 \pm 2.7 \times 1.5$} & $-29.6$ \\
&                    $\sqrt{\Gamma_{N {\overline K}} \Gamma_{\Lambda \eta}}$ &
 \underline{$-6.9 \pm 3.1$} & $-4.8$ \\
&                    $\Gamma_{\Sigma(1385) \pi}$ & \underline{$3.7 \pm 2.4$} & 5.9 \\
                 & $\Gamma_{\Sigma \pi}$ & & 19.8 \\
& & & \\
$\Sigma(1920)$ & $\Gamma_{N {\overline K}}$ & \underline{$3.9 \pm 2.6$} & 4.9 \\
$\Gamma=130 \pm 10$ & $\sqrt{\Gamma_{N {\overline K}} \Gamma_{\Sigma \pi}}$ & 
\underline{$-24.7 \pm 4.3 \times 1.5$} & $-13.6$ \\
& $\sqrt{\Gamma_{N {\overline K}} \Gamma_{\Lambda \pi}}$ & \underline{$-11.7 \pm 4.0$} & $-11.4$ \\

&  $\Gamma_{\Sigma \pi}$ & & 37.4 \\
&  $\Gamma_{\Lambda \pi}$ & & 26.3 \\
& & & \\
$\Xi(2025)$ & $\Gamma_{\Sigma {\overline K}}$ & & 46.9 \\
$\Gamma=21 \pm 6$ & $\Gamma_{\Xi \pi}$ & &  4.1 \\
\hline
\end{tabular}
\caption{SU(3) analysis of (\textbf{8}, $5/2^+$)=(1680, 1820, 1915, 2030).}
\label{table:m15}
\end{center}
\end{table}

Note that the predicted large $\Gamma_{\Sigma {\overline K}}$ of 
$\Xi(2030)$ is in agreement with the experiments~\cite{Eidelman:2004wy},
 which indicate that this decay rate
is the largest. While the predicted
value of $\Gamma_{\Sigma {\overline K}}$ appears to be larger than
$\Gamma_{{\rm tot}}$, the total width of $\Xi(2030)$ is poorly known
and varies from $\Gamma_{{\rm tot}}=16 \pm 5$ MeV to
$\Gamma_{{\rm tot}}=60 \pm 24$ MeV~\cite{Eidelman:2004wy}.

Based on our results presented in Table~\ref{table:m15} and
Eq.~(\ref{eq:m15}) we conclude that SU(3) works  well for the
considered octet.

\subsection{Multiplet 12: (\textbf{8}, $5/2^-$)=(1675, 1830, 1775, 1950)}

This is an established octet present both in Tables~\ref{table:before}
and \ref{table:pdg}: 
the $N$, $\Lambda$  and $\Sigma$ members of the octet have a four-star rating
in the RPP; the $\Xi$ state has a three-star rating, but its spin 
and parity are unknown.

The masses of all the states are known with high accuracy. Taking the RPP
estimates for the masses, the mismatch between the left and right hand 
sides of the GMO mass formula is tiny, $\Delta M=-3.8$ MeV.
The Weldon's relation among the total widths is satisfied exactly
\begin{equation}
\frac{1}{2}\left(\Gamma_N+\Gamma_{\Xi} \right)=110 \ {\rm MeV} \quad  vs. \quad
 \frac{1}{4}\left(3\, \Gamma_{\Lambda}+\Gamma_{\Sigma} \right)=109
\ {\rm MeV} \,.
\label{eq:weldon_m12}
\end{equation}
\begin{table}[h]
\begin{center}
\begin{tabular}{|c c c c |}
\hline
 Mass and width (MeV) & \ Observables \ & Experiment (MeV) & SU(3) pred. (MeV) \\
\hline
$N(1676)$            & $\Gamma_{N \pi}$  &   \underline{$74.7 \pm 4.6$} &  73.3 \\
$\Gamma=159 \pm 7$ &   $\Gamma_{\Delta \pi}$  & \underline{$83.2 \pm 5.2$} & 78.3 \\
&       $\Gamma_{N \eta}$ & $0.00 \pm 0.01$  & 3.9 \\
&       $ \Gamma_{\Lambda K}$ & & 0.02 \\              
& & & \\
$\Lambda(1831)$ &    $\Gamma_{N {\overline K}}$ & \underline{$8.0 \pm 3.1$} & 4.0 \\
 $\Gamma=100 \pm 10$ & $\sqrt{\Gamma_{N {\overline K}} \Gamma_{\Sigma \pi}}$ & 
\underline{$-17.0 \pm 3.5$} & $-18.5$ \\
&                    $\Gamma_{\Sigma(1385) \pi}$ & \underline{$24.9 \pm 10.8$} & 58.6 \\
                 & $\Gamma_{\Sigma \pi}$ & & 86.0 \\
                 & $\Gamma_{\Lambda \eta}$ & & 5.0 \\
& & & \\
$\Sigma(1775)$ & $\Gamma_{N {\overline K}}$ & \underline{$54.8 \pm 4.9$} & 55.9 \\
$\Gamma=137 \pm 10$ & $\sqrt{\Gamma_{N {\overline K}} \Gamma_{\Sigma \pi}}$ & 
\underline{$17.8 \pm 3.0$} & $14.9$ \\
& $\sqrt{\Gamma_{N {\overline K}} \Gamma_{\Lambda \pi}}$ & \underline{$-38.4 \pm 5.0$} & $-42.3$ \\
&                    $\Gamma_{\Sigma(1385) \pi}$ & \underline{$11.7 \pm 1.7 \times 2$} & 6.7 \\
&  $\Gamma_{N {\overline K}}$ & & 55.9 \\
&  $\Gamma_{\Lambda \pi}$ & & 32.0 \\
& & & \\
$\Xi(1950)$ & $\Gamma_{\Sigma {\overline K}}$ & & 21.9 \\
$\Gamma=60 \pm 20$ & $\Gamma_{\Xi \pi}$ & &  84.3 \\
&  $\Gamma_{\Xi(1530) \pi}$ & & 14.2 \\
\hline
\end{tabular}
\caption{SU(3) analysis of (\textbf{8}, $5/2^-$)=(1675, 1830, 1775, 1950).}
\label{table:m12}
\end{center}
\end{table}

For this octet, the barrier factor is calculated with
$l=2$ for the ${\bf 8} \to {\bf 8}+{\bf 8}$ and 
${\bf 8} \to {\bf 10}+{\bf 8}$ decays,
see Eq.~(\ref{eq:phase_volume}).

Table~\ref{table:m12} summarizes the results of our $\chi^2$ fit to
 nine observables of the considered octet. 
As one can see from Table~\ref{table:m15}, the absolute values and
the relative signs of the measured ${\bf 8} \to {\bf 8}+{\bf 8}$
decay rates are reproduced well.
At the same time, SU(3) significantly overestimates the
$\Gamma_{\Lambda(1830) \to \Sigma(1385) \pi}$, which results in a very
large $\chi^2$, even after we have increased the error on another fitted
partial decay width, $\Gamma_{\Sigma(1775) \to \Sigma(1385) \pi}$, by the
factor of two.

Let us examine this in some more detail. The value of $\Gamma_{\Lambda(1830) \to \Sigma(1385) \pi}$
in Table~\ref{table:m12} was obtained by combining the results of two different
analyses:
 $\sqrt{\Gamma_{\Lambda(1830) \to \Sigma(1385) \pi}\Gamma_{\Lambda(1830) \to N \overline{K}}}$ 
from~\cite{Cameron:1978en} with $Br(\Lambda(1830) \to N \overline{K})$ and
$\Gamma_{\Lambda(1830)}$ from~\cite{Gopal:1980ur}. Since these are two completely different
analyses, one can imagine that the central value of $\Gamma_{\Lambda(1830) \to \Sigma(1385) \pi}$
is much more uncertain than indicated by our estimate of its experimental error.
Note also that the same situation was encountered in the analysis of 
Samios {\it et al.}~\cite{Samios:1974tw}:
the SU(3) prediction $\Gamma_{\Lambda(1830) \to \Sigma(1385) \pi}=55.1$ MeV was significantly 
larger than the experimental value $\Gamma_{\Lambda(1830) \to \Sigma(1385) \pi}=27 \pm 26$ MeV.
However, the 100\% experimental error resulted in an acceptably low  $\chi^2$.

The $\chi^2$ fits to the six ${\bf 8}  \to {\bf 8}+{\bf 8}$
and three ${\bf 8}  \to {\bf 10}+{\bf 8}$ decays presented in 
Table~\ref{table:m12} give
\begin{eqnarray}
A_8&=&26.8 \pm 0.7 \,, \quad \alpha=-0.23 \pm 0.02 \,, \quad \chi^2/{\rm d.o.f.}=3.59/4 \,, \nonumber\\
A_8^{\prime}&=&158.7 \pm 4.9 \,, \quad  \chi^2/{\rm d.o.f.}=12.66/2 \,.
\label{eq:m12}
\end{eqnarray}
The obtained values of the coupling constants are close to those obtained 
in~\cite{Samios:1974tw}. 
The value of $\alpha$ is lower than the SU(6) prediction
$\alpha=-0.5$, see Table~\ref{table:alpha}, which was also observed in the
analysis of~\cite{Samios:1974tw}.

Note that the sum of the predicted partial decay widths of 
$\Xi(1950)$ is larger than the RPP estimate for the total width
of this hyperon.  However, $\Gamma_{{\rm tot}}$ of $\Xi(1950)$
is not known well and could be much larger than the
RPP estimate~\cite{Eidelman:2004wy}. Therefore, with the present
level of accuracy, SU(3) predictions for $\Xi(1950)$ cannot be ruled out.

In conclusion, based on our results presented in Table~\ref{table:m12} and
Eq.~(\ref{eq:m12}) we conclude that SU(3) works rather well for the
${\bf 8} \to {\bf 8}+{\bf 8}$ decays of the
considered octet. At the same time, SU(3) fails to describe 
the $\Gamma_{\Lambda(1830) \to \Sigma(1385) \pi}$
partial decay width. However, since the experimental value of 
$\Gamma_{\Lambda(1830) \to \Sigma(1385) \pi}$ is extracted by 
combining two different partial wave analyses, it is intrinsically very
uncertain, which might be the cause of the inconsistency.

\subsection{Multiplets 8 and 7: (\textbf{8}, $3/2^-$)=(1520, 1690, 1670, 1820)
and (\textbf{1}, $3/2^-$)=$\Lambda$(1520)}

It has been known since the early 70's that the $\Lambda(1520)$ and
$\Lambda(1405)$ hyperons, which were thought to be SU(3) 
singlets~\cite{Tripp:1969rd}, in reality are not pure singlets but are mixed with 
$\Lambda$ hyperons
from octets with the corresponding quantum numbers.
The direct evidence for the mixing exists only for $\Lambda(1520)$, 
which decays into the $\Sigma(1385) \pi$ final state. Since a singlet
cannot decay into the ${\bf 10}+{\bf 8}$ final states, the decay can take 
place only through the mixing, presumably with $\Lambda(1690)$ from 
octet 8.

The considered octet is well-established  and is present both 
in Tables~\ref{table:before} and \ref{table:pdg}: 
the $N$, $\Lambda$  and $\Sigma$ members of the octet have a four-star rating
in the RPP; the $\Xi$ state has a three-star rating.

The masses are known with high accuracy. Taking the RPP
estimates for the masses, the mismatch between the left and right hand 
sides of the GMO mass formula is $\Delta M=-15$ MeV, i.e. it less than 1\%
of the involved masses. This illustrates that while the GMO mass relation
is satisfied with very high accuracy, the mixing might still place. In other
words, the GMO mass formulas are not sensitive to small mixing, which is the
case in the considered example.

The Weldon's relation among the total widths should be modified in the presence 
of mixing. The final result is analogous to Eq.~(\ref{eq:mixing:masses}). The accuracy of
the modified Weldon's relation is fair (we use the value of the mixing angle 
$\theta$ from Eq.~(\ref{eq:m8}) that follows)
\begin{equation}
\frac{1}{2}\left(\Gamma_N+\Gamma_{\Xi} \right)=74 \ {\rm MeV} \ vs. \
 \frac{1}{4}\left[3\, \left(\Gamma_{\Lambda(1690)} \cos^2 \theta +\Gamma_{\Lambda(1520)} \sin^2 \theta\right)+\Gamma_{\Sigma} \right]=59
\ {\rm MeV} \,.
\label{eq:weldon_m8}
\end{equation}

For the considered case of $J=3/2^-$,  
the barrier factor is calculated with
$l=2$ for the ${\bf 8} \to {\bf 8}+{\bf 8}$ decays and with 
$l=0$ for the ${\bf 8} \to {\bf 10}+{\bf 8}$ decays. 
Because of the mixing, we simultaneously fit the octet and singlet decay 
rates. Therefore, we have five free parameters: $A_8$, $\alpha$, $A_1$
and the mixing angle $\theta$ are determined from the $\chi^2$ fit to the
${\bf 8} \to {\bf 8}+{\bf 8}$ and ${\bf 1} \to {\bf 8}+{\bf 8}$ decay rates;
$A_8^{\prime}$ is determined from the fit to the ${\bf 8} \to {\bf 10}+{\bf 8}$
 decay rates. Note that the $\Lambda(1520) \to \Sigma(1385) \pi$ decay
is not allowed kinematically and, hence, cannot be used in our fit.
\begin{table}[h]
\begin{center}
\begin{tabular}{|c c c c |}
\hline
 Mass and width (MeV) & \ Observables \ & Experiment (MeV) & SU(3) pred. (MeV) \\
\hline
$N(1520)$            & $\Gamma_{N \pi}$  &   \underline{$73.2 \pm 6.0$} &  72.7 \\
$\Gamma=124 \pm 8$ &   $\Gamma_{\Delta \pi}$  & \underline{$18.6 \pm 2.8$} & 19.6 \\
&       $\Gamma_{N \eta}$ & $0.00 \pm 0.01$  & 0.2 \\              
& & & \\
$\Lambda(1690)$ &    $\Gamma_{N {\overline K}}$ & \underline{$14.3 \pm 2.2$} & 13.0 \\
 $\Gamma=61 \pm 5$ & $\sqrt{\Gamma_{N {\overline K}} \Gamma_{\Sigma \pi}}$ & 
\underline{$-15.3 \pm 2.2 \times 1.5$} & $-20.0$ \\
&                    $\Gamma_{\Sigma(1385) \pi}$ & \underline{$19.5 \pm 6.3$} & 14.1 \\
                 & $\Gamma_{\Sigma \pi}$ & & 30.7 \\
& & & \\
$\Lambda(1520)$ & $\Gamma_{N {\overline K}}$ & \underline{$6.9 \pm 0.8$} & 6.7 \\
$\Gamma=15.5 \pm 1.6$ & $\Gamma_{\Sigma \pi}$  & \underline{$6.6 \pm 0.7$} & 6.6 \\
& & & \\
$\Sigma(1682)$ & $\Gamma_{N {\overline K}}$ & \underline{$7.9 \pm 2.57$} & 6.5 \\
$\Gamma=79 \pm 10$ & $\sqrt{\Gamma_{N {\overline K}} \Gamma_{\Sigma \pi}}$ & 
\underline{$16.6 \pm 2.6$} & $18.0$ \\
& $\sqrt{\Gamma_{N {\overline K}} \Gamma_{\Lambda \pi}}$ & \underline{$7.9 \pm 1.9$} & $3.6$ \\
&                    $\Gamma_{\Sigma(1385) \pi}$ & \underline{$9.6 \pm 6.1$} & 3.1 \\
&  $\Gamma_{\Sigma \pi}$ & & 49.7 \\
& & & \\
$\Xi(1820)$ & $\Gamma_{\Sigma {\overline K}}$ & & 9.3 \\
$\Gamma=24 \pm 6$ & $\Gamma_{\Xi \pi}$ & &  8.2 \\
&  $\Gamma_{\Lambda {\overline K}}$ & & 12.3 \\
\hline
\end{tabular}
\caption{SU(3) analysis of (\textbf{8}, $3/2^-$)=(1520, 1690, 1670, 1820)
and (\textbf{1}, $3/2^-$)=$\Lambda$(1520).}
\label{table:m8}
\end{center}
\end{table}

Table~\ref{table:m8} summarizes the results of our $\chi^2$ fit to
 eleven observables of the considered octet and singlet. 
As one can see from Table~\ref{table:m8}, the absolute values and
the relative signs of the measured decay rates are reproduced well.

An examination shows that the value of the $\chi^2$ function is dominated by
the $\sqrt{\Gamma_{N {\overline K}} \Gamma_{\Sigma \pi}}$ of $\Lambda(1690)$ and
by the $\sqrt{\Gamma_{N {\overline K}} \Gamma_{\Lambda \pi}}$ of $\Sigma(1670)$.
In this respect we note that the modern value of the 
strength of the $\Sigma(1670) \to \Lambda \pi$
 decay is  smaller than the
experimental value used in the analysis of Samios {\it et al.}~\cite{Samios:1974tw}.
Also, the experimental value of 
$\sqrt{\Gamma_{N {\overline K}} \Gamma_{\Sigma \pi}}$ of $\Lambda(1690)$, which is
reported by our standard source~\cite{Gopal:1976gs} and which we used in our analysis,
is the smallest compared to all other measurements~\cite{Eidelman:2004wy}.
In order to take the theoretical uncertainty in the measurement of 
$\sqrt{\Gamma_{N {\overline K}} \Gamma_{\Sigma \pi}}$ for $\Lambda(1690)$ into
account, we increase the experimental error on this observable by $1.5$.
Note also that there is large ambiguity in the value of $\Gamma_{N \to \Delta \pi}$.
In our analysis, we use the $S$-wave value ($l=0$) for $\Gamma_{N \to \Delta \pi}$
of~\cite{Vrana:1999nt}.

The $\chi^2$ fits to the observables underlined in Table~\ref{table:m8} give
\begin{eqnarray}
A_8&=&42.7 \pm 1.5 \,, \quad \alpha=0.74 \pm 0.03 \,, \nonumber\\
A_1& = & 175.4 \pm 7.0 \,, \quad \theta=(26 \pm 2)^0 \,, \quad \chi^2/{\rm d.o.f.}=8.28/4 \,, \nonumber\\
A_8^{\prime}&=&12.7 \pm 0.9 \,, \quad  \chi^2/{\rm d.o.f.}=2.03/2 \,.
\label{eq:m8}
\end{eqnarray}
The obtained values of the coupling constants and the mixing angle are very close to those obtained 
in~\cite{Samios:1974tw}. 
The value of $\alpha$ is in fair agreement with the SU(6) prediction
$\alpha=0.625$, see Table~\ref{table:alpha}.

One should note that, in principle,  there is another candidate for the $\Sigma$ member of the
considered octet -- $\Sigma(1580)$. We studied this possibility and found that the 
inclusion of $\Sigma(1580)$ into octet 8 leads to a larger value of $\chi^2$
than in Eq.~(\ref{eq:m8}): $\chi^2/{\rm d.o.f.}=12.0/4$.
Therefore, we take $\Sigma(1670)$ as the $\Sigma$ member of octet 8.

Taking the mixing between $\Lambda(1690)$ and $\Lambda(1520)$ into account, the accuracy
of the GMO mass formula improves. The mismatch between the left and right hand sides of
Eq.~(\ref{eq:mixing:masses}) reduces to $\Delta M=8.3$ MeV.

Based on the results presented in Table~\ref{table:m8} and the acceptably low
value of $\chi^2$, see
Eq.~(\ref{eq:m8}), we conclude that SU(3) works rather well for octet 8, whose $\Lambda(1690)$
hyperon is mixed with  $\Lambda(1520)$.

\subsection{Multiplets 9 and 6: (\textbf{8}, $1/2^-$)=(1535, 1670, 1560, \underline{1620-1725})
and (\textbf{1}, $1/2^-$)=$\Lambda$(1405)}

The particle content of octet 9 is not established. In the analysis of Samios {\it et al.}~\cite{Samios:1974tw},
it was assumed that the $\Sigma$ member of the considered octet is $\Sigma(1750)$. However,
 the decay rates of $\Sigma(1750)$ were not used in the $\chi^2$ analysis. 
As we shall argue, the inclusion of $\Sigma(1750)$ in octet 9 leads to very large $\chi^2$.
Therefore, one cannot assign
 $\Sigma(1750)$ to octet 9 without mistrusting the data on the $\Sigma(1750)$ decays.
In Table~\ref{table:pdg}, the $\Sigma$ member of octet 9 is assumed to be $\Sigma(1620)$. This 
choice also gives unacceptably large $\chi^2$, $\chi^2/d.o.f. > 5$. 
 The good description of the decays of the considered
octet is achieved only by assigning $\Sigma(1560)$ to octet 9.

The $\Xi$ member of the octet is unknown. Using the modified GMO mass formula,
we predict its mass to lie in the interval $1620 < m_{\Xi} < 1725$ MeV.
The decay rates of $\Xi$ are predicted using
SU(3), see Table~\ref{table:m9}.

The $\Lambda(1405)$ has a 100\% branching ratio into the $\Sigma \pi$ final state. Therefore,
the mixing between $\Lambda(1405)$ and $\Lambda(1670)$ from octet 9 can be established
only indirectly, by noticing that the mixing provides a better simultaneous $\chi^2$ fit to the decays of
$\Lambda(1405)$ and octet~9.

Since the $N$ and $\Lambda$ states of the considered octet are established, 
we first perform a $\chi^2$ fit to the decays of $N(1535)$, $\Lambda(1670)$ and $\Lambda(1405)$.
An examination shows that if for the $N(1535)$ decay rates  one takes the results of~\cite{Manley:1992yb},
the $\chi^2$ fit fails because the large partial widths of $N(1535)$ are incompatible with the rather narrow
$\Lambda(1670)$. Therefore, in order to obtain a sensible result, we use the RPP estimates for the total width
and $Br(N \pi)$ of $N(1535)$. Note that the value of $\Gamma_{N \to N \pi}$ used in the analysis of
Samios {\it al.}~\cite{Samios:1974tw} is two times smaller than its modern value~\cite{Manley:1992yb}.

Next we begin trying different candidates for the $\Sigma$ state of the considered octet.
The $\Sigma(1750)$ does not fit this octet because SU(3) predicts the $\Gamma_{\Sigma(1750) \to N {\overline K}}$,
which is several times larger than the experimental value. The $\Sigma(1620)$ cannot be assigned
to the octet because of the same reason. Finally, $\Sigma(1690)$ cannot be a member of the
considered octet because SU(3) predicts 
$\Gamma_{\Sigma(1690) \to N {\overline K}}/\Gamma_{\Sigma(1690) \to \Lambda \pi} \approx 2.5$, which 
seriously contradicts the experimental value 
$\Gamma_{\Sigma(1690) \to N {\overline K}}/\Gamma_{\Sigma(1690) \to \Lambda \pi} \approx 0.4 \pm 0.25$ or 
even smaller~\cite{Eidelman:2004wy}.

In the appropriate mass range, the last remaining candidate for the $\Sigma$ member of octet~9 is
$\Sigma(1560)$ with the two-start rating and unknown spin and parity. 
The only measured observable of $\Sigma(1560)$, $\Gamma_{\Sigma \pi}/(\Gamma_{\Sigma \pi}+\Gamma_{\Lambda \pi})$, fits nicely the decays of octet 9 so that we assign $\Sigma(1560)$ to octet 9.

Table~\ref{table:m9} summarizes the results of our $\chi^2$ fit to
the eight observables of the considered octet and singlet. 
The absolute values and
the relative signs of the measured decay rates are reproduced well. The barrier factor 
is calculated with
$l=0$ for the ${\bf 8} \to {\bf 8}+{\bf 8}$ decays and with 
$l=2$ for the ${\bf 8} \to {\bf 10}+{\bf 8}$ decays. 

\begin{table}[h]
\begin{center}
\begin{tabular}{|c c c c |}
\hline
 Mass and width (MeV) & \ Observables \ & Experiment (MeV) & SU(3) pred. (MeV) \\
\hline
$N(1535)$            & $\Gamma_{N \pi}$  &   \underline{$67.5 \pm 27.0$} &  46.2 \\
$\Gamma=150 \pm 50$ &   $\Gamma_{\Delta \pi}$  & \underline{$0 \pm 6$} & 3.9 \\
&       $\Gamma_{N \eta}$ & $63.8 \pm 28.3$  & 20.2 \\              
& & & \\
$\Lambda(1670)$ &    $\Gamma_{N {\overline K}}$ & \underline{$5.5 \pm 1.3$} & 5.2 \\
 $\Gamma=29 \pm 5$ & $\sqrt{\Gamma_{N {\overline K}} \Gamma_{\Sigma \pi}}$ & 
\underline{$-9.0 \pm 1.8$} & $-9.7$ \\
& $\Gamma_{\Lambda \eta}$ & \underline{$8.7 \pm 2.8$} & 6.1 \\
&                    $\Gamma_{\Sigma(1385) \pi}$ & \underline{$4.6 \pm 3.6$} & 1.9 \\
                 & $\Gamma_{\Sigma \pi}$ & & 18.1 \\
& & & \\
$\Lambda(1405)$ & & & \\
$\Gamma=50 \pm 2$ & $\Gamma_{\Sigma \pi}$  & \underline{$50 \pm 2$} & 50.1 \\
& & & \\
$\Sigma(1560)$ &  $\Gamma_{\Sigma \pi}/(\Gamma_{\Sigma \pi}+\Gamma_{\Lambda \pi})$ & \underline{$0.35 \pm 0.12$} & 0.38 \\
$\Gamma=15 - 79$ & $\Gamma_{\Sigma \pi}$ & & 22.1 \\ 
 & $\Gamma_{\Lambda \pi}$ & & 36.4 \\
&  $\Gamma_{N {\overline K}}$ & & 78.9 \\
& $\Gamma_{\Sigma(1385) \pi}$ &  & 0.6 \\
& & & \\
\underline{$\Xi(1620-1725)$} & $\Gamma_{\Xi \pi}$ & & $\approx 100$ \\
&  $\Gamma_{\Lambda {\overline K}}$ & & $\approx 15$ \\
\hline
\end{tabular}
\caption{SU(3) analysis of (\textbf{8}, $1/2^-$)=(1535, 1670, 1560, \underline{1620-1725})
and (\textbf{1}, $1/2^-$)=$\Lambda$(1405)}
\label{table:m9}
\end{center}
\end{table}

The $\chi^2$ fits to the observables underlined in Table~\ref{table:m9} give
\begin{eqnarray}
A_8 & = & 7.1 \pm 1.1 \,, \quad \alpha=-0.53 \pm 0.15 \,, \nonumber\\
A_1 & = & 13.1 \pm 4.9 \,, \quad \theta=(-48 \pm 9)^0 \,, \quad \chi^2/{\rm d.o.f.}=1.67/2 \,, \nonumber\\
A_8^{\prime}&=&93.9 \pm 56.0 \,, \quad  \chi^2/{\rm d.o.f.}=0.99/1 \,.
\label{eq:m9}
\end{eqnarray}
The obtained values of the coupling constants and the mixing angle are very different from those
obtained in~\cite{Samios:1974tw}
\begin{eqnarray}
A_8 & = & 5.2 \pm 0.5 \,, \quad \alpha=-0.28 \pm 0.06 \,, \nonumber\\
A_1 & = & 26.2 \pm 1.8 \,, \quad |\theta|=(16 \pm 5)^0 \,, \quad \chi^2/{\rm d.o.f.}=3.2/1 \,, 
\nonumber\\
A_8^{\prime}&=&262 \pm 58 \,, \quad  \chi^2=0 \,.
\label{eq:m9_samios}
\end{eqnarray}
 The large difference between our analysis and that of~\cite{Samios:1974tw}  is the result
 of very significant changes in the measured
decay rates of the considered octet since 1974.
Our value of $\alpha$ is in good agreement with the SU(6) prediction
$\alpha=-0.5$, see Table~\ref{table:alpha}.

The obtained mixing angle is very large. The reason for this is as follows.
The small $\Gamma_{\Lambda(1670)\to N \overline{K}}$ forces $\Gamma_{N(1535)\to N \pi}$ to be also
small, which is in conflict with the experimental measurement.  The mixing of $\Lambda(1670)$ with
the fairly wide $\Lambda(1405)$ enables one to simultaneously have sufficiently large 
$A_8$ and small $g_{\Lambda(1670)\to N \overline{K}}$, but the resulting mixing angle has to
be large in order 
to provide the sufficient compensation of the octet and singlet contributions to
$g_{\Lambda(1670)\to N \overline{K}}$, see Eq.~(\ref{eq:mixing:cc}). Note also that quark model
calculations support our finding (which contradicts the analysis of~\cite{Samios:1974tw})
 that the mixing angle associated with 
$\Lambda(1405)$ is larger than the mixing angle associated with 
$\Lambda(1520)$~\cite{Isgur:1978xj,Loring:2001ky}.
One should keep in mind, however, 
that none of the existing quark model calculations is able to 
reproduce the small mass of  $\Lambda(1405)$.
Finally, as we explained above, the value of $\theta$ is 
indirectly affected
by the $\Gamma_{N(1535)\to N \pi}$, which is known with almost 50\% uncertainty. 
 Therefore,  the error on the obtained value
of the mixing angle $\theta$ is most likely much larger than we quote in~Eq.~(\ref{eq:m9}).

% not needed
% The nature of the well-established $\Lambda(1405)$ is a subject of ongoing discussion
% and controversy, see the note on ''The $\Lambda(1405)$'' in the 2000 edition of the 
% RPP~\cite{RPP:2000}. The two extreme possibilities discussed are that  
% $\Lambda(1405)$ is a three-quark $L=1$ flavor singlet, which mixed with the 
% octet $\Lambda(1670)$, 
% or that $\Lambda(1405)$ is an unstable $N \overline{K}$ bound state. In our analysis, we
% assumed the former scenario. If the bound-state hypothesis is correct, then there should
% be another $\Lambda$ with $J^P=1/2^-$ and the mass near the $N \overline{K}$ threshold,
%  which does not appear to be supported by the data.

Since the $\Xi$ member of octet 9 is not known, one can predict its mass using the modified
Gell-Mann--Okubo mass formula, Eq.~(\ref{eq:mixing:masses}). Using $\theta=-48^0$ in 
Eq.~(\ref{eq:mixing:masses}) leads to $m_{\Xi}=1534$ MeV, which is probably too small.
This is another reason to believe that the true mixing angle is smaller than given by
our $\chi^2$ fit.
Also, from the theoretical point of view, the large value of the mixing angle is not welcome
because this means that our basic assumption of small SU(3) violations proportional to the mass
of the strange quark is not legitimate. 
If one uses smaller values of $\theta$ in Eq.~(\ref{eq:mixing:masses}), for instance 
$15^0 < |\theta| < 35^0$, then one obtains larger values for the mass of the $\Xi$,
 $1620 < m_{\Xi} < 1725$ MeV. We assume that this is the true range, where the mass
of the missing $\Xi$ baryon lies. The decay rates of $\Xi$ in Table~\ref{table:m9} were 
predicted assuming $m_{\Xi} = 1650$ MeV.

Based on the results presented in Table~\ref{table:m9} and the  low
value of $\chi^2$, see Eq.~(\ref{eq:m9}), we conclude that SU(3) works rather well for octet 9,
 whose $\Lambda(1670)$ baryon is mixed with $\Lambda(1405)$.
Based on the modified Gell-Mann--Okubo mass formula and taking into account significant ambiguities
in the extraction of the mixing angle from the $\chi^2$ fit to the decay rates of octet 9, we 
predict the mass of the missing $\Xi$ state of octet 9 in the range
$1620 < m_{\Xi} < 1725$ MeV. This $\Xi$ hyperon 
should have $J^P=1/2^-$ and a very large partial decay
width in the $\Xi \pi$ final state, see Table~\ref{table:m9}.

Note also that the Weldon's relation among the total widths of the considered octet is badly 
violated: the $\Lambda(1670)$ mixed with $\Lambda(1405)$ and the $\Sigma(1560)$ (even if we use
$\Gamma_{\Sigma(1560)}=80$ MeV) are too narrow for the $\Gamma_{N(1535)}=150$ MeV.
A possible explanation is that the octet is too light for the Weldon's relation to work
(see the relevant discussion in Sect.~\ref{sec:intro}).

\subsection{Multiplet 3: (\textbf{8}, $1/2^+$)=(1440, 1600, 1660, 1690)}

The $N$, $\Lambda$  and $\Sigma$ members of the considered octet are well-established:
$N(1440)$ has a four-star rating in the RPP and
 $\Lambda(1600)$ and $\Sigma(1660)$ have a three-star rating.
Ignoring the one-star $\Xi(1620)$, the RPP contains only one candidate for the $\Xi$ member
of octet 3, the $\Xi(1690)$ with a three-star rating and unknown spin and parity.
Therefore, we assume that $\Xi(1690)$ belongs to the considered octet, see 
also~\cite{Diakonov:2003jj}.

The masses of the baryons from octet 3 are known with a significant uncertainty since 
different analyses reporting these baryons give rather different predictions for the masses. 
Taking the average RPP values, 
the mismatch between the left and right hand 
sides of the GMO mass formula is $\Delta M=-50$ MeV, which is $\approx 3\%$ of the mass of
$N(1440)$. If one takes the actual measured values for the masses,
$m_{N}=1462 \pm 10$ MeV~\cite{Manley:1992yb}, $m_{\Lambda}=1568 \pm 20$ MeV and
$m_{\Sigma}=1670 \pm 10$ MeV~\cite{Gopal:1980ur}, and $m_{\Xi}=1690$ (the RPP average),
the the mismatch becomes only $\Delta M=-17.5$ MeV.
Therefore, while the precision of the GMO mass formula for octet 3 is worse 
than for the previously considered cases of octets 1, 12 and 15, it is still at a few
percent level, i.e.~rather high.

The Weldon's relation among the total widths is at most qualitative
\begin{equation}
\frac{1}{2}\left(\Gamma_N+\Gamma_{\Xi} \right)=211 \ {\rm MeV} \quad  vs. \quad
 \frac{1}{4}\left(3\, \Gamma_{\Lambda}+\Gamma_{\Sigma} \right)=125
\ {\rm MeV} \,.
\label{eq:weldon_m12}
\end{equation}
Again, like in the case of octet 9, light masses of the considered octet might explain
the failure of the Weldon's relation.

Based on the observation that the GMO mass formula works only with a $\approx 3\%$
accuracy for octet 3, it was argued by Diakonov and Petrov~\cite{Diakonov:2003jj}
 that this serves as an indication of  the mixing between octet 3 and the antidecuplet
 (the mixing takes place for $N$ and $\Sigma$ states of the octet and the antidecuplet).
While, indeed, mixing with the antidecuplet, whose $N$ and $\Sigma$ states are heavier than
$N(1440)$ and $\Sigma(1660)$, respectively, might improve the accuracy of the generalized
GMO mass formula,   the $\chi^2$ analysis of the decays of octet 3 is very weakly  affected 
by the possible mixing.

As we shall show in Sect.~\ref{sec:anti10}, the current data on the decays of the
 antidecuplet strongly favor mixing of the antidecuplet with some non-exotic or
exotic multiplet. Assuming that the antidecuplet
has $J^P=1/2^+$ as predicted in the chiral quark soliton model~\cite{Diakonov:1997mm},
 the antidecuplet can mix with 
octets 1, 3 and 4 (decuplet 19 is probably too heavy to be mixed with the antidecuplet).
In Sect.~\ref{sec:anti10}, we consider the scenario that the antidecuplet mixes with octet 3. 
It is important to emphasize that since the resulting mixing angle is small and the intrinsic SU(3) 
coupling constant for the antidecuplet is expected to be small, mixing with the antidecuplet does not affect
the decay rates of octet 3. Therefore,
in the following analysis, we consider decays of octet 3 as if it were unmixed.
\begin{table}[h]
\begin{center}
\begin{tabular}{|c c c c |}
\hline
 Mass and width (MeV) & \ Observables \ & Experiment (MeV) & SU(3) pred. (MeV) \\
\hline
$N(1462)$            & $\Gamma_{N \pi}$  &   \underline{$269.8 \pm 26.2 \times 2.5$} &  152.1 \\
$\Gamma=391 \pm 34$ &   $\Gamma_{\Delta \pi}$  & \underline{$86.0 \pm 12.3 \times 1.5$} & 86.0 \\           
& & & \\
$\Lambda(1568)$ &    $\Gamma_{N {\overline K}}$ & \underline{$26.7 \pm 6.5$} & 28.7 \\
 $\Gamma=116 \pm 20$ & $\sqrt{\Gamma_{N {\overline K}} \Gamma_{\Sigma \pi}}$ & 
\underline{$-18.6 \pm 5.6 \times 2.5$} & $-34.8$ \\
&                    $\Gamma_{\Sigma \pi}$ & & 42.3 \\
                 & $\Gamma_{\Sigma(1385) \pi}$ & & 70.4 \\
& & & \\
$\Sigma(1670)$ & $\Gamma_{N {\overline K}}$ & \underline{$18.2 \pm 5.1$} & 18.8 \\
$\Gamma=152 \pm 20$ & $\sqrt{\Gamma_{N {\overline K}} \Gamma_{\Sigma \pi}}$ & 
\underline{$-24.3 \pm 5.6$} & $-20.3$ \\
& $|\sqrt{\Gamma_{N {\overline K}} \Gamma_{\Lambda \pi}}|$ & $< 6.1$ & $-27.5$ \\
&  $\Gamma_{\Sigma \pi}$ & & 21.8 \\
&  $\Gamma_{\Lambda \pi}$ & & 40.1 \\
& $\Gamma_{\Sigma(1385) \pi}$ & & 43.4 \\
& & & \\
$\Xi(1690)$ & $\Gamma_{\Xi \pi}$ & & 11.5  \\
$\Gamma < 30$ & $\Gamma_{\Xi(1530) \pi}$ & & 2.7  \\
\hline
\end{tabular}
\caption{SU(3) analysis of (\textbf{8}, $1/2^+$)=(1440, 1600, 1660, 1690).}
\label{table:m3}
\end{center}
\end{table}

Using the experimental values and errors for the decay rates of $N(1440)$~\cite{Manley:1992yb} and
$\Lambda(1600)$~\cite{Gopal:1976gs}, we find that the $\chi^2$ fit fails because it is impossible to
simultaneously accommodate a very wide $N(1440)$ with a large branching into the
$N \pi$ final state with  moderate partial decay widths of $\Lambda(1600)$. 
In order to obtain sensible results of the $\chi^2$ fit, we increase the experimental errors
on $\Gamma_{N(1440) \to N \pi}$ and $\sqrt{\Gamma_{\Lambda(1660) \to N {\overline K}}\Gamma_{\Lambda(1660) \to \Sigma \pi}}$ by the factor $2.5$.

While the exact value of the multiplication factor is somewhat arbitrary,
it is clear that the experimental information on the decays of $N(1440)$ and $\Lambda(1600)$
is ambiguous~\cite{Eidelman:2004wy}. In particular, different analyses of $N(1440)$ reporting
its total width and the branching ratio into the $N \pi$ final state give conflicting results;
the experimental value of $\sqrt{\Gamma_{\Lambda(1660) \to N {\overline K}}\Gamma_{\Lambda(1660) \to \Sigma \pi}}$, which comes from our standard source~\cite{Gopal:1976gs} and which we used in our
analysis, is much smaller than all other measurements. One way
to take the mentioned experimental inconsistency into account is to introduce the multiplication
 factor as we did, see also~\cite{Bukhvostov:1997nf}. Another possibility would be to replace the
$\chi^2$ criterion of the SU(3) testing by a new parameter-testing criterion, which would also
help to eliminate ''bugs'' in experiments and their analysis~\cite{Collins:2001es}.

Table~\ref{table:m3} summarizes the results of our $\chi^2$ fit to
 six observables of the considered octet. The barrier factor 
is calculated with
$l=1$ for the ${\bf 8} \to {\bf 8}+{\bf 8}$ and  ${\bf 8} \to {\bf 10}+{\bf 8}$ decays. 
As one can see from Table~\ref{table:m3}, SU(3) describes the decay rates of octet 3
fairly well, except for the $\Gamma_{N(1440) \to N \pi}$ and 
$|\sqrt{\Gamma_{N {\overline K}} \Gamma_{\Lambda \pi}}|$ of $\Sigma(1660)$.
However, one has to admit that the latter observable is poorly known~\cite{Eidelman:2004wy}.

The $\chi^2$ fits to the underlined observables in 
Table~\ref{table:m3} give
\begin{eqnarray}
A_8 & = & 32.4 \pm 2.5 \,, \quad \alpha=0.27 \pm 0.03 \,, \quad \chi^2/{\rm d.o.f.}=4.75/3 \,, \nonumber\\
A_8^{\prime}&=&229. \pm 16.4  \,, \quad  \chi^2=0 \,.
\label{eq:m3}
\end{eqnarray}
The value of $\alpha$ qualitatively agrees with  the SU(6) prediction
$\alpha=0.4$, see Table~\ref{table:alpha}.

We conclude that SU(3) works sufficiently well for octet 3.

\subsection{Multiplet 4: (\textbf{8}, $1/2^+$)=(1710, 1810, 1880, \underline{1950})}

The $N$, $\Lambda$  and $\Sigma$ members of the considered octet are established,
see Table~\ref{table:pdg}. The
$N(1710)$ and $\Lambda(1810)$ have a three-star rating and
$\Sigma(1880)$ has a two-star rating and the known spin and parity.
The $\Xi$ member of octet 4 is missing. We shall estimate its mass using the GMO mass 
formula.

The total width of $N(1710)$ is known with a large ambiguity. 
In order to use the same experimental sources for all multiplets, we use
for $\Gamma_{{\rm tot}}=480 \pm 230$ MeV~\cite{Manley:1992yb}, which is much larger than
the values of $\Gamma_{{\rm tot}}$ from other experiments and the RPP average.

Table~\ref{table:m4} summarizes the results of our $\chi^2$ fit to
 six observables of octet 4. The barrier factor 
is calculated with
$l=1$ for the ${\bf 8} \to {\bf 8}+{\bf 8}$ and  ${\bf 8} \to {\bf 10}+{\bf 8}$ decays. 
\begin{table}[h]
\begin{center}
\begin{tabular}{|c c c c |}
\hline
 Mass and width (MeV) & \ Observables \ & Experiment (MeV) & SU(3) pred. (MeV) \\
\hline
$N(1717)$            & $\Gamma_{N \pi}$  &   \underline{$43.2 \pm 28.2$} &  80.0 \\
$\Gamma=480 \pm 230$ &   $\Gamma_{\Delta \pi}$  & \underline{$187.2 \pm 97.6$} & 56.4 \\
     & $\Gamma_{N \eta}$  &   $28.8 \pm 14.6$ &  0.3 \\                         
& & & \\
$\Lambda(1841)$ &    $\Gamma_{N {\overline K}}$ & \underline{$39.4 \pm 8.1$} & 37.9 \\
 $\Gamma=164 \pm 20$ & $\sqrt{\Gamma_{N {\overline K}} \Gamma_{\Sigma \pi}}$ & 
\underline{$-39.4 \pm 8.1$} & $-34.6$ \\
& $\Gamma_{\Sigma(1385) \pi}$ &\underline{$22.1 \pm 25.1$} & 36.0 \\
&                    $\Gamma_{\Sigma \pi}$ & & 31.6 \\
&                  $\Gamma_{\Lambda \eta}$ & & 3.9 \\
& & & \\
$\Sigma(1826)$ & $\Gamma_{N {\overline K}}$ & \underline{$5.1 \pm 1.9$} & 5.0 \\
$\Gamma=85 \pm 15$ & $\Gamma_{\Sigma \pi}$ & & 13.6 \\
&  $\Gamma_{\Lambda \pi}$ & & 13.0 \\
& $\Gamma_{\Sigma(1385) \pi}$ & & 7.3 \\
& & & \\
\underline{$\Xi(1950)$} & $\Gamma_{\Xi \pi}$ & & 5.9  \\
& $\Gamma_{\Sigma {\overline K}}$ & & 32.4 \\
& $\Gamma_{\Xi(1530) \pi}$ & & 9.1  \\
\hline
\end{tabular}
\caption{SU(3) analysis of (\textbf{8}, $1/2^+$)=(1710, 1810, 1880, \underline{1950}).}
\label{table:m4}
\end{center}
\end{table}

The $\chi^2$ fits to the underlined observables in 
Table~\ref{table:m4} give
\begin{eqnarray}
A_8 & = & 14.9 \pm 1.1 \,, \quad \alpha=0.32 \pm 0.03 \,, \quad \chi^2/{\rm d.o.f.}=1.73/2 \,, \nonumber\\
A_8^{\prime}&=&44.9 \pm 14.4  \,, \quad  \chi^2/{\rm d.o.f.}=2.02/1 \,.
\label{eq:m4}
\end{eqnarray}
The value of $\alpha$ compares well to the SU(6) prediction
$\alpha=0.4$, see Table~\ref{table:alpha}.

Using the GMO mass formula for the considered octet, we find that the mass of the missing
$\Xi$ state of the octet should be  around 1950 MeV (this value is obtained using either the average RPP values
for the masses or the values used in Table~\ref{table:m4}).
Note that the $\Xi(1950)$, which exists in the RPP and which we assigned to octet 12, does not
fit octet 4 because, for instance, SU(3) predicts a very large 
$\Gamma_{\Sigma {\overline K}}/\Gamma_{\Lambda {\overline K}}$, which contradicts the 
data on the $\Xi(1950)$~\cite{Eidelman:2004wy}.

Our attempt to estimate the total width of the predicted $\Xi(1950)$ using
the Weldon's relation failed: the central value of the $N(1710)$ total width,
$\Gamma_{N(1710)}=480$ MeV, is way too large compared to the total widths of 
$\Lambda(1810)$ and $\Sigma(1880)$. However, since the total width of 
$N(1710)$ is very uncertain, the failure of the Weldon's relation does not mean that
the particle assignment for the octet is wrong -- it rather means that the large 
uncertainties in the measured total widths preclude the use of the Weldon's formula.

\subsection{Multiplet 17: (\textbf{8}, $3/2^+$)=(1720, 1890, 1840, \underline{2035})}

The $N$ and $\Lambda$  members of the considered octet are established,
see Table~\ref{table:pdg}. They have a four-star rating in the RPP.
An examination of the list of available $\Sigma$ baryons shows that the only candidate for
the $\Sigma$ member of the considered octet is $\Sigma(1840)$ with a one-star rating
and with the proper $J^P=3/2^+$.
The $\Xi$ member of octet 17 is missing. We shall estimate its mass using the GMO mass 
formula.

The masses and total widths of the baryons in this octet are known rather poorly. 
In our $\chi^2$ analysis, for $N(1720)$ we use the results of~\cite{Manley:1992yb}
and for $\Lambda(1890)$ we use the results of~\cite{Gopal:1980ur}.

Table~\ref{table:m17} summarizes the results of our $\chi^2$ fit to
 three observables of octet 4. The barrier  factor 
is calculated with
$l=1$ for the ${\bf 8} \to {\bf 8}+{\bf 8}$ decays. Since the 
${\bf 8} \to {\bf 10}+{\bf 8}$ decays of  $N(1720)$ and $\Lambda(1890)$ are known very poorly,
we do not attempt to fit them. Therefore, we do not determine the
$A_8^{\prime}$ coupling constant.
\begin{table}[h]
\begin{center}
\begin{tabular}{|c c c c |}
\hline
 Mass and width (MeV) & \ Observables \ & Experiment (MeV) & SU(3) pred. (MeV) \\
\hline
$N(1717)$            & $\Gamma_{N \pi}$  &   \underline{$49.4 \pm 30.1$} &  16.2 \\
$\Gamma=380 \pm 180$ &   $\Gamma_{N \eta}$  &    &  1.0 \\                         
& & & \\
$\Lambda(1897)$ &    $\Gamma_{N {\overline K}}$ & \underline{$14.8 \pm 2.5$} & 14.9 \\
 $\Gamma=74 \pm 10$ & $\sqrt{\Gamma_{N {\overline K}} \Gamma_{\Sigma \pi}}$ & 
\underline{$-6.7 \pm 2.4$} & $-7.0$ \\
&                    $\Gamma_{\Sigma \pi}$ & & 3.2 \\
& & & \\
$\Sigma(1840)$ & $\Gamma_{N {\overline K}}$ &  & 0.1 \\
$\Gamma=65-120$ & $\Gamma_{\Sigma \pi}$ & & 8.5 \\
&  $\Gamma_{\Lambda \pi}$ & & 1.2 \\
& & & \\
\underline{$\Xi(2035)$} & $\Gamma_{\Lambda {\overline K}}$ & & 2.2  \\
& $\Gamma_{\Sigma {\overline K}}$ & & 9.8 \\
\hline
\end{tabular}
\caption{SU(3) analysis of (\textbf{8}, $3/2^+$)=(1720, 1890, 1840, \underline{2035}).}
\label{table:m17}
\end{center}
\end{table}

The $\chi^2$ fits to three observables of octet 17 gives
\begin{equation}
A_8  =  6.7 \pm 0.7 \,, \quad \alpha=0.56 \pm 0.12 \,, \quad \chi^2/{\rm d.o.f.}=1.23/1 \,.
\label{eq:m17}
\end{equation}
The value of $\alpha$ compares well to the SU(6) prediction
$\alpha=0.4$, see Table~\ref{table:alpha}.

It is important to note that since the resulting value of $\alpha$ is so close to 0.5, where 
the coupling constant $g_{\Sigma \to N {\overline K}}$ vanishes, small variations in $\alpha$
lead to large variations in the predicted $\Gamma_{\Sigma \to N {\overline K}}$,
see the relevant discussion in~\cite{Samios:1974tw}.
 Therefore, the SU(3) predictions for $\Gamma_{\Sigma \to N {\overline K}}$ 
in Table~\ref{table:m17} should not be taken too literally.

Substituting $m_N=1720$ MeV, $m_{\Lambda}=1890$ MeV and $m_{\Sigma}=1840$ MeV in
the GMO mass formula, we determine the mass of the missing $ \Xi$ member of the considered
octet, $m_{\Xi}=2035$ MeV.

As a consequence of small fitted partial decay widths of the considered octet, 
SU(3) predicts that all partial decay widths in Table~\ref{table:m17} are rather
small. As a result, the total width of the predicted $ \Xi(2035)$ appears to be of the order of
10 MeV. However, if ${\bf 8} \to {\bf 10}+{\bf 8}$ decays contribute significantly 
to the large total width of $N(1720)$ (the present experimental information is uncertain),
then this would automatically imply that the ${\bf 8} \to {\bf 10}+{\bf 8}$ partial
decay widths of all baryons in the considered octet are large. Naturally, this would
significantly increase our estimate of the total width of the predicted $ \Xi(2035)$.
Until the situation with the ${\bf 8} \to {\bf 10}+{\bf 8}$ decays is settled,
we hypothesize that $ \Xi(2035)$ is rather narrow.
One should also mention that the Weldon's relation among the total widths is violated:
$N(1720)$ appears to be too wide for the octet. Note, however, that $\Gamma_{N(1720)}$ 
is very uncertain~\cite{Eidelman:2004wy} and, hence, because of this,
the Weldon's formula cannot be used.

\subsection{Multiplet 14: (\textbf{8}, $1/2^-$)=(1650, 1800, 1620, \underline{1860-1915})}

The $N$ and $\Lambda$ members of the considered octet are well-established and 
well-studied experimentally: $N(1650)$ and $\Lambda(1800)$ have a four-star and 
three-star rating in the RPP, respectively.
For the $\Sigma$ member of the considered octet, the RPP contains several candidates in
the appropriate mass range. Table~\ref{table:pdg} assigns the $\Sigma(1750)$ to octet 14.
However, our $\chi^2$ analysis demonstrates that $\Sigma(1750)$ cannot belong to octet 14
because SU(3) predicts the positive $\sqrt{\Gamma_{N {\overline K}}\Gamma_{\Sigma \pi}}$,
 while the experiment gives a negative value
for this observable. We show that a good $\chi^2$ value is obtained only if the
$\Sigma$ member of the considered octet is identified with the $\Sigma(1620)$ with a
two-star rating and the known spin and parity.
Since the $\Xi$ state of octet 14 is missing, we predict its mass using the GMO mass formula.

We begin our $\chi^2$ analysis by fitting the four underlined in Table~\ref{table:m4} 
decay rates of $N(1650)$ and $\Lambda(1800)$. Since the important for the fit branching ratio of
the $N \to N \pi$ decay is known with large experimental uncertainty, we use the RPP estimates for
$\Gamma_{{\rm tot}}$ and $Br(N \pi)$ of $N(1650)$.
After the initial $\chi^2$ fit is successful, we test the candidates for the 
$\Sigma$ member of octet 14 by adding their decay rates to the $\chi^2$ fit.
The best result is obtained with the $\Sigma(1620)$ when its decay properties are taken 
from~\cite{Langbein:1973if}.

 As to the other two candidates, SU(3)
significantly overestimates the $\Gamma_{N {\overline K}}/\Gamma_{\Lambda \pi}$ and
$\Gamma_{\Sigma \pi}/\Gamma_{\Lambda \pi}$ of $\Sigma(1690)$ and predicts the positive 
 $\sqrt{\Gamma_{N {\overline K}}\Gamma_{\Sigma \pi}}$ for $\Sigma(1750)$, which results in
an unacceptably large value of $\chi^2$ because the experiment gives a negative 
$\sqrt{\Gamma_{N {\overline K}}\Gamma_{\Sigma \pi}}$ with a sufficiently small error.

Table~\ref{table:m14} summarizes the results of our $\chi^2$ fit to
 eight observables of the considered octet, when $\Sigma(1620)$ is used.
 The barrier factor 
is calculated with
$l=0$ for the ${\bf 8} \to {\bf 8}+{\bf 8}$ decays and with $l=2$ for the
  ${\bf 8} \to {\bf 10}+{\bf 8}$ decays. 
\begin{table}[h]
\begin{center}
\begin{tabular}{|c c c c |}
\hline
 Mass and width (MeV) & \ Observables \ & Experiment (MeV) & SU(3) pred. (MeV) \\
\hline
$N(1659)$            & $\Gamma_{N \pi}$  &   \underline{$108.8 \pm 26.8$} &  68.2 \\
$\Gamma=150 \pm 10$ &   $\sqrt{\Gamma_{N \pi} \Gamma_{\Lambda K}}$  &  \underline{$-33.0 \pm 4.4$} &
  $-31.7$ \\
& $\Gamma_{\Delta \pi}$  & \underline{$2.8 \pm 1.9$} & 2.4 \\
  & $\Gamma_{N \eta}$  & $12.1 \pm 3.1$ & 22.7 \\         
& & & \\
$\Lambda(1841)$ &    $\Gamma_{N {\overline K}}$ & \underline{$82.1 \pm 11.6$} & 92.8 \\
 $\Gamma=228 \pm 20$ & $\sqrt{\Gamma_{N {\overline K}} \Gamma_{\Sigma \pi}}$ & 
\underline{$-18.2 \pm 11.5$} & $-17.7$ \\
& $\Gamma_{\Sigma(1385) \pi}$ & \underline{$2.0 \pm 2.0$} & 2.4 \\
&                    $\Gamma_{\Sigma \pi}$ & & 3.4 \\
& $\Gamma_{\Xi K}$ & & 15.2 \\
& & & \\
$\Sigma(1620)$ & $\Gamma_{N {\overline K}}$ & \underline{$14.3 \pm 4.59$} & 10.4 \\
$\Gamma=65 \pm 20$ & $|\sqrt{\Gamma_{N {\overline K}} \Gamma_{\Sigma \pi}}|$ & 
\underline{$26.0 \pm 8.9$} & $27.9$ \\
& $\Gamma_{\Sigma \pi}$ & & 75.0 \\
&  $\Gamma_{N \overline{K}}$ & & 10.4 \\
& & & \\
\underline{$\Xi(1860-1915)$} & $\Gamma_{\Xi \pi}$ & & 19.9  \\
& $\Gamma_{\Lambda \overline{K}}$  & & 31.1  \\
& $\Gamma_{\Sigma \overline{K}}$  & & 53.8  \\
& $\Gamma_{\Xi \eta}$  & & 27.7  \\
\hline
\end{tabular}
\caption{SU(3) analysis of (\textbf{8}, $1/2^-$)=(1650, 1800, 1620, \underline{1860-1915}).}
\label{table:m14}
\end{center}
\end{table}

The $\chi^2$ fits to the underlined observables in 
Table~\ref{table:m14} give
\begin{eqnarray}
A_8 & = & 8.3 \pm 0.5 \,, \quad \alpha=0.79 \pm 0.05 \,, \quad \chi^2/{\rm d.o.f.}=3.99/4
 \,, \nonumber\\
A_8^{\prime}&=&30.4 \pm 8.8  \,, \quad  \chi^2/{\rm d.o.f.}=0.08/1 \,.
\label{eq:m14}
\end{eqnarray}
The value of $\alpha$ qualitatively agrees with the SU(6) prediction
$\alpha=0.625$, see Table~\ref{table:alpha}.

The mass of the missing $\Xi$ member of the octet is estimated using the GMO mass formula. Using
the RPP estimates for the masses of $N$, $\Lambda$ and $\Sigma$ states,
 we obtain $m_{\Xi}=1860$ MeV. However, the mass of
 $\Lambda(1800)$ is known with a large uncertainty. Using for the masses the values used
in Table~\ref{table:m14}, we obtain $m_{\Xi}=1913$ MeV. Therefore, based on our SU(3) analysis
of octet 14, we predict the existence of a new $\Xi$ resonance with $J^P=1/2^-$, the mass 
in the 1860-1915 MeV range and large branching ratios to all allowed decay channels,
see Table~\ref{table:m14}.
In addition, using the Weldon's relation for the total widths, one can estimate the total 
width of the predicted $\Xi$. Using the central values for
the total widths listed in Table~\ref{table:m14}, we derive from Eq.~(\ref{eq:weldon1}) that
 $\Gamma_{\Xi(1860-1915)} \approx 220$ MeV.

\subsection{Multiplet 11: (\textbf{8}, $3/2^-$)=(1700, \underline{1850}, 1940, \underline{2045})}

Octet 11 is remarkable because this is the only octet in our scheme, which is missing
the $\Lambda$ member. All other eleven $\Lambda$ resonances, which are
 required to complete the picture of light SU(3) multiplets in 
Table~\ref{table:before}, are known very well and have three and four-star
ratings in the RPP.

The octet opens with the well-established three-star $N(1700)$. 
In addition, in the required mass range, one can offer a candidate for the 
$\Sigma$ member of the considered octet -- the three-star
$\Sigma(1940)$.
Since octet 11 lacks two states, we cannot use the GMO mass formula to
estimate the mass of the missing $\Lambda$. 
Instead, we notice that for the seven unmixed octets considered so far, 
the mass difference between the $N$ and $\Lambda$ baryons is on average 150 MeV.
 Therefore, we {\it assume} that the mass of the missing $\Lambda$ hyperon
of octet 11 is 1850 MeV. Note that the mass of $N(1700)$, which we use
as a reference point, is itself
known with a large uncertainty: $m_{N(1700)}=1650-1750$ MeV according to the 
RPP estimate~\cite{Eidelman:2004wy}. Therefore, the uncertainty in the predicted 
mass of $\Lambda(1850)$ is approximately 50 MeV.
 
The mass of the missing $\Xi$ member of the octet is then estimated using the
 GMO mass formula. Using
$m_N=1700$ MeV, $m_{\Sigma}=1940$ MeV and $m_{\Lambda}=1850$ MeV, 
 we obtain $m_{\Xi}=2045$ MeV.

Next we turn to two-body hadronic decays.
Since for $N(1700)$ the total width and the important branching into the $N \pi$ final state
are known with large ambiguity, we use the RPP estimates in our $\chi^2$ analysis. 
Note that according to the analysis of~\cite{Gopal:1976gs}, both
$\sqrt{\Gamma_{N {\overline K}}\Gamma_{\Lambda \pi}}$ and  
$\sqrt{\Gamma_{N {\overline K}}\Gamma_{\Sigma \pi}}$ of $\Sigma(1940)$ are negative.
This contradicts SU(3): expecting that $\alpha$ 
will be close to its SU(6) prediction $\alpha=-1/2$, we notice that SU(3) requires that the
signs of $\sqrt{\Gamma_{N {\overline K}}\Gamma_{\Lambda \pi}}$ and 
$\sqrt{\Gamma_{N {\overline K}}\Gamma_{\Sigma \pi}}$ should be opposite,
see Table~\ref{table:octets}.
SU(3) also requires the opposite signs, if $\Sigma(1940)$ belongs to a decuplet.
Therefore, we reverse the sign of $\sqrt{\Gamma_{N {\overline K}}\Gamma_{\Sigma \pi}}$.
This is consistent with the analysis~\cite{Martin:1977md}, which 
reports the positive value for $\sqrt{\Gamma_{N {\overline K}}\Gamma_{\Sigma \pi}}$,
which is somewhat larger (no errors are given) than the value from the 
analysis~\cite{Gopal:1976gs}.

Table~\ref{table:m11} summarizes the results of our $\chi^2$ fit to
 three observables of the considered octet.
 The barrier factor 
is calculated with $l=2$.
\begin{table}[h]
\begin{center}
\begin{tabular}{|c c c c |}
\hline
 Mass and width (MeV) & \ Observables \ & Experiment (MeV) & SU(3) pred. (MeV) \\
\hline
$N(1700)$             & $\Gamma_{N \pi}$  &   \underline{$10.0 \pm 7.1$} &  7.9 \\
$\Gamma=100 \pm 50$   &  $\Gamma_{N \eta}$  & $0 \pm 1$ & 2.0 \\
                     &  $\Gamma_{\Delta \pi}$, $D$-wave & \underline{$14.4 \pm 17.0$} & 17.9 \\
& & & \\
\underline{$\Lambda(1850)$} &    $\Gamma_{N {\overline K}}$ & & 0.2 \\
 & $\Gamma_{\Sigma \pi}$ & & $17.8$ \\
& $\Gamma_{\Lambda \eta}$ & & 1.2 \\
& $\Gamma_{\Sigma(1385) \pi}$ & & $12.8$ \\
& & & \\
$\Sigma(1940)$ &  $|\sqrt{\Gamma_{N {\overline K}} \Gamma_{\Sigma \pi}}|$ & 
\underline{$24.0 \pm 13.6$} & $18.4$ \\
$\Gamma=300 \pm 80$ & $\sqrt{\Gamma_{N {\overline K}} \Gamma_{\Lambda \pi}}$ & 
\underline{$-18.0 \pm 10.2$} & $-22.2$ \\
&                    $\Gamma_{\Delta \overline{K}}$, $D$-wave & \underline{$47.2 \pm 42.0$} & 10.6 \\
& $\Gamma_{\Sigma \pi}$ & & 9.1 \\
& $\Gamma_{\Lambda \pi}$ & & 13.1 \\
&  $\Gamma_{N \overline{K}}$ & & 37.4 \\
&  $\Gamma_{\Sigma(1385)\pi}$ & & 5.4 \\
& & & \\
\underline{$\Xi(2045)$} & $\Gamma_{\Xi \pi}$ & & 39.5  \\
& $\Gamma_{\Lambda \overline{K}}$  & & 12.3  \\
& $\Gamma_{\Sigma \overline{K}}$  & & 4.8  \\
& $\Gamma_{\Xi(1530) \pi}$  & & 6.7  \\
& $\Gamma_{\Sigma(1385) \overline{K}}$  & & 2.6  \\
\hline
\end{tabular}
\caption{SU(3) analysis of (\textbf{8}, $3/2^-$)=(1700, \underline{1850}, 1940, \underline{2045}).}
\label{table:m11}
\end{center}
\end{table}

The $\chi^2$ fit to the five underlined observables in 
Table~\ref{table:m11} gives
\begin{eqnarray}
A_8  & =&   8.3 \pm 3.5 \,, \quad \alpha=-0.70 \pm 0.54 \,, \quad \chi^2/{\rm d.o.f.}=0.42/1 \,, 
\nonumber\\
A_8^{\prime}& =& 67.2 \pm 31.0 \,, \quad \chi^2/{\rm d.o.f.}=0.8/1 \,. 
\label{eq:m11}
\end{eqnarray}
The central value of $\alpha$ compares well to its SU(6) prediction
$\alpha=-1/2$. However, since we used only two fitted observables,
 which depend on $\alpha$, the error on the obtained value of 
$\alpha$ is large.

Note that in order to convert the experimentally measured 
$\sqrt{Br(N {\overline K}) Br(\Delta \overline{K})}$ 
of $\Sigma(1940)$ into the corresponding $\Gamma_{\Delta \overline{K}}$ 
 used in the fit, we used the SU(3) prediction 
$\Gamma_{\Sigma(1940)\to N {\overline K}}=37.4$ MeV. 
Also, we chose to fit only the $D$-wave ($l=2$) ${\bf 8} \to {\bf 10}+{\bf 8}$ decays
because the $S$-wave $N(1700) \to \Delta \pi$ branching ratio is rather uncertain and
is smaller than the corresponding $D$-wave branching~\cite{Eidelman:2004wy}.

An examination of SU(3) predictions in Table~\ref{table:m11} shows that the sum of
predicted two-body hadronic decay widths significantly underestimates the known total
widths of $N(1700)$ and $\Sigma(1940)$: $\Gamma_{N(1700)}^{{\rm SU(3)}}=28$ MeV 
vs.~$\Gamma_{N(1700)}=100 \pm 50$ MeV and
$\Gamma_{\Sigma(1940)}^{{\rm SU(3)}}=77$ MeV 
vs.~$\Gamma_{\Sigma(1940)}=300 \pm 80$ MeV. In both cases, the central value of the total width
is underestimated by the factor of  $3.5-4$. Therefore,
in order to obtain a realistic estimate for the total width of the 
predicted $\Lambda(1850)$,
we simply multiply the sum of the SU(3) predictions for the two-body hadronic decays
 by the factor four
\begin{equation}
\Gamma_{\Lambda(1850)}=4 \times 32 \ {\rm MeV} \approx 128 \ {\rm MeV} \,.
\end{equation}

We can estimate the total
width of the predicted $\Xi(2045)$ in a similar way
\begin{equation}
\Gamma_{\Xi(2045)}=4 \times 66 \ {\rm MeV} = 264 \ {\rm MeV} \,.
\end{equation}

Now that all total widths are in place, one can check how accurately Weldon's
relations for the total widths are satisfied. Substituting
the central values of the total widths into Eq.~(\ref{eq:weldon1}), 
we obtain
\begin{equation}
\frac{1}{2}\left(\Gamma_N+\Gamma_{\Xi} \right)=182 \ {\rm MeV} \quad  vs. \quad
 \frac{1}{4}\left(3\, \Gamma_{\Lambda}+\Gamma_{\Sigma} \right)=171 
\ {\rm MeV} \,.
\label{eq:agreement}
\end{equation}
While the nice agreement seen in Eq.~(\ref{eq:agreement}) should not be 
taken literally because of large uncertainties in the measured total widths
of $N(1700)$ and $\Sigma(1940)$, it still illustrates that our method of
estimating the total widths of $\Lambda(1850)$ and $\Xi(2045)$ has
a certain merit.

A remarkable property of $\Lambda(1850)$ is its vanishingly small coupling
to the $N \overline{K}$ state, see Table~\ref{table:m11}. This is a consequence
of the fact that $\alpha=-0.70 \pm 0.54$, which strongly suppresses the
$g_{\Lambda(1850) \to N \overline{K}}$ coupling constant,
$g_{\Lambda \to N \overline{K}}=\sqrt{2/3} (2 \alpha +1) A_8$, 
see Table~\ref{table:octets}. In the SU(6) limit, $\alpha=-1/2$, 
see Table~\ref{table:alpha}, which leads to 
$g_{\Lambda \to N \overline{K}}=0$.
Therefore, our prediction that $\Lambda(1850)$ very weakly couples to the
$N \overline{K}$ final state is rather model-independent.

As follows from Table~\ref{table:m11}, SU(3) predicts that the $\Lambda(1850)$ 
has significant branching ratios into the
$\Sigma \pi$ and $\Sigma(1385) \pi$ final states. This suggests that
one should experimentally search for the $\Lambda(1850)$
in production reactions using the $\Sigma \pi$ and $\Sigma(1850) \pi$ invariant
mass spectrum.

The existence of a new $\Lambda$ hyperon with $J^P=3/2^-$ was predicted in
different constituent quark models. 
In 1978, Isgur and Karl predicted that the new $\Lambda$
has the mass 1880 MeV and a very small coupling to the $N \overline{K}$ state. The latter
fact is a consequence of SU(6) selection rules and explains why this state was not
observed in the $N \overline{K}$ partial wave 
analyses~\cite{Gopal:1976gs,Cameron:1977jr,Gopal:1980ur}.
In a subsequent analysis, Isgur and Koniuk explicitly calculated
the partial decay widths of the  $\Lambda(1880)$ and found that 
$\Gamma_{N \overline{K}}$ is small, while $\Gamma_{\Sigma \pi}$ and
$\Gamma_{\Sigma(1385) \pi}$ are dominant~\cite{Koniuk:1979vy}.

More recent calculations within the  constituent quark model framework 
also predict the existence of a new $\Lambda$ with $J^P=3/2^-$, but with somewhat different
masses: the analysis of L\"oring, Metsch and Petry~\cite{Loring:2001ky} (model A)   
gives 1775 MeV; the analysis of  Glozman, Plessas, Varga and Wagenbrunn~\cite{Glozman:1997ag}
gives $\approx 1780$ MeV. Note also that the analysis~\cite{Loring:2001ky}
predicts that the $\Lambda$ very weakly couples to the  $N \overline{K}$ state.

We would like to emphasize that while many results concerning the new $\Lambda$ were
previously derived in specific constituent quark models with various
assumptions about the quark dynamics, we demonstrate that they
are actually model-independent and follow directly from
flavor SU(3) symmetry.
 
In conclusion, the existence of a new $\Lambda$ hyperon with $J^P=3/2^-$ is
required by the general principle of the flavor SU(3) symmetry of strong 
interactions. Our SU(3) analysis predicts that its mass is $\approx 1850$ MeV
 and the total width is $\approx 130$ MeV. We predict that 
$\Lambda(1850)$ has a very small coupling to the $N \overline{K}$ state and 
large branching ratios into the $\Sigma \pi$ and $\Sigma(1385) \pi$ final states.   
Therefore, $\Lambda(1850)$ can be searched for in production reactions by
studying the $\Sigma \pi$ and $\Sigma(1385) \pi$ invariant mass spectra. 
The fact that the total width of
$\Lambda(1850)$ is not larger than $\approx 130$ MeV makes the experimental
 search feasible. In addition, in order to have  the complete octet, we predict the existence
of a new $\Xi$ baryon with $J^P=3/2^-$,
the mass $\approx 2045$ MeV and the total width 
$\approx 265$ MeV.

\section{SU(3) classification of decuplets}
\label{sec:decuplets}

In this section, we perform the SU(3) classification of eight decuplets
of Table~\ref{table:before}. Since the $\Delta$ baryons that open those decuplets
are well-established, SU(3) requires the existence of the corresponding
 $\Sigma$, $\Xi$ and $\Omega$
members of the considered decuplets. In many cases those states are missing 
-- SU(3) then makes predictions for  their spin and parity
and estimates of their  masses and decay widths.
 Similarly to the analysis of the octets
presented above, the main tools
of our analysis are the equal spacing rule for the mass splitting in 
a given decuplet, Eq.~(\ref{eq:gmo10}), and 
SU(3) predictions for the partial decay widths. 

The bulk of the experimental information on two-body hadronic decays of 
decuplets comes from ${\bf 10} \to {\bf 8}+{\bf 8}$ decays. In the SU(3)
 limit, the $B_1 \to B_2 \,P$ coupling constants are parameterized in terms of
a single universal constant $A_{10}$
\begin{equation}
g_{B_1 B_2 P }=A_{10}\ \left(
\begin{array}{cc}
8 & 8 \\
Y_2 T_2 & Y_P T_P
\end{array}\right|\left.\begin{array}{c}
          10 \\Y_1 T_1
          \end{array}\right) \,.
\label{eq:1088}
\end{equation}
In addition, there is less precise and complete information on
${\bf 10} \to {\bf 10}+{\bf 8}$ decays, whose coupling constants are 
parameterized in terms of a constant $A_{10}^{\prime}$
\begin{equation}
g_{B_1 B_2 P }=A_{10}^{\prime}\ \left(
\begin{array}{cc}
10 & 8 \\
Y_2 T_2 & Y_P T_P
\end{array}\right|\left.\begin{array}{c}
          10 \\Y_1 T_1
          \end{array}\right) \,.
\label{eq:10108}
\end{equation}
The coupling constants for all possible decay channels of decuplets
are summarized in Table~\ref{table:decuplets}. 
\begin{table}[h]
\begin{center}
\begin{tabular}{|c c c |}
\hline
 & ${\bf 10} \to {\bf 8}+{\bf 8}$ & ${\bf 10} \to {\bf 10}+{\bf 8}$  \\
Decay mode & $g_{B_1 B_2 P}$ &  $g_{B_1 B_2 P}$ \\
\hline
$\Delta \to N \pi$ & $-(\sqrt{2}/2)\, A_{10}$ &  \\
$ \to \Sigma K$ & $(\sqrt{2}/2)\, A_{10}$ &  \\
$ \to \Delta \pi$ & & $(\sqrt{10}/4)\,A_{10}^{\prime}$ \\
$ \to \Delta \eta$ & & $-(\sqrt{2}/4)\,A_{10}^{\prime}$ \\
$ \to \Sigma^{\ast} K$ & & $1/2\,A_{10}^{\prime}$ \\
\hline
$\Sigma  \to   \Lambda \pi$ & $-1/2\,A_{10}$ & \\
$\to \Sigma \pi, \ \Xi K$ & $(\sqrt{6}/6)\,A_{10}$ & \\
$ \to N {\overline K}$ & $-(\sqrt{6}/6)\,A_{10}$ & \\
$\to \Sigma \eta$ & $1/2\,A_{10}$ & \\
$\to \Sigma^{\ast} \pi, \ \Xi^{\ast} K, \, \Delta {\overline K} $ &
 & $(\sqrt{3}/6)\,A_{10}^{\prime}$  \\
$\to \Sigma ^{\ast}\eta $ & & $0$  \\
\hline
$\Xi \to \Xi \pi, \ \Xi \eta, \Sigma {\overline K}$ & $1/2\,A_{10}$ & \\
$ \to \Lambda {\overline K}$ & $-1/2\,A_{10}$ & \\
$ \to \Xi^{\ast} \pi, \ \Xi^{\ast} \eta$ & & $(\sqrt{2}/4)\,A_{10}^{\prime}$  \\
$ \to \Sigma^{\ast} {\overline K}$ & & $(\sqrt{2}/2)\,A_{10}^{\prime}$  \\
$ \to \Omega K$ & & $1/2\,A_{10}^{\prime}$  \\
\hline
$\Omega \to \Xi {\overline K}$ & $A_{10}$ & \\
$\to \Xi^{\ast} {\overline K}, \ \Omega \eta$ & & $(\sqrt{2}/2)\,A_{10}^{\prime}$  \\
\hline
\end{tabular}
\caption{The SU(3) universal coupling constants for ${\bf 10} \to {\bf 8}+{\bf 8}$ and ${\bf 10} \to {\bf 10}+{\bf 8}$ decays.}
\label{table:decuplets}
\end{center}
\end{table}

The SU(3) classification of decuplets is somewhat  more ambiguous than
for the octets~\cite{Samios:1974tw}. Since there are
fewer free parameters for the $\chi^2$ fit to the decays rates into the
${\bf 8}+{\bf 8}$ final state
($A_{10}$ for decuplets vs.~$A_8$ and $\alpha$ for octets), we expect that 
the $\chi^2$ fit should be somewhat less successful. We begin our analysis by first 
considering the decuplets already analyzed in~\cite{Samios:1974tw}.

\subsection{Multiplet 2: (\textbf{10}, $3/2^+$)=(1232, 1385, 1530, 1672)}

Naturally, the ground-state decuplet (decuplet 2) is very well
 established (all its states 
have a four-star rating in the RPP) and  its decay rates are known with
very high precision. Because of the small experimental errors on the measured
two-body hadronic partial decay widths, an attempt of the $\chi^2$ fit to 
the decay rates returns an unacceptably large $\chi^2$. At this point, 
we have two options. First, one can increase the experimental errors in
several times because one should not expect that SU(3) works with a few percent
 accuracy. Second, one can try other models for the kinematic phase space
factor. As shown by Samios {\it et al.}~\cite{Samios:1974tw}, there are models
for the phase space factor, which give a much smaller $\chi^2$ with the same
experimental input. In our analysis, we used the former approach and
increased the experimental errors on all four used decay rates of the ground-state
decuplet by factor two.

The Gell-Mann--Okubo mass formula (the equal splitting rule) works very well for
decuplet~2. Using Eq.~(\ref{spld2}), one finds that the standard 
deviation from the average spacing, $\Delta z$, is much smaller than the average
spacing $\langle z \rangle$, $\Delta z=5.7$ MeV and $\langle z \rangle=146.7$ MeV.

The Weldon's relation for the total widths of the ground-state decuplet
reads~\cite{Weldon:1977wf}
\begin{equation}
\frac{1}{4} \Gamma_{\Delta}+\frac{3}{4} \Gamma_{\Xi}=
\frac{1}{4} \Gamma_{\Omega}+\frac{3}{4} \Gamma_{\Sigma} \,.
\label{eq:weldon_gs}
\end{equation}
The use of the RPP values for the total widths gives
\begin{equation}
\frac{1}{4} \Gamma_{\Delta}+\frac{3}{4} \Gamma_{\Xi}=36.3 \ {\rm MeV} \quad
vs.  \quad \frac{1}{4} \Gamma_{\Omega}+\frac{3}{4} \Gamma_{\Sigma}=26.8 \ {\rm MeV} \,.
\end{equation}
Note the values of the total widths have changed since 1977, when the 
agreement of Eq.~(\ref{eq:weldon_gs}) was much better~\cite{Weldon:1977wf}.

\begin{table}[h]
\begin{center}
\begin{tabular}{|c c c c |}
\hline
 Mass and width (MeV) & \ Observables \ & Experiment (MeV) & SU(3) pred. (MeV) \\
\hline
$\Delta(1232)$            & $\Gamma_{N \pi}$  &   \underline{$118 \pm 4 \times 2$} & 100.0  \\
$\Gamma=118 \pm 4$ &   &  & \\         
& & & \\
$\Sigma(1385)$ &    $\Gamma_{\Lambda \pi}$ & \underline{$31.5 \pm 1.0 \times 2$}  & 33.1 \\
$\Gamma=35.8 \pm 0.8$ & $\Gamma_{\Sigma \pi}$ & \underline{$4.30 \pm 0.72 \times 2$}  & 5.1 \\
& & & \\
$\Xi(1530)$ &  $\Gamma_{\Xi \pi}$ & \underline{$9.1 \pm 0.5 \times 2$}  & 10.4 \\
$\Gamma=9.1 \pm 0.5$ & & & \\
& & & \\
$\Omega(1672)$ & & & \\
 $\Gamma \approx 0$ & & & \\
\hline
\end{tabular}
\caption{SU(3) analysis of (\textbf{10}, $3/2^+$)=(1232, 1385, 1530, 1672).}
\label{table:m2}
\end{center}
\end{table}

Table~\ref{table:m2} summarizes the results of our $\chi^2$ fit.
 The barrier factor is calculated with $l=1$.
The $\chi^2$ fit gives
\begin{equation}
A_{10}  =  142.1 \pm 3.0 \,,  \quad \chi^2/{\rm d.o.f.}=7.59/3 \,.
\label{eq:m2}
\end{equation}

As discussed in the beginning of this subsection, the goodness of the $\chi^2$ fit can be 
improved by using a different phase space factor. For instance, the $\chi^2$ fit 
 with the phase space factor of Eq.~(\ref{eq:phase_volume}) 
multiplied by the $M/M_1$ factor, where $M_1$ is the mass of the decaying baryon
and $M=1000$ MeV,
gives a much lower $\chi^2$
\begin{equation}
A_{10}  =  166.2 \pm 3.5 \,,  \quad \chi^2/{\rm d.o.f.}=1.46/3 \,.
\label{eq:m2_b}
\end{equation}
This conclusion was first obtained in~\cite{Samios:1974tw}.
Note also that the improvement of the $\chi^2$ value by modifying the phase space factor 
is so dramatic only for decuplet 2. For other decuplets, whose spin and parity
are different and decay rates are not known with such good precision, the change of the
phase space factor does not systematically lead to the improvement of the fit. Therefore, 
throughout our analysis, we shall use the phase space factor as given by Eq.~(\ref{eq:phase_volume}).

In summary, SU(3) works very well for decuplet 2.

\subsection{Multiplet 20: (\textbf{10}, $7/2^+$)=(1950, 2030, 2120, 2250)}

Multiplet 20 was considered by Samios {\it et al.}~\cite{Samios:1974tw} in 1974,
when only the $\Delta(1950)$ and $\Sigma(2030)$ members of the decuplet were
known. Using the equal spacing rule, the new $\Xi(2120)$ and $\Omega(2250)$
resonances were predicted (we use the modern RPP average masses). Later those baryons were discovered:
$\Xi(2120)$ has a one-star rating in the RPP and $\Omega(2250)$ has a 
three-star rating.
The spin and parity of the both baryons are not known.

Using the RPP estimates for the masses, we find that the equal spacing rule works with
mediocre accuracy for decuplet 20: $\Delta z=26.5$ MeV and $\langle z \rangle=100$ MeV.
Note that the equal spacing rule for the total widths predicted by the Weldon's formula 
does not hold. 

Table~\ref{table:m20} summarizes the results of our $\chi^2$ fit to the six underlined
observables. The barrier factor is calculated with $l=3$.
\begin{table}[h]
\begin{center}
\begin{tabular}{|c c c c |}
\hline
 Mass and width (MeV) & \ Observables \ & Experiment (MeV) & SU(3) pred. (MeV) \\
\hline
$\Delta(1950)$            & $\Gamma_{N \pi}$  &   \underline{$114 \pm 4$} & 115.8  \\
$\Gamma=300 \pm 7$ & $\Gamma_{\Delta \pi}$  & \underline{$57.6 \pm 8.8$}  & 58.4 \\
& $\sqrt{\Gamma_{N \pi}\Gamma_{\Sigma K}}$ & $-15.9 \pm 1.6$ & $-21.2$ \\          
& & & \\
$\Sigma(2030)$ &    $\Gamma_{N {\overline K}}$ & \underline{$32.7 \pm 5.5$}  & 24.9 \\
$\Gamma=172 \pm 10$ & $\sqrt{\Gamma_{N {\overline K}}\Gamma_{\Lambda \pi}}$
 & \underline{$31.0 \pm 3.9$}  & 30.4 \\
&  $\sqrt{\Gamma_{N {\overline K}}\Gamma_{\Sigma \pi}}$
 & \underline{$-25.8 \pm 5.4$}  & $-19.9$ \\
& $\Gamma_{\Delta {\overline K}}$ & \underline{$23.2 \pm 9.6$}  & $2.7$ \\
& $\Gamma_{\Lambda \pi}$ & & 31.1 \\
& & & \\
$\Xi(2120)$ &  $\Gamma_{\Xi \pi}$ &   & 18.8 \\
$\Gamma=25 \pm 12$ & $\Gamma_{\Lambda {\overline K}}$  & & 23.1  \\
& $\Gamma_{\Sigma {\overline K}}$  & & 13.1  \\
& & & \\
$\Omega(2250)$ & & & \\
$\Gamma=55 \pm 18$ & $\Gamma_{\Xi {\overline K}}$  & & 55.7  \\
 & & & \\
\hline
\end{tabular}
\caption{SU(3) analysis of (\textbf{10}, $7/2^+$)=(1950, 2030, 2120, 2250).}
\label{table:m20}
\end{center}
\end{table}

The $\chi^2$ fit to the underlined observables in 
Table~\ref{table:m20} gives
\begin{eqnarray}
A_{10} & = & 60.7 \pm 1.0 \,,  \quad \chi^2/{\rm d.o.f.}=4.02/3 \,, \nonumber\\
A_{10}^{\prime} & = & 94.1 \pm 7.1 \,,  \quad \chi^2/{\rm d.o.f.}=4.63/1 \,. 
\label{eq:m20}
\end{eqnarray}

Since the $\Xi(2120)$ has only a one-star rating in the RPP, we do not consider the fact
 that the sum of the predicted partial decays widths
is greater than the total widths of $\Xi(2120)$ as a contradiction.
If the decay rates of $\Xi(2120)$ are measured in the future, they can be included in the $\chi^2$ fit,
which will adjust the resulting $A_{10}$ and $A_{10}^{\prime}$ to the experiment.

%Based on the SU(3) analysis of decuplet 20 summarized in Table~\ref{table:m20},
%we conclude that SU(3) works rather well for the considered decuplet.

\subsection{Multiplet 18: (\textbf{10}, $5/2^+$)=(1905, 2070, 2250, 2380)}

The content of multiplet 18 is not established: only the obvious four-star $\Delta(1905)$ member
is listed in Tables~\ref{table:before} and \ref{table:pdg}. However, 
an examination of the RPP shows
 that
one can offer candidates within a suitable mass range for all members of the considered decuplet.
 These are the one-star $\Sigma(2070)$ with $J^P=5/2^+$, the two-star $\Xi(2250)$ with unknown
spin and parity and the two-star $\Omega(2380)$ with unknown
spin and parity. 
Assuming that multiplet 18 consists of $\Delta(1905)$, $\Sigma(2070)$, $\Xi(2250)$ and 
$\Omega(2380)$, we find that
the equal spacing rule for the mass splitting works with
fair accuracy: $\Delta z=25.7$ MeV and $\langle z \rangle=158$ MeV.
For the total widths, the equal spacing rule is qualitatively fulfilled, see the values in
Table~\ref{table:m18}.

Table~\ref{table:m18} summarizes the results of our $\chi^2$ fit to the underlined
observables. The barrier factor is calculated with $l=3$ for the
${\bf 10} \to {\bf 8}+{\bf 8}$ decays and with  $l=1$ ($P$-wave) for the
${\bf 10} \to {\bf 10}+{\bf 8}$ decays. Note that in order to have a successful fit, 
we increased the experimental error on the $\Gamma_{\Sigma(2070) \to N {\overline K}}$ by
the factor two (the resulting experimental error is still smaller than
30\%, see the relevant discussion in Sect.~\ref{sec:octets}).
 Note also that SU(3) predicts the negative sign for 
$\sqrt{\Gamma_{N {\overline K}}\Gamma_{\Sigma \pi}}$ of any $\Sigma$ member of  decuplets, while
$\sqrt{\Gamma_{N {\overline K}}\Gamma_{\Sigma \pi}}> 0$ for $\Sigma(2070)$ 
experimentally~\cite{Kane:1972qa}.
Since the analysis of~\cite{Kane:1972qa} is not our standard source of information on 
hyperons~\cite{Gopal:1976gs,Cameron:1977jr,Gopal:1980ur}, we ignore this
inconsistency.

\begin{table}[h]
\begin{center}
\begin{tabular}{|c c c c |}
\hline
 Mass and width (MeV) & \ Observables \ & Experiment (MeV) & SU(3) pred. (MeV) \\
\hline
$\Delta(1881 \pm 18)$            & $\Gamma_{N \pi}$  &   \underline{$39.2 \pm 11.6$} & 45.1  \\
$\Gamma=327 \pm 51$ &  $\sqrt{\Gamma_{N \pi}\Gamma_{\Sigma K}}$ &
\underline{$-4.9 \pm 1.2$} & $-5.7$  \\
& $\Gamma_{\Delta \pi}$  & \underline{$75.2 \pm 12.2$}  & 75.2 \\          
& & & \\
$\Sigma(2051 \pm 25)$ &    $\Gamma_{N {\overline K}}$ & \underline{$24.0 \pm 2.6 \times 2$}  & 16.3 \\
$\Gamma=300 \pm 30$ & $\sqrt{\Gamma_{N {\overline K}}\Gamma_{\Sigma \pi}}$ & $31.2 \pm 6.8$ &
$-13.1$ \\
& $\Gamma_{\Lambda \pi}$ & & 24.1 \\
& $\Gamma_{\Sigma \pi}$ & & 10.5 \\
& $\Gamma_{\Sigma(1385) \pi}$ & & 10.4 \\
& & & \\
$\Xi(2250)$ &  $\Gamma_{\Xi \pi}$ &   & 25.6 \\
$\Gamma=60 - 150$& $\Gamma_{\Lambda {\overline K}}$  & & 32.4  \\
& $\Gamma_{\Sigma {\overline K}}$  & & 20.9 \\
& $\Gamma_{\Sigma(1385) {\overline K}}$  & & 63.3 \\
& & & \\
$\Omega(2380)$ & $\Gamma_{\Xi {\overline K}}$ & & 89.6 \\
$\Gamma=26 \pm 23$ & $\Gamma_{\Xi(1385) {\overline K}}$  & & 58.4  \\
 & & & \\
\hline
\end{tabular}
\caption{SU(3) analysis of (\textbf{10}, $5/2^+$)=(1905, 2070, 2250, 2380).}
\label{table:m18}
\end{center}
\end{table}

As can be seen from Table~\ref{table:m18}, SU(3) predicts large partial decay rates 
of $\Omega(2380)$, whose sum exceeds the
estimate for the total width of $\Omega(2380)$. However, until the decay rates of 
$\Omega(2380)$ are measured and used in the $\chi^2$ fit, one cannot conclude that 
our predictions for $\Omega(2380)$ are ruled out
(see also 
the appropriate discussion in the previous subsection).

The $\chi^2$ fit to the underlined observables in 
Table~\ref{table:m18} gives
\begin{eqnarray}
A_{10} & = & 45.9 \pm 3.4 \,,  \quad \chi^2/{\rm d.o.f.}=2.89/2 \,, \nonumber\\
A_{10}^{\prime} & = & 39.7 \pm 3.2 \,,  \quad \chi^2=0 \,. 
\label{eq:m18}
\end{eqnarray}
Based on the sufficient accuracy of the Gell-Mann--Okubo mass formula and the results
presented in Table~\ref{table:m18} and Eq.~(\ref{eq:m18}), we conclude that SU(3) works 
well for decuplet 18. However, one should keep in mind the problem with the sign
of $\sqrt{\Gamma_{N {\overline K}}\Gamma_{\Sigma \pi}}$ for $\Sigma(2070)$.

\subsection{Multiplet 16: (\textbf{10}, $3/2^+$)=(1920, 2080, \underline{2240}, 2470)}

The content of multiplet 16 is known even worse than that of decuplet 18: decuplet 16
opens with a three-star $\Delta(1920)$, see Table~\ref{table:before}, and the other members
 are not established. Assuming that the mass splitting for the decuplet is around 150 MeV
(recall $\langle z \rangle \approx 150$ MeV for decuplets 2 and 18)
and using the RPP estimate for the mass of the $\Delta(1920)$ (its mass is rather uncertain),
we find in the RPP a candidate for the $\Sigma$ member of the decuplet -- the two-star $\Sigma(2080)$
with the appropriate $J^P=3/2^+$. 
Using the 160 MeV splitting between the $\Delta(1920)$ and $\Sigma(2080)$, we estimate that
the mass of the $\Xi$ member of the decuplet could be around 2240 MeV and the mass
of the $\Omega$ member of the decuplet could be around 2400 MeV. 
While the RPP contains no $\Xi$ baryons around 2240 MeV, there is a two-star $\Omega$ with the mass
2470 MeV. In order to use up as many baryons from the RPP as possible, we assume that
$\Omega(2470)$ belongs to decuplet 16.

The total widths for this decuplet are known with very large uncertainty. Therefore,
the Weldon's equal spacing rule can give only a very rough estimate for the total width
of the predicted $\Xi(2240)$: we estimate that $\Gamma_{\Xi(2240)} \approx 100-150$ MeV.

The $\chi^2$ analysis of the decay rates of decuplet 16 can be hardly performed: 
 the decay rates of $\Delta(1920)$ are known with large ambiguity and the data on the only 
measured observable of $\Sigma(2080)$~\cite{Corden:1976sc} does  not come from our standard 
source of information on 
hyperons~\cite{Gopal:1976gs,Cameron:1977jr,Gopal:1980ur}. Moreover,
SU(3) predicts the positive value for 
$\sqrt{\Gamma_{N {\overline K}}\Gamma_{\Lambda \pi}}$ for $\Sigma(2080)$, which conflicts
with the data. 
Therefore, we perform the $\chi^2$ fit using 
the RPP estimates for the mass, total width and branching ratios of
$\Delta(1920)$ and ignore the $\sqrt{\Gamma_{N {\overline K}}\Gamma_{\Lambda \pi}}$ of 
$\Sigma(2080)$. Consequently, the resulting SU(3) predictions should not be taken too seriously: we use
only two observables for the fit, which have very large experimental errors.

Table~\ref{table:m16} summarizes the results of our $\chi^2$ fit. 
The barrier factor is calculated with $l=1$ for the
${\bf 10} \to {\bf 8}+{\bf 8}$ decays. We chose not to give predictions for
the ${\bf 10} \to {\bf 10}+{\bf 8}$ decays because they would have been based on
the poorly measured $\Gamma_{\Delta(1920) \to \Delta \pi}$ partial decay width.

\begin{table}[h]
\begin{center}
\begin{tabular}{|c c c c |}
\hline
 Mass and width (MeV) & \ Observables \ & Experiment (MeV) & SU(3) pred. (MeV) \\
\hline
$\Delta(1920)$            & $\Gamma_{N \pi}$  &   \underline{$25 \pm 15$} & 23.2   \\
$\Gamma=200$ &  $\sqrt{\Gamma_{N \pi}\Gamma_{\Sigma K}}$ &
\underline{$-10.4 \pm 3.0$} & $-10.6$  \\          
& & & \\
$\Sigma(2080)$ & $\sqrt{\Gamma_{N {\overline K}}\Gamma_{\Lambda \pi}}$ & $-18.6 \pm 7.4$ &
$9.2$ \\
$\Gamma=186 \pm 48$ & $\Gamma_{N {\overline K}}$ & & 7.5 \\
& $\Gamma_{\Lambda \pi}$ & & 11.2 \\
& & & \\
\underline{$\Xi(2240)$} &  $\Gamma_{\Xi \pi}$ &   & 10.0 \\
& $\Gamma_{\Lambda {\overline K}}$  & & 11.0  \\
& $\Gamma_{\Sigma {\overline K}}$  & & 9.1 \\
& & & \\
$\Omega(2470)$ & $\Gamma_{\Xi {\overline K}}$ & & 47.0 \\
$\Gamma=72 \pm 33$ &  & &  \\
 & & & \\
\hline
\end{tabular}
\caption{SU(3) analysis of (\textbf{10}, $3/2^+$)=(1920, 2080, \underline{2240}, 2470).}
\label{table:m16}
\end{center}
\end{table}
The $\chi^2$ fit to the two underlined observables in Table~\ref{table:m16} gives
\begin{equation}
A_{10}  =  15.4 \pm 2.0 \,,  \quad \chi^2/{\rm d.o.f.}=0.02/1 \,. 
\label{eq:m16}
\end{equation}
In summary, we propose that decuplet 16 contains the known $\Delta(1920)$, $\Sigma(2080)$
and $\Omega(2470)$ baryons and the unknown $\Xi$ resonance with the mass around 2240 MeV,
the sum of two-body partial decay widths $\approx 30$ MeV and the total width
$\approx 100-150$ MeV.
It is important to emphasize that unlike for all previously considered SU(3) 
multiplets, the SU(3) analysis of the partial decay widths of decuplet 16 fails because SU(3)
predicts the positive $\sqrt{\Gamma_{N {\overline K}}\Gamma_{\Lambda \pi}}$ for $\Sigma(2080)$,
 which contradicts all the available data on this observable~\cite{Eidelman:2004wy}.
Therefore, the particle assignment in decuplet 16 is essentially done using only the GMO
mass formula and it driven by the desire to use up as many baryons with the appropriate
spin, parity and mass as possible.

\subsection{Multiplet 10: (\textbf{10}, $1/2^-$)=(1620, 1750, \underline{1900}, \underline{2050})}

Decuplet 10 opens with a four-star $\Delta(1620)$, see Tables~\ref{table:before} and
\ref{table:pdg}. The RPP contains  two potential candidates for the $\Sigma$ member of the
decuplet: $\Sigma(1690)$ and  $\Sigma(1750)$. We assume that decuplet 10 contains 
$\Sigma(1750)$ because it gives the larger mass splitting (we always keep in mind the
approximately 150 MeV mass splitting observed in the best studied case of  decuplet 2) and because 
decuplet 10 is the only place where $\Sigma(1750)$ can be fitted in.
The $\Xi$ and $\Omega$ members of the decuplet are missing. We predict their masses
assuming the 150 MeV mass splitting.

Table~\ref{table:m10} summarizes the results of our $\chi^2$ fit to the available
decay rates of decuplet 10.
Since the decay rates of $\Delta(1620)$ and $\Sigma(1750)$ are rather ambiguous, we use
the RPP estimates for their total widths and branching ratios.
The barrier factor is calculated with $l=0$ for the
${\bf 10} \to {\bf 8}+{\bf 8}$ decays and with $l=2$ ($D$-wave) for
the ${\bf 10} \to {\bf 10}+{\bf 8}$ decays.

\begin{table}[h]
\begin{center}
\begin{tabular}{|c c c c |}
\hline
 Mass and width (MeV) & \ Observables \ & Experiment (MeV) & SU(3) pred. (MeV) \\
\hline
$\Delta(1620)$            & $\Gamma_{N \pi}$  &   \underline{$37.5 \pm 7.5$} & 25.1   \\
$\Gamma=150$ &  $\Gamma_{\Delta \pi}$ &
\underline{$61.5 \pm 19.5$} & $61.9$  \\          
& & & \\
$\Sigma(1750)$ & $\Gamma_{N {\overline K}}$ & \underline{$22.5 \pm 13.5$} & 7.1 \\
$\Gamma=90$ & $|\sqrt{\Gamma_{N {\overline K}}\Gamma_{\Lambda \pi}}|$ & 
\underline{$3.6 \pm 2.7$} & 8.9 \\
& $\sqrt{\Gamma_{N {\overline K}}\Gamma_{\Sigma \pi}}$ & 
\underline{$-8.1 \pm 4.5$} & $-6.9$ \\
& $\Gamma_{\Sigma(1385) \pi}$ & \underline{$20.7 \pm 35.1$} & 5.9 \\
& & & \\
\underline{$\Xi(1900)$} &  $\Gamma_{\Xi \pi}$ &   & 9.7 \\
& $\Gamma_{\Lambda {\overline K}}$  & & 9.7  \\
& $\Gamma_{\Sigma {\overline K}}$  & & 8.3 \\
& $\Gamma_{\Xi(1530) \pi}$ &   & 9.1 \\
& & & \\
\underline{$\Omega(2050)$} & $\Gamma_{\Xi {\overline K}}$ & & 32.8 \\
& $\Gamma_{\Xi(1530) {\overline K}}$ & & 0.4  \\
 & & & \\
\hline
\end{tabular}
\caption{SU(3) analysis of (\textbf{10}, $1/2^-$)=(1620, 1750, \underline{1900}, \underline{2050}).}
\label{table:m10}
\end{center}
\end{table}

The $\chi^2$ fit to the underlined observables in Table~\ref{table:m10} gives
\begin{eqnarray}
A_{10}   & = &  12.4 \pm 1.2 \,,  \quad \chi^2/{\rm d.o.f.}=7.99/3 \,, \nonumber\\
A_{10}^{\prime}   & = &  221.4 \pm 34.8 \,,  \quad \chi^2/{\rm d.o.f.}=0.18/1 \,. 
\label{eq:m10}
\end{eqnarray}

Note that an even better value of $\chi^2$ can be obtained by using $\Sigma(1690)$ as the
$\Sigma$ member of decuplet 10. However, as we explained above, the assignment of $\Sigma(1690)$
to decuplet 10 would spoil the overall picture of SU(3) multiplets by leaving out $\Sigma(1750)$.

In order to have a complete multiplet, we predict the existence of two new strange
resonances with $J^P=1/2^-$: a $\Xi$ baryon with the mass around 1900 MeV and an $\Omega$ baryon
 with the mass around
2050 MeV. In the estimate of their masses, we assumed that 
$m_{\Xi}-m_{\Sigma(1750)}=m_{\Omega}-m_{\Xi}=150$ MeV. As follows from Table~\ref{table:m10},
the sum of two-body partial decay widths of $\Xi(1900)$ and $\Omega(2050)$ is at the level of
40 MeV. Also, judging by the central values of the $\Delta(1620)$ and $\Sigma(1750)$ total
widths, the total widths of the predicted $\Xi$ and $\Omega$ should be $\approx 50-60$ MeV.

\subsection{Multiplet 13: (\textbf{10}, $3/2^-$)=(1700, \underline{1850}, \underline{2000}, 
\underline{2150})}

Decuplet 13 is the least established SU(3) multiplet we have considered so far: it opens with a
four-star $\Delta(1700)$, but other members are missing and the RPP contains no candidates.
By analogy with our previous analysis, we assume that the mass difference between the members
of decuplet 13 is 150 MeV. This allows us to estimate the masses of the missing $\Sigma$, $\Xi$ and
$\Omega$ states: $m_{\Sigma} \approx 1850$ MeV, $m_{\Xi} \approx 2000$ MeV and 
$m_{\Omega} \approx 2150$ MeV. The decay rates of the missing resonances are predicted by fitting
the SU(3) predictions to the measured decay rates of $\Delta(1700)$, see Table~\ref{table:m13}.

Table~\ref{table:m13} presents the results of our $\chi^2$ fit to the 
decay rates of $\Delta(1700)$. Since our standard source of information on non-strange 
particles~\cite{Manley:1992yb} reports the total width of $\Delta(1700)$, which is much larger than
the values obtained in many other analyses, we use the RPP estimates for the total width
and branching ratios of $\Delta(1700)$.
The barrier factor is calculated with $l=2$ for the
${\bf 10} \to {\bf 8}+{\bf 8}$ decays and with $l=0$ for
the ${\bf 10} \to {\bf 10}+{\bf 8}$ decays.

\begin{table}[h]
\begin{center}
\begin{tabular}{|c c c c |}
\hline
 Mass and width (MeV) & \ Observables \ & Experiment (MeV) & SU(3) pred. (MeV) \\
\hline
$\Delta(1700)$            & $\Gamma_{N \pi}$  &   \underline{$45.0 \pm 15.0$} & 45.0   \\
$\Gamma=300$ &  $\Gamma_{\Delta \pi}$ & \underline{$126.0 \pm 57.0$} & $126.0$  \\          
& & & \\
\underline{$\Sigma(1850)$} & $\Gamma_{N {\overline K}}$ &  & 12.2 \\
& $\Gamma_{\Lambda \pi}$ &  & 20.2 \\
& $\Gamma_{\Sigma \pi}$ &  & 8.8 \\
& $\Gamma_{\Sigma(1385) \pi}$ &  & 15.6 \\
& $\Gamma_{\Delta {\overline K}}$ &  & 12.1 \\
& & & \\
\underline{$\Xi(2000)$} &  $\Gamma_{\Xi \pi}$ &   & 14.9 \\
& $\Gamma_{\Lambda {\overline K}}$  & & 16.3  \\
& $\Gamma_{\Sigma {\overline K}}$  & & 9.4 \\
& $\Gamma_{\Xi(1530) \pi}$ &   & 22.0 \\
& $\Gamma_{\Sigma(1385) \overline{K}}$ &   & 67.2 \\
& & & \\
\underline{$\Omega(2150)$} & $\Gamma_{\Xi {\overline K}}$ & & 45.5 \\
& $\Gamma_{\Xi(1530) {\overline K}}$ & & 64.8  \\
 & & & \\
\hline
\end{tabular}
\caption{SU(3) analysis of (\textbf{10}, $3/2^-$)=(1700, \underline{1850}, \underline{2000}, \underline{2150}).}
\label{table:m13}
\end{center}
\end{table}

The coupling constants $A_{10}$ and $A_{10}^{\prime}$, which are used to make the predictions summarized
in  Table~\ref{table:m13}, are determined from the $\chi^2$ fit
to the two underlined observables in Table~\ref{table:m13}
\begin{equation}
A_{10} =   48.3 \pm 8.0 \,, \quad A_{10}^{\prime}  =   29.8 \pm 6.7  \,. 
\label{eq:m13}
\end{equation}
Note that since the partial decay width $\Gamma_{\Delta(1700) \to \Delta \pi}$ is large, SU(3) predicts
large ${\bf 10} \to {\bf 10}+{\bf 8}$ decays widths for other members of decuplet 13. Based on the
results in Table~\ref{table:m13}, we can estimate the sum of two-body partial decay
widths of the the predicted baryons:
$\Gamma_{\Sigma(1850)}^{{\rm 2-body}} \approx 70$ MeV, 
$\Gamma_{\Xi(2000)}^{{\rm 2-body}} \approx 130$ MeV and
$\Gamma_{\Omega(2150)}^{{\rm 2-body}} \approx 110$ MeV.

\subsection{Multiplet 19: (\textbf{10}, $1/2^+$)=(1910, \underline{2060}, \underline{2210}, 
\underline{2360})}

Decuplet 19 is very similar to previously considered decuplet 13: only the four-star 
$\Delta(1910)$ is known, while the other members are missing in the RPP.
 Their masses are estimated using
the equal spacing rule with the 150 MeV mass difference: 
$m_{\Sigma} \approx 2060$ MeV, $m_{\Xi} \approx 2210$ MeV and 
$m_{\Omega} \approx 2360$ MeV. The decay rates of the missing resonances are predicted by fitting
the SU(3) predictions to the RPP estimates for the 
decay rates of $\Delta(1910)$, see Table~\ref{table:m19}.

Table~\ref{table:m19} summarizes our results for decuplet 19.
The barrier factor is calculated with $l=1$.
\begin{table}[h]
\begin{center}
\begin{tabular}{|c c c c |}
\hline
 Mass and width (MeV) & \ Observables \ & Experiment (MeV) & SU(3) pred. (MeV) \\
\hline
$\Delta(1910)$            & $\Gamma_{N \pi}$  &   \underline{$56.3 \pm 18.8$} & 56.3   \\
$\Gamma=250$ &  $\Gamma_{\Delta \pi}$ & \underline{$4.0 \pm 1.2$} & $4.0$  \\
& $\Gamma_{\Sigma K}$ & & 11.1 \\          
& & & \\
\underline{$\Sigma(2060)$} & $\Gamma_{N {\overline K}}$ &  & 17.8 \\
& $\Gamma_{\Lambda \pi}$ &  & 26.4 \\
& $\Gamma_{\Sigma \pi}$ &  & 14.7 \\
& $\Gamma_{\Sigma(1385) \pi}$ &  & 0.5 \\
& & & \\
\underline{$\Xi(2210)$} &  $\Gamma_{\Xi \pi}$ &   & 22.9 \\
& $\Gamma_{\Lambda {\overline K}}$  & & 25.2  \\
& $\Gamma_{\Sigma {\overline K}}$  & & 20.6 \\
& $\Gamma_{\Xi(1530) \pi}$ &   & 0.8 \\
& $\Gamma_{\Sigma(1385) \overline{K}}$ &   & 2.5 \\
& & & \\
\underline{$\Omega(2360)$} & $\Gamma_{\Xi {\overline K}}$ & & 87.1 \\
& $\Gamma_{\Xi(1530) {\overline K}}$ & & 2.6  \\
 & & & \\
\hline
\end{tabular}
\caption{SU(3) analysis of (\textbf{10}, $1/2^+$)=(1910, \underline{2060}, \underline{2210}, \underline{2360}).}
\label{table:m19}
\end{center}
\end{table}
The coupling constants $A_{10}$ and $A_{10}^{\prime}$, which are used to make the predictions summarized
in  Table~\ref{table:m19}, are 
\begin{equation}
A_{10} =   24.2 \pm 4.0 \,, \quad A_{10}^{\prime}  =   8.7 \pm 1.4  \,. 
\label{eq:m19}
\end{equation}
Summing the SU(3) predictions for the partial decays widths, we obtain the following estimates:
$\Gamma_{\Sigma(2060)}^{{\rm 2-body}} \approx 75$ MeV,
 $\Gamma_{\Xi(2210)}^{{\rm 2-body}} \approx 85$ MeV and
$\Gamma_{\Omega(2360)}^{{\rm 2-body}} \approx 90$ MeV. Note that the total widths could be 
significantly larger, for instance, by a factor $1.5$.

\subsection{Multiplet 5: (\textbf{10}, $3/2^+$)=(1600, 1690, \underline{1900}, 
\underline{2050})}

In the picture of SU(3) multiplets presented in Table~\ref{table:before},
decuplet 5 can be interpreted as a radial excitation of the ground-state
decuplet. Decuplet 5 opens with a well-established three-star $\Delta(1600)$.
The RPP contains one potential candidate for the $\Sigma$ member of 
decuplet 5: the two-star $\Sigma(1690)$ with unmeasured spin and parity.
Note that the mass difference between $\Sigma(1690)$ and  $\Delta(1600)$ is
only 90 MeV (if one uses the RPP estimates for the masses), which is smaller than
the typical 150 MeV mass difference in decuplets. However, 
if decuplet 5 is mixed, for example with decuplet 2, this renders 
 the equal spacing rule and the 150 MeV mass difference estimate inapplicable
to decuplet 5. We shall not consider the mixing option here.
We choose to assign $\Sigma(1690)$ to decuplet 5 because of the following 
two reasons. First, the decay rates of $\Sigma(1690)$ fit extremely well 
decuplet 5. In more detail, fixing the coupling constants $A_{10}$ and 
$A_{10}^{\prime}$ from the decay rates of $\Delta(1600)$, SU(3) predictions 
for the measured ratios of the decay rates of $\Sigma(1690)$ are in a very good
agreement with the experiment, see Table~\ref{table:m5}. Second, decuplet 5 is the only SU(3) 
multiplet, where the two-star $\Sigma(1690)$ can be placed, and we intend to systematize as many
baryons existing in the RPP as possible.
\begin{table}[h]
\begin{center}
\begin{tabular}{|c c c c |}
\hline
 Mass and width (MeV) & \ Observables \ & Experiment (MeV) & SU(3) pred. (MeV) \\
\hline
$\Delta(1600)$            & $\Gamma_{N \pi}$  &   \underline{$61.3 \pm 26.3$} & 61.3  \\
$\Gamma=350$ &  $\Gamma_{\Delta \pi}$ & \underline{$178.5 \pm 84.0$} & $178.5$  \\          
& & & \\
$\Sigma(1690)$ & $\Gamma_{N {\overline K}}/\Gamma_{\Lambda \pi}$ & $0.4 \pm 0.25$  or small  & 0.54 \\
$\Gamma=100-240$ & $\Gamma_{\Sigma \pi}/\Gamma_{\Lambda \pi}$ & $0.3 \pm 0.3$  or small  & 0.46 \\
& $\Gamma_{\Sigma(1385) \pi}/\Gamma_{\Lambda \pi}$ & $< 0.5$  & 0.58 \\
& $\Gamma_{N {\overline K}}$ & & 11.4 \\
& $\Gamma_{\Lambda \pi}$ &  & 21.28 \\
& $\Gamma_{\Sigma \pi}$ &  & 9.8 \\
& $\Gamma_{\Sigma(1385) \pi}$ &  & 12.4 \\
& & & \\
\underline{$\Xi(1900)$} &  $\Gamma_{\Xi \pi}$ &   & 20.8 \\
& $\Gamma_{\Lambda {\overline K}}$  & & 20.5  \\
& $\Gamma_{\Sigma {\overline K}}$  & & 12.9 \\
& $\Gamma_{\Xi(1530) \pi}$ &   & 32.5 \\
& $\Gamma_{\Sigma(1385) \overline{K}}$ &   & 6.2 \\
& & & \\
\underline{$\Omega(2050)$} & $\Gamma_{\Xi {\overline K}}$ & & 58.2 \\
& $\Gamma_{\Xi(1530) {\overline K}}$ & & 8.8  \\
 & & & \\
\hline
\end{tabular}
\caption{SU(3) analysis of (\textbf{10}, $3/2^+$)=(1600, 1690, \underline{1900}, \underline{2050}).}
\label{table:m5}
\end{center}
\end{table}

The $\Xi$ and $\Omega$ members of decuplet 5 are missing. We assume that the missing 
$\Omega$ is 450 MeV heavier than the $\Delta(1600)$ (the equal spacing with the 150 MeV 
mass difference)
and that $m_{\Omega}-m_{\Xi}=150$ MeV. This gives $m_{\Xi}=1900$ MeV and  $m_{\Omega}=2050$ MeV.

In the $\chi^2$ fit,  we used two decay rates of $\Delta(1600)$. We used the average RPP values 
for the mass, total width and branching ratios of $\Delta(1600)$. Note that when we use
$m_{\Delta(1600)}=1600$ MeV, the measured $\Delta(1600) \to \Sigma K$ decay
is prohibited by kinematics and cannot be used in the fit.
Since for $\Sigma(1690)$ only ratios of the partial decay widths are known, it does not make sense
to use the $\Sigma(1690)$ measured observables in the fit. 
Table~\ref{table:m5} summarizes our results for decuplet 5.
The barrier factor is calculated with $l=1$.

The coupling constants $A_{10}$ and $A_{10}^{\prime}$, which are used to make the predictions summarized
in  Table~\ref{table:m5}, are 
\begin{equation}
A_{10} =   38.2 \pm 8.2 \,, \quad A_{10}^{\prime}  =   129.3 \pm 30.4  \,. 
\label{eq:m5}
\end{equation}
Summing the SU(3) predictions for the partial decays widths, we estimate:
$\Gamma_{\Xi(1900)}^{{\rm 2-body}} \approx 95$ MeV and
$\Gamma_{\Omega(2050)}^{{\rm 2-body}} \approx 70$ MeV.

\section{SU(3) analysis of the antidecuplet}
\label{sec:anti10}

As discussed in the Introduction, the recent wave of interest in the
baryon spectroscopy was triggered by experimental indications of the
 existence of the exotic $\Theta^+$ baryon~\cite{Nakano:2003qx,Barmin:2003vv,Stepanyan:2003qr,Kubarovsky:2003fi,Asratyan:2003cb,Barth:2003es,Airapetian:2003ri,Aleev:2004sa,Abdel-Bary:2004ts,Chekanov:2004kn,Aleev:2005yq}
and the exotic $\Xi^{--}$ baryon~\cite{Alt:2003vb}. These baryons are called exotic because,
in the language of quark models, their
quantum numbers cannot be obtained from three constituent quarks, i.e.~the minimal Fock 
component of $\Theta^+$ and $\Xi^{--}$ contains four quarks and one antiquark. Both  
$\Theta^+$ and $\Xi^{--}$ are members of the multiplet called the antidecuplet (an 
${\bf \at}$ SU(3) representation).
 The existence of the 
antidecuplet and the masses, total widths and decay rates of its members were predicted 
within the chiral quark soliton model by Diakonov, Petrov and Polyakov~\cite{Diakonov:1997mm}.
In addition to the $\Theta^+$ and $\Xi^{--}$ reports, there was the observation of
an exotic anti-charmed baryon state~\cite{Aktas:2004qf}.

At the moment of writing of this report, the existence of $\Theta^+$ and $\Xi^{--}$ is
 uncertain because there is a number
of experiments and analyses, which do not see these states~\cite{Bai:2004gk,Schael:2004nm,Aubert:2004bm,Abt:2004tz,Litvintsev:2004yw,Stenson:2004yz,Abe:2004wf,Pinkenburg:2004ux,Dzierba:2003cm,Dzierba:2004db,Fischer:2004qb,Zavertyaev:2005yf,Battaglieri:2005er,Hicks:2005jf}.
However, we expect that in the near future, there will appear thorough
analyses trying to understand in detail why some some experiments see the $\Theta^+$ and
some do not, see e.g.~\cite{Titov:2005kf,Chekanov:2005uz}.

We assume that the antidecuplet exists and has spin and parity as predicted in the
chiral quark soliton model, $J^P=1/2^+$. The $\Theta^+$ is the lightest members of the antidecuplet with
the mass approximately 1540 MeV and the $\Xi_{\at}$
($\Xi^{--}$ is the member of the isoquadruplet referred to as $\Xi_{\at}$)
 is the heaviest member of the antidecuplet with
the mass 1862 MeV, see Fig.~\ref{fig:antidecuplet}. Only these two members of the antidecuplet can be considered established.
Applying the equal spacing rule to the antidecuplet, see Eq.~(\ref{eq:gmoanti10}),
we find that the mass spacing for the antidecuplet is 107 MeV. This means that the
$N_{\at}$ member of the antidecuplet should have the mass around 1650 MeV and the  
$\Sigma_{\at}$ member of the antidecuplet should have the mass around 1760 MeV.

While no information is available about the $\Sigma_{\at}$ member of the antidecuplet, candidates for the 
$N_{\at}$ member
were recently discussed in the literature. The partial wave analysis (PWA) of pion-nucleon scattering,
which was modified for the search of narrow resonances, presented two candidate for the $N_{\at}$ 
with masses 1680 MeV and 1730 MeV~\cite{Arndt:2003ga}. In both cases, 
$\Gamma_{N_{\at} \to N \pi} < 0.5$ MeV (the
resonance is highly inelastic) and $\Gamma_{{\rm tot}} < 30$ MeV.

Experimental evidence for a new nucleon resonance with the mass near 1670 MeV was recently 
obtained by the GRAAL collaboration~\cite{Kuznetsov:2004gy}. The fact that the resonance peak is 
seen in the $\gamma n \to n \eta$ process
and is absent in the  $\gamma p \to p \eta$ process supplies a strong piece of evidence that the resonance
belongs to the antidecuplet because photoproduction of the antidecuplet is strongly suppressed on the 
proton target~\cite{Polyakov:2003dx}, see also the relevant discussion using the $U$-spin
argument in Sect.~\ref{sec:intro}.
The position of the peak is very close to the 1680 MeV solution of~\cite{Arndt:2003ga}.

There is another candidate for the $N_{\at}$ member of the antidecuplet, which corresponds to the higher mass
solution of the analysis of~\cite{Arndt:2003ga}. In gold-gold collisions at RHIC, 
the STAR collaboration observes a narrow peak at approximately 1734 MeV in 
the $\Lambda K^0_s$ 
invariant mass.

Therefore, the present experimental information on the properties of the antidecuplet can be
summarized as follows. The $\Theta^+$ and $\Xi_{\at}$  members  are observed 
experimentally. These states have exotic quantum numbers and, hence, cannot mix with non-exotic baryons.
The $\Theta^+$ and $\Xi_{\at}$ are narrow. The $N_{\at}$ member of ${\bf \at}$ is not established, but
there are at least two candidates for it.
 The present experimental data suggests that the pattern of the $N_{\at}$ decays is the
following: $\Gamma_{N_{\at}\to N \pi} < 0.5$ MeV, $\Gamma_{N_{\at} \to N \eta}$ is sizable (measurable)
 and
 $\Gamma_{N_{\at} \to \Lambda K}$ is non-vanishing. There exists no information on
 the $\Sigma_{\at}$ member
of ${\bf \at}$, except for an estimate based on the equal spacing rule, 
$m_{\Sigma_{\at}} \approx 1760$ MeV.

This section is organized as follows. In Subsect.~\ref{subsec:decays}, we collect the SU(3) predictions
for the antidecuplet decays and show that they are inconsistent with the antidecuplet decay pattern 
summarized above. This strongly implies that the non-exotic members of the antidecuplet,
$N_{\at}$ and $\Sigma_{\at}$, are mixed with $N$ and $\Sigma$ from non-exotic multiplets. Another option, which 
we shall not consider here, is to assume that the antidecuplet is mixed with 
the 27-plet~\cite{Ellis:2004uz,Praszalowicz:2004dn,Bijker:2003pm}.
 Among several possible scenarios of the mixing, in Subsect.~\ref{subsec:mixing} we
 examine the possibility that the 
$N_{\at}$ and $\Sigma_{\at}$ members of the antidecuplet mix with $N(1440)$ and $\Sigma(1660)$ of octet 3.
We show that this can accommodate in a simple way all experimental information on the antidecuplet
decays.

\subsection{The antidecuplet decays: no mixing}
\label{subsec:decays}

Similarly to the ${\bf 10} \to {\bf 8}+{\bf 8}$ decays, the ${\bf \at} \to {\bf 8}+{\bf 8}$ decays
of the antidecuplet are described in terms of one free parameter $A_{\at}$
\begin{equation}
g_{B_1 B_2 P }=-A_{\at} \frac{1}{\sqrt{5}} \ \left(
\begin{array}{cc}
8 & 8 \\
Y_2 T_2 & Y_P T_P
\end{array}\right|\left.\begin{array}{c}
           \at \\Y_1 T_1
          \end{array}\right) \,.
\label{eq:anti1088}
\end{equation}
The coupling constants for all possible decay channels of the antidecuplet
are summarized in Table~\ref{table:antidecuplet}. 

\begin{table}[h]
\begin{center}
\begin{tabular}{|c c |}
\hline
 & ${\bf \at} \to {\bf 8}+{\bf 8}$ \\
Decay mode & $g_{B1 B_2 P}$  \\
\hline
$\Theta^+ \to N K$ & $(1/\sqrt{5})\,A_{\at}$ \\
\hline
$N_{\at} \to N \pi$ & $1/(2 \sqrt{5})\, A_{\at}$   \\
$ \to N \eta$ & $-1/(2 \sqrt{5})\, A_{\at}$   \\
$ \to \Lambda K$ & $1/(2 \sqrt{5})\, A_{\at}$   \\
$ \to \Sigma K$ & $1/(2 \sqrt{5})\, A_{\at}$   \\
\hline
$\Sigma_{\at} \to N {\overline K}$ & $-(1/(\sqrt{30})\, A_{\at}$ \\
$\to \Sigma \pi$ & $(1/(\sqrt{30})\, A_{\at}$ \\
$\to \Sigma \eta$ & $-1/(2\sqrt{5})\, A_{\at}$ \\
$\to \Lambda \pi$ & $1/(2\sqrt{5})\, A_{\at}$ \\
$\to \Xi K$ & $1/(\sqrt{30})\, A_{\at}$ \\
\hline
$\Xi_{\at} \to \Xi \pi$ & $(1/\sqrt{10})\,A_{10}$  \\
$ \to \Sigma {\overline K}$ & $-(1/\sqrt{10})\,A_{10}$  \\
\hline
\end{tabular}
\caption{The SU(3) universal coupling constants for ${\bf \at} \to {\bf 8}+{\bf 8}$ decays.}
\label{table:antidecuplet}
\end{center}
\end{table}

Using the SU(3) universal coupling constants of the $N_{\at}$, it is 
easy to show that the emerging pattern of the $N_{\at}$ decay rates is inconsistent with the trend
discussed above. Indeed, treating the total width of the $\Theta^+$ 
as an input, one can readily determine $A_{\at}$, which allows to unambiguously 
predict the ${\bf \at}$ decay rates. Note that while the experimental determination of
 $\Gamma_{\Theta^+}$ is limited by the detector resolution and the experimental upper limit is
$\Gamma_{\Theta^+} < 10$ MeV, many theoretical analyses suggest that $\Gamma_{\Theta^+}$ is
even smaller, of the order of several MeV or even less than 1 MeV~\cite{Arndt:2003xz,Casher:2003ep,Haidenbauer:2003rw,Cahn:2003wq,Gibbs:2004ji,Sibirtsev:2004bg,Diakonov:2005ib}.

\begin{figure}[h]
\begin{center}
\epsfig{file=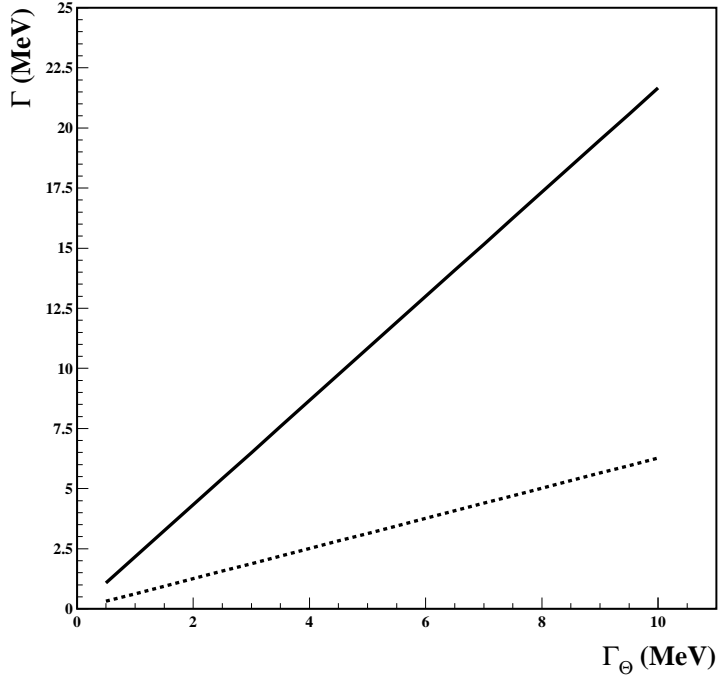,width=10cm,height=10cm} 
\caption{The unmixed antidecuplet. The $\Gamma_{N_{\at}\to N \pi}$ (the solid curve) and
 $\Gamma_{N_{\at} \to N \eta}$ (the dashed curve) partial decay widths
as functions of the total width of the $\Theta^+$, $\Gamma_{\Theta^+}$.
}
\label{fig:nomixing}
\end{center}
\end{figure}

In the following, we assume that $m_{N_{\at}}=1670$ MeV. 
 Figure~\ref{fig:nomixing} presents
$\Gamma_{N_{\at} \to N \pi}$ (the solid curve) and $\Gamma_{N_{\at} \to N \eta}$ (the dashed curve)
as functions of $\Gamma_{\Theta^+}$. 
As can be seen from the figure, SU(3) predicts that $\Gamma_{N_{\at} \to N \pi} > \Gamma_{N_{\at} \to N \eta}$. 
In addition, SU(3) predicts  that $\Gamma_{N_{\at} \to N \pi} < 0.5$ MeV only for
  $\Gamma_{\Theta^+}< 0.25$ MeV. These two results contradict
the analysis of~\cite{Arndt:2003ga} and the GRAAL result~\cite{Kuznetsov:2004gy}.
The only way to alter the pattern of the $N_{\at}$ decays and to have consistency with the data
 is to introduce mixing. In the following subsection, we consider the scenario that 
the antidecuplet is mixed octet 3.

\subsection{Mixing of the antidecuplet with octet 3}
\label{subsec:mixing}

One possible way to suppress $\Gamma_{N_{\at} \to N \pi}$ while
enhancing $\Gamma_{N_{\at} \to N \eta}$  is to mix the $N_{\at}$ 
 with $N(1440)$ of octet 3. 
This automatically implies that the $\Sigma_{\at}$ is mixed with $\Sigma(1660)$ of octet 3~\cite{Diakonov:2003jj}.
Naturally, because of their exotic quantum numbers, $\Theta^+$ and $\Xi_{\at}$ do not mix with non-exotic baryons.
Parameterizing the mixing in terms of the mixing angle $\theta$, the relevant 
$N_{\at}$  coupling constants read (see Tables~\ref{table:octets} and \ref{table:antidecuplet})
\begin{eqnarray}
g_{N_{\at}\to N \pi} & = & -\sin \theta \sqrt{3} A_8 +\cos \theta \frac{1}{2 \sqrt{5}} A_{\at} \,, 
\nonumber\\
g_{N_{\at} \to N \eta} & = & -\sin \theta \frac{(4\, \alpha-1)}{\sqrt{3}} A_8 -
\cos \theta \frac{1}{2 \sqrt{5}} A_{\at} \,, 
\label{eq:mix1}
\end{eqnarray}
where $A_8=32.4$ and $\alpha=0.27$, see Eq.~(\ref{eq:m3}).
Because of the relative minus sign between the terms proportional to $A_8$ and $A_{\at}$ in the
expression for $g_{N_{\at} \to N \pi}$ and the positive relative sign between the
two contributions to $g_{N_{\at} \to N \eta}$, 
is is possible to simultaneously suppress $g_{N_{\at} \to N \pi}$ and to keep
$g_{N_{\at} \to N \eta}$ sizable by a suitable choice of the mixing angle $\theta$.

\begin{figure}[h]
\begin{center}
\epsfig{file=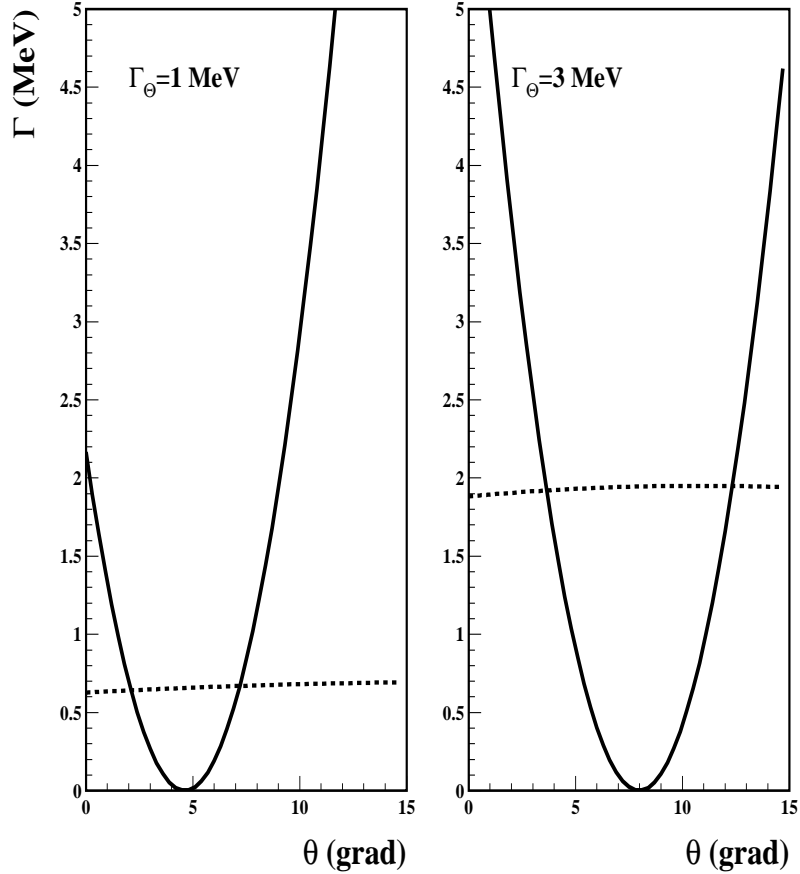,width=11cm,height=13cm} 
\caption{The $\Gamma_{N_{\at} \to N \pi}$ (solid curves) and
 $\Gamma_{N_{\at} \to N \eta}$ (dashed curves) partial decay widths
as functions of the mixing angle $\theta$ for
$\Gamma_{\Theta^+}=1$ MeV (left panel) and $\Gamma_{\Theta^+}=3$ MeV (right panel).
}
\label{fig:mixing}
\end{center}
\end{figure}

This is illustrated in Fig.~\ref{fig:mixing}, where $\Gamma_{N_{\at} \to N \pi}$ (solid curves) and
 $\Gamma_{N_{\at} \to N \eta}$ (dashed curves) are plotted as functions of the mixing angle $\theta$ for
two values of the total width of the $\Theta^+$, 
$\Gamma_{\Theta^+}=1$ MeV (left panel) and $\Gamma_{\Theta^+}=3$ MeV (right panel). Note that 
$\Gamma_{N_{\at} \to N \eta}$ very weakly depends on $\theta$ for small $\theta$ because the non-exotic
contribution to $g_{N_{\at} \to N \eta}$ is numerically very small.

As can be seen from Fig.~\ref{fig:mixing}, the conditions $\Gamma_{N_{\at} \to N \pi} < 0.5$ 
MeV~\cite{Arndt:2003ga} and $\Gamma_{N_{\at} \to N \eta}$ is sizable~\cite{Kuznetsov:2004gy} are 
provided, if $3^0 < \theta < 7^0$ for $\Gamma_{\Theta^+}=1$ MeV and
$6^0 < \theta < 10^0$ for $\Gamma_{\Theta^+}=3$ MeV.
It is important to emphasize that when the mixing angle $\theta$ is as small as we find,
mixing with the antidecuplet does not affect the coupling constants of octet 3. This
allows us to use the values of the parameters $A_8$, $\alpha$ and $A_8^{\prime}$, which are
determined for the unmixed octet 3.

The mixing angle $\theta_{\Sigma}$ for the $\Sigma_{\at}$ and $\Sigma(1660)$ states is related to 
$\theta$~\cite{Diakonov:2003jj}
\begin{equation} 
\theta_{\Sigma} (m_{\Sigma_{\at}}-m_{\Sigma(1660)})=\theta (m_{N_{\at}}-m_{N(1440)}) \,.
\label{eq:theta_sigma}
\end{equation}
Using $m_{N(1440)}=1440$ MeV, $m_{\Sigma(1660)}=1660$ MeV and $m_{\Theta^+}=1540$ MeV and
assuming that $m_{\Sigma_{\at}} \approx 1760$ MeV (the equal spacing rule estimate), we find that
$\theta_{\Sigma} \approx \theta$.

Table~\ref{table:antidecuplet:mixing} summarizes SU(3) predictions for the decay rates of the 
antidecuplet mixed with octet 3, provided that the mixing angle $\theta$ is chosen such that
the $\Gamma_{N_{\at} \to N \pi} < 0.5$ MeV.
\begin{table}[h]
\begin{center}
\begin{tabular}{|c c c|}
\hline
Partial decay widths & $\Gamma_{\Theta^+}=1$ MeV & $\Gamma_{\Theta^+}=3$ MeV \\
(MeV) &  $3^0 < \theta < 7^0$ &    $6^0 < \theta < 10^0$ \\       
\hline
$\Gamma_{N_{\at} \to N \pi}$ &  $ < 0.5$ & $ < 0.5$   \\
$\Gamma_{N_{\at} \to N \eta}$ & $0.65-0.67$ & $1.94-1.95$   \\
$\Gamma_{N_{\at} \to \Lambda K}$ & $0.16-0.29$ & $0.56-0.76$   \\
$\Gamma_{N_{\at} \to \Delta \pi}$ & $2.6-15.6$ & $12.9-34.8$   \\
& & \\
$\Gamma_{\Sigma_{\at} \to N {\overline K}}$ & $0.11-0.50$ & $0.49-1.18$ \\
$\Gamma_{\Sigma_{\at} \to \Sigma \pi}$ & $0.02-2.64$ & $0.57-5.00$ \\
$\Gamma_{\Sigma_{\at} \Sigma \eta}$ & $0.04-0.08$ & $0.15-0.20$ \\
$\Gamma_{\Sigma_{\at}\to \Lambda \pi}$ & $0.15-0.81$ & $0.72-1.90$ \\
$\Gamma_{\Sigma_{\at}\to \Sigma(1385) \pi}$ & $0.33-1.96$ & $1.6-4.3$ \\
& & \\
$\Gamma_{\Xi_{\at} \to \Xi \pi}$ & 1.98 & 5.94  \\
$ \Gamma_{\Xi_{\at}\to \Sigma {\overline K}}$ & 1.08 & 3.23 \\
\hline
\end{tabular}
\caption{SU(3) predictions for the decays rates of the antidecuplet mixed with octet 3.}
\label{table:antidecuplet:mixing}
\end{center}
\end{table}

The results presented in Table~\ref{table:antidecuplet:mixing} deserve a discussion. The pattern
of the $N_{\at}$ decays complies with the following picture supported by experiments:
 the $N_{\at} \to N \pi$ decay is suppressed~\cite{Arndt:2003ga}, the 
$N_{\at} \to N \eta$ decay is measurable~\cite{Kuznetsov:2004gy} and 
$N_{\at} \to \Lambda K$ decay is non-vanishing~\cite{Kabana:2004hh}.
Note that while we assumed that $m_{N_{\at}}=1670$ MeV, the qualitative picture of the
$N_{\at}$ decays does not change, if $m_{N_{\at}}=1734$ MeV~\cite{Kabana:2004hh} is used.

The $N_{\at} \to \Delta \pi$ and $\Sigma_{\at} \to \Sigma(1385) \pi$ decays are possible
only due to the mixing because the ${\bf \at} \to {\bf 10}+{\bf 8}$ decays are 
not allowed. The corresponding coupling  constants are
\begin{eqnarray}
&&g_{N_{\at} \to \Delta \pi}  =  -\sin \theta\, \frac{2}{\sqrt{5}} A_8^{\prime} \,, \nonumber\\
&&g_{\Sigma_{\at} \to \Sigma(1385) \pi}  =  \sin \theta \,\frac{\sqrt{30}}{15} A_8^{\prime} \,.
\end{eqnarray}
Despite the small mixing angle $\theta$, the large value of $A_8^{\prime}$, 
see Eq.~(\ref{eq:m3}), provides large $\Gamma_{N_{\at} \to \Delta \pi}$.

It can be seen from Table~\ref{table:antidecuplet:mixing} that the branching ratio of
$\Sigma_{\at}$ into the $N {\overline K}$ final state is sizable, $Br(N {\overline K}) \approx 15-20$\%.
This is important for the potential search for $\Sigma_{\at}$. Among the experiments reporting 
the $\Theta^+$ signal, there are four experiments, where the $\Theta^+$ is observed as a peak in the
$p K_S$ invariant mass and strangeness is not tagged~\cite{Asratyan:2003cb,Airapetian:2003ri,Aleev:2004sa,Chekanov:2004kn}. Since $\Sigma_{\at}$ has a sizable branching ratio into the same final state, the four
experiments can give information on the $\Sigma_{\at} \to N {\overline K}$ decay.
We shall consider this in detail.

The analysis of neutrino-nuclear (mostly neon) interaction data~\cite{Asratyan:2003cb}
 clearly reveals  the $\Theta^+$ peak as well as a number of other peaks in
 the $1650 < M_{p \, K_S} < 1850$ MeV mass region, which cannot be 
suppressed by the random-star elimination procedure, see Fig.~3 of~\cite{Asratyan:2003cb}. 
Any of the peaks  in the 1700-1800 MeV mass range is a good candidate  for  $\Sigma_{\at}$.

Similar conclusions apply to the SVD collaboration result~\cite{Aleev:2004sa}. Before
the cuts aimed to enhance the $\Theta^+$ signal are imposed, 
the $p \, K_S$ invariant
 mass spectrum contains at least two prominent peaks in the 1700-1800 MeV mass
 range (see Fig.~5 of~\cite{Aleev:2004sa}), each of which can be interpreted as
$\Sigma_{\at}$. 

The HERMES~\cite{Airapetian:2003ri} and ZEUS~\cite{Chekanov:2004kn} $p \, K_S$ invariant 
mass spectra extend only up to
1.7 MeV and, therefore, do not allow to make any conclusions about the 
$\Sigma_{\at}$. 

In addition to the $p \, K_S$ invariant mass spectrum, the HERMES 
collaboration also presents the $\Lambda \, \pi$ invariant mass spectrum
in order to see if the observed peak in the $p \, K_S$ final state is indeed 
generated by the $\Theta^+$ and not by some yet unknown $\Sigma^{\ast}$ 
resonance~\cite{Lorenzon:2004rw}. The  $\Lambda \, \pi$ invariant mass spectrum
has no resonance structures except for the prominent $\Sigma(1385)$ peak. 
According to our analysis, the $\Gamma_{\Sigma_{\at} \to \Lambda \pi}$
 partial decay width is small, when $\Gamma_{\Sigma_{\at} \to N {\overline K}}$ is 
large (this correlation is not explicitly indicated in Table~\ref{table:antidecuplet:mixing}).
 This correlation 
 seems to be exactly what is needed to 
comply with the non-observation of $\Sigma_{\at}$ in the HERMES 
$\Lambda \, \pi$ invariant mass spectrum.

In summary, the~\cite{Asratyan:2003cb,Aleev:2004sa} data contain an indication 
 for a narrow 
$\Sigma_{\at}$ member of the antidecuplet in the 1700-1800 MeV mass range
 and the \cite{Airapetian:2003ri,Chekanov:2004kn,Lorenzon:2004rw} data do not rule out 
its existence.
 Obviously, a dedicated search for the $\Sigma_{\at}$ signal in the
 $p \, K_S$ and $\Lambda \, \pi$ invariant mass spectra is needed in
 order to address several key issues surrounding this least known member
 of the antidecuplet. 

It is curios that one can offer a candidate $\Sigma_{\at}$ state, 
the one-star $\Sigma(1770)$ with $J^P=1/2^+$, which has been known for almost 
three decades~\cite{Eidelman:2004wy,Gopal:1976gs}.
We would like to emphasize that the $\Sigma(1770)$ was not used 
in any of the non-exotic multiplets we considered in Sects.~\ref{sec:octets} and
\ref{sec:decuplets}.
The $\Sigma(1770)$ has the $14 \pm 4$ \% branching ratio in the $N \, \overline{K}$ final state and 
poorly known but still probably rather small branching ratios into the
$\Lambda \, \pi$ and $\Sigma \, \pi$ final states.
The total width of the $\Sigma(1770)$ is $\Gamma_{{\rm tot}}=72 \pm 10$ MeV,
which is much larger than the sum of the partial decay widths in Table~\ref{table:antidecuplet:mixing}.
However, if other decay channels contribute significantly, this will reduce the 
inconsistency between the SU(3) predictions for $\Gamma_{{\rm tot}}$ and the experimental value.
The possibility of large three-body partial decay widths of  $\Sigma(1770)$ was studied in~\cite{Hosaka:2004mv}.

Turning to $\Xi_{\at}$ we observe that, although in the considered mixing scenario
 $\Xi_{\at}$ cannot mix,
its partial decay rates are rather large for the antidecuplet because of the large
phase space factor.

We considered a possible scenario of the antidecuplet mixing with octet 3, which allowed us
to have a phenomenologically consistent picture of the ${\bf \at}$ decays.
Mixing of the antidecuplet with octet 3 was also considered in~\cite{Diakonov:2003jj},
 where the decays were not considered and the mixing angles $\theta$ and $\theta_{\Sigma}$
 were determined using the modified Gell-Mann--Okubo mass formulas for the octet and the 
antidecuplet.
We demonstrated in a simple way
 that even very insignificant mixing  dramatically affects the antidecuplet decays.
Of course, since the antidecuplet decays are virtually unknown,
many more mixing scenarios are possible and were considered in the literature.
For instance, the antidecuplet can mix with octet 4 
(the emerging picture is similar to the case of mixing with octet 3), with the ground-state
octet~\cite{Diakonov:1997mm,Arndt:2003ga}, simultaneously with the ground-state octet and
exotic ${\bf 27}$-plet and $\overline{{\bf 35}}$ plet~\cite{Ellis:2004uz,Praszalowicz:2004dn}, 
simultaneously with the
ground-state octet and octets~3 and 4~\cite{Guzey:2005mc}.

\section{Conclusions and discussion}
\label{sec:conclusions}

We have analyzed 20 multiplets of baryons with the mass less than
approximately 2000-2200 MeV indicated in Table~\ref{table:before}
and the antidecuplet using the Gell-Mann--Okubo mass formulas and SU(3)
predictions for partial decay widths. We confirm the main conclusion 
of the 1974 analysis by Samios, Goldberg and Meadows~\cite{Samios:1974tw}
that the simple scheme based on flavor SU(3) symmetry of the strong interaction
describes remarkably well  the mass splitting and the decay rates of all considered 
multiplets. 
The main result of our analysis is the
final list of SU(3) multiplets, which
 is presented in Table~\ref{table:final}.
 The underlined entries in the table
are predictions of new particles, which are absent in the Review of Particle 
Physics. 

\begin{table}[t]
\begin{center}
\begin{tabular}{|c|c|c|c|}
\hline
  & 1 & $(8,\frac{1}{2}^+)$ & (939, 1115, 1189, 1314) \\
 & 2 & $(10,\frac{3}{2}^+)$ & (1232, 1385, 1530, 1672) \\ 
$(56, L=0)$  & 3 & $(8,\frac{1}{2}^+)$ & (1440, 1600, 1660, 1690) \\
& 4 &$(8,\frac{1}{2}^+)$ & (1710, 1810, 1880, \underline{1950}) \\ 
& 5 & $(10,\frac{3}{2}^+)$ &  (1600, 1690, \underline{1900}, \underline{2050}) \\
\hline \hline
& 6 & $(1,\frac{1}{2}^-)$ & $\Lambda(1405)$ \\
& 7 & $(1,\frac{3}{2}^-)$ & $\Lambda(1520)$ \\
& 8 & $(8,\frac{3}{2}^-)$ & (1520, 1690, 1670, 1820) \\
$(70, L=1)$& 9 & $(8,\frac{1}{2}^-)$ & (1535, 1670, 1560, \underline{1620-1725}) \\
& 10 & $(10,\frac{1}{2}^-)$ & (1620, 1750, \underline{1900}, \underline{2050}) \\
 & 11 & $(8,\frac{3}{2}^-)$ & (1700, \underline{1850}, 1940, \underline{2045}) \\
& 12 & $(8,\frac{5}{2}^-)$ & (1675, 1830, 1775, 1950) \\
& 13 & $(10,\frac{3}{2}^-)$ & (1700, \underline{1850}, \underline{2000}, \underline{2150}) \\
& 14 & $(8,\frac{1}{2}^-)$ & (1650, 1800, 1620, \underline{1860-1915}) \\
\hline \hline
& 15 & $(8,\frac{5}{2}^+)$ & (1680, 1820, 1915, 2030) \\
& 16 & $(10,\frac{3}{2}^+)$ & (1920, 2080, \underline{2240}, 2470) \\
$(56, L=2)$ & 17 & $(8,\frac{3}{2}^+)$ & (1720, 1890, 1840, \underline{2035}) \\
& 18 & $(10,\frac{5}{2}^+)$ & (1905, 2070, 2250, 2380) \\
& 19 & $(10,\frac{1}{2}^+)$ & (1910, \underline{2060}, \underline{2210},
 \underline{2360} ) \\
& 20 & $(10,\frac{7}{2}^+)$ & (1950, 2030, 2120, 2250) \\
\hline \hline
& 21 & $(\at,\frac{1}{2}^+)$ & (1540, 1670, \underline{1760}, 1862) \\
\hline
\end{tabular}
\caption{The final list  of SU(3) multiplets.}
\label{table:final}
\end{center}
\end{table}

An examination of the RPP baryon listing shows that we have cataloged 
all four and three-star baryons with the mass less than approximately 2000-2200 MeV and
almost all baryons with the weaker rating in the same mass region.
In the following, we shall discuss the baryons, for which we could not find a place in 
Table~\ref{table:final}. We will ignore baryons with the mass greater than 1900 MeV,
 which can be interpreted as radial excitations of the corresponding
lighter baryons presented Table~\ref{table:final} or as states opening higher
SU(6)$\times$O(3) supermultiplets~\cite{Samios:1974tw}. One example is the $J^P=1/2^-$ nonet
consisting of the $N(2190),\Lambda(?),\Sigma(?),\Xi(?)$ octet\footnote{
Note that the $\Lambda$, $\Sigma$ and $\Xi$ members of the considered octet have disappeared
or have changed their masses since 1974 such that their RPP candidates cannot be easily
established.} mixed with
the $\Lambda(2100)$ singlet, which
can be thought to belong to the $({\bf 70}, L=3)$ supermultiplet~\cite{Samios:1974tw}.
The remaining unused baryons are the one-star $\Delta(1750)$, the one-star $\Sigma(1480)$
and $\Sigma(1770)$, the two-star $\Sigma(1580)$ and the one-star $\Xi(1620)$, which we shall 
discuss below.

It was suggested in~\cite{Azimov:2003bb} that $\Sigma(1480)$ and $\Xi(1620)$ might be members
of a new light octet, whose $N$  member (called $N^{\prime}$) is predicted to have the mass 
around 1100 MeV and the vanishingly small total width. 
In addition, the $g_{N^{\prime} N \pi}$ coupling constant was predicted to be strongly suppressed
compared to the usual $NN \pi$ coupling constant,
$g_{N^{\prime} N \pi}/g_{N N \pi} \leq 0.01$. In our notation, this corresponds to
the vanishingly small $A_8$ coupling constant. However,
the experimental information on the decays of
$\Sigma(1480)$ and $\Xi(1620)$ is too sketchy to perform a $\chi^2$ analysis of the decay
rates.
 It is interesting to observe that
the 1972 edition of the Review of Particle Physics~\cite{RPP:1972} contained a possible candidate
for the $\Lambda$ member of this superlight octet, $\Lambda(1330)$, which disappeared in the 2004 
edition of the RPP.

As we discussed in Sect.~\ref{sec:anti10}, the $J^P=1/2^+$ and the mass of 
 $\Sigma(1770)$ make it a potential candidate for the $\Sigma_{\at}$ member of
the antidecuplet.

Experimental evidence for the $\Delta(1750)$ is too weak to make any hypothesis concerning
its place in our SU(3) scheme. The same applies to $\Sigma(1580)$. While it has a two-star
status, this state is not seen in the analyses, which we use as 
our primary sources of information on hyperons~\cite{Gopal:1976gs,Cameron:1977jr,Gopal:1980ur}.

In order to have a complete picture of unitary multiplets, we predict 
 a number of new strange baryons, whose properties 
are summarized in Table~\ref{table:pred}. In the table, besides the predicted spin and parity,
we also give estimates for the mass, the sum of two-body partial decay widths 
$\Gamma^{{\rm 2-body}}$ and the total width $\Gamma_{{\rm tot}}$.
The latter quantity is given only for the cases, where a meaningful estimate could be done.
The last column lists final states with
large branching ratios.
\begin{table}[h]
\begin{center}
\begin{tabular}{|c c c c c c |}
\hline
Particle & $J^P$ (multiplet) & Mass (MeV) & $\Gamma^{{\rm 2-body}}$ (MeV) & 
$\Gamma^{{\rm tot}}$ (MeV) & Large branchings \\
\hline
$\Lambda$ & $3/2^-$ \ (11)  & 1850     &   32  & 130  & $\Sigma \pi$, $\Sigma^{\ast} \pi$  \\
& & & & & \\
$\Sigma$ & $1/2^+$ \ (21)  & 1760     &   10  &  & $\Sigma \pi$, $\Sigma^{\ast} \pi$  \\
$\Sigma$ & $3/2^-$ \ (13)  & 1850     &   70   & & $\Lambda \pi$, $N \overline{K}$, $\Sigma^{\ast} \pi$  \\
$\Sigma$ & $1/2^+$ \ (19)  & 2060     &   75   & &  $\Lambda \pi$, $N \overline{K}$, $\Sigma \pi$  \\
& & & & & \\
$\Xi$ & $1/2^-$ \ (9)  & 1620-1725     &   115   & &  $\Xi \pi$, $\Lambda \overline{K}$  \\
$\Xi$ & $1/2^-$ \ (14)  & 1860-1915     &   135  & 220 & $\Sigma \overline{K}$, 
 $\Lambda \overline{K}$, $\Xi \eta$ \\
$\Xi$ & $1/2^-$ \ (10)  & 1900     &   33   & $50-60$& $\Xi \pi$, 
$\Sigma \overline{K}$, $\Lambda \overline{K}$, $\Xi^{\ast}\pi$ \\
$\Xi$ & $3/2^+$ \ (5)  & 1900     &   95  &  & $\Xi \pi$, 
$\Lambda \overline{K}$, $\Xi^{\ast}\pi$ \\
$\Xi$ & $1/2^+$ \ (4)  & 1950     &   50   & & $\Sigma \overline{K}$, $\Xi^{\ast}\pi$ \\
$\Xi$ & $3/2^-$ \ (13)  & 2000     &   130 &  & $\Sigma^{\ast} \overline{K}$,
$\Xi^{\ast}\pi$, 
$\Lambda \overline{K}$, $\Xi \pi$ \\
$\Xi$ & $3/2^+$ \ (17)  & 2035     &   15   & & $\Sigma \overline{K}$ \\
$\Xi$ & $3/2^-$ \ (11)  & 2045     &   66   & 264 &  $\Xi \pi$, $\Lambda \overline{K}$ \\
$\Xi$ & $1/2^+$ \ (19)  & 2210     &   85  & & $\Lambda \overline{K}$, $\Xi \pi$,  $\Sigma \overline{K}$ \\
$\Xi$ & $3/2^+$ \ (16)  & 2240     &   35  & $100-150$ &  $\Lambda \overline{K}$, $\Xi \pi$,  $\Sigma \overline{K}$ \\
& & & & & \\
$\Omega$ & $1/2^-$ \ (10)  & 2050     &   35  & $\leq 50-60$ & $\Xi \overline{K}$ \\
$\Omega$ & $3/2^+$ \ (5)  & 2050     &   70  & &  $\Xi \overline{K}$, $\Xi^{\ast} \overline{K}$  \\
$\Omega$ & $3/2^-$ \ (13)  & 2150     &   110  & & $\Xi \overline{K}$, $\Xi^{\ast} \overline{K}$  \\
$\Omega$ &  $1/2^+$ \ (19)  & 2360     &   90  & & $\Xi \overline{K}$ \\
\hline
\end{tabular}
\caption{Predicted baryons.}
\label{table:pred}
\end{center}
\end{table}

In Table~\ref{table:pred}, the most remarkable prediction is the existence of the 
$\Lambda$ hyperon with $J^P=3/2^-$, the
mass around 1850 MeV, $\Gamma^{{\rm 2-body}} \approx 32$ MeV, 
$\Gamma^{{\rm tot}} \approx 130$ MeV
and very small coupling to the $N \overline{K}$ state. 
This is the only missing $\Lambda$ resonance, which is needed
to complete octet 11 -- all other eleven $\Lambda$ hyperons, see
Table~\ref{table:before}, are known very well and have three and four-star
ratings in the RPP. Note that the existence of a $\Lambda$ hyperon with $J^P=3/2^-$
in the 1775-1880 mass range,
which almost decouples from the $N \overline{K}$ state, is also predicted in the
constituent quark model~\cite{Isgur:1978xj,Loring:2001ky,Glozman:1997ag}.
Our analysis suggests that the missing $\Lambda$ baryon can be searched for in 
production reactions by studying 
 the $\Sigma \pi$ and $\Sigma(1385) \pi$
invariant mass spectra.

As can be seen from Table~\ref{table:pred}, we predict the existence of ten new $\Xi$ baryons.
In this respect, one should mention the recent interest in double-strangeness baryon 
spectroscopy~\cite{Xi:workshop}.

In addition to twenty multiplets of 
Table~\ref{table:before}, we have an additional twenty-first multiplet in 
Table~\ref{table:final} -- the antidecuplet.
While the existence of the antidecuplet is under debate, 
we assumed that the antidecuplet does
exist, has $J^P=1/2^+$ and contains the $\Theta^+(1540)$ and $\Xi^{--}(1862)$ as its lightest and
heaviest states. The nucleon-like member of ${\bf \at}$, $N_{\at}$, is identified with the
new nucleon resonance at 1670 MeV seen by GRAAL~\cite{Kuznetsov:2004gy} and
 predicted by the PWA analysis
of pion-proton scattering~\cite{Arndt:2003ga}. The mass of the $\Sigma$-like member of
${\bf \at}$, $\Sigma_{\at}$, is predicted using the equal spacing rule, 
$m_{\Sigma_{\at}} \approx 1760$ MeV.

Using the scarce information on the antidecuplet decays, we showed 
that $N_{\at}$ must mix with the $N$ member of another multiplet. 
We examine the scenario that the 
$N_{\at}$ and $\Sigma_{\at}$ members of the antidecuplet mix with $N(1440)$ and $\Sigma(1660)$ of octet 3. We showed that this can accommodate in a simple way all experimental information
 on the antidecuplet decays. Prediction for the unmeasured ${\bf \at}$ decays were made.

One should not have the impression that our SU(3) analysis of two-body baryon decays was
flawless. In several cases, we had to increase experimental errors by hand in order to
claim that the $\chi^2$ fit was successful.  
Also, we encountered three cases, when SU(3) predicted incorrectly the sign of the interference
observables. These are $\sqrt{\Gamma_{N \overline{K}}\Gamma_{\Sigma \pi}}$ for $\Sigma(2070)$ 
(decuplet 18),  $\sqrt{\Gamma_{N \overline{K}}\Gamma_{\Sigma \pi}}$ for $\Sigma(1940)$ 
(octet 11) and $\sqrt{\Gamma_{N \overline{K}}\Gamma_{\Lambda \pi}}$ for $\Sigma(2080)$
 (decuplet 16).

\bigskip
\bigskip

{\bf \large Acknowledgments}

We thank Ya.I. Azimov, P. Pobilitsa, Fl. Stancu and
I. Strakovsky for useful discussions.
This work is supported by the Sofia Kovalevskaya Program of the Alexander
 von Humboldt Foundation.
 \newpage
\bibliographystyle{h-elsevier}
\bibliography{su3_report}

\end{document}